\documentclass[preprint,11pt]{aastex}

\shorttitle{Perseus Disks}
\shortauthors{Tobin et al.}

\newcommand{\kms}{\mbox{km s$^{-1}$}}

\begin{document}

\title{A Sub-arcsecond Survey Toward Class 0 Protostars in Perseus: Searching for Signatures of Protostellar Disks}
\author{John J. Tobin\altaffilmark{1,2,11}, Leslie W. Looney\altaffilmark{3}, David J. Wilner\altaffilmark{4},
Woojin Kwon\altaffilmark{5}, Claire J. Chandler\altaffilmark{6}, Tyler L. Bourke\altaffilmark{7,4}, Laurent Loinard\altaffilmark{8}, 
Hsin-Fang Chiang\altaffilmark{9}, Scott Schnee\altaffilmark{1}, Xuepeng Chen\altaffilmark{10}}

\altaffiltext{1}{National Radio Astronomy Observatory, Charlottesville, VA 22903}
\altaffiltext{2}{Leiden Observatory, Leiden University, P.O. Box 9513, 2300-RA Leiden, The Netherlands; tobin@strw.leidenuniv.nl}
\altaffiltext{3}{Department of Astronomy, University of Illinois, Urbana, IL 61801}
\altaffiltext{4}{Harvard-Smithsonian Center for Astrophysics, Cambridge, MA 02138}
\altaffiltext{5}{SRON Netherlands Institute for Space Research, Landleven 12, 9747 AD, Groningen, The Netherlands}
\altaffiltext{6}{National Radio Astronomy Observatory, Socorro, NM 87801}
\altaffiltext{7}{Square Kilometer Array, Jodrell Bank, UK}
\altaffiltext{8}{Centro de Radioastronom{\'\i}a y Astrof{\'\i}sica, UNAM, Apartado Postal 3-72 (Xangari), 58089 Morelia, Michoac\'an, M\'exico}
\altaffiltext{9}{Institute for Astronomy and NASA Astrobiology Institute, University of Hawaii at Manoa, Hilo, HI 96720}
\altaffiltext{10}{Purple Mountain Observatory, Chinese Academy of Sciences, 2 West Beijing Road, Nanjing 210008, China}
\altaffiltext{11}{Hubble Fellow}

\begin{abstract}
We present a CARMA 1.3 mm continuum survey toward 9 Class 0 protostars 
in the Perseus molecular cloud at $\sim$0\farcs3 (70 AU)
resolution. This study approximately doubles the number of Class 0 protostars
observed with spatial resolutions $<$ 100 AU at millimeter wavelengths, enabling the 
presence of protostellar disks and proto-binary systems to be probed.
We detect flattened structures with radii $>$ 100 AU 
around 2 sources (L1448 IRS2 and Per-emb-14) and these sources may be strong disk candidates. 
Marginally-resolved structures with position angles within 30\degr\ of
 perpendicular to the outflow are found toward 3 protostars 
(L1448 IRS3C, IRAS 03282+3035, and L1448C) and are considered
disk candidates. Two others (L1448 IRS3B and IRAS 03292+3039) have resolved
structure, possibly indicative of massive inner envelopes or disks; 
L1448 IRS3B also has a companion separated by 0\farcs9 ($\sim$210 AU).
IC348-MMS does not have well-resolved structure
and the candidate first hydrostatic core L1451-MMS is marginally resolved on 1\arcsec\ 
scales. The strong disk candidate sources were followed-up 
with C$^{18}$O ($J=2\rightarrow1$) observations, detecting velocity 
gradients consistent with rotation, but it is unclear if the rotation
is Keplerian. We compare the observed visibility 
amplitudes to radiative transfer models, finding that
visibility amplitude ratios suggest a compact component 
(possibly a disk) is necessary for 5 of 9 Class 0 sources;
envelopes alone may explain the other 4 systems.
We conclude that there is evidence for the formation of large disks in the
Class 0 phase with a range of radii and masses dependent upon their 
initial formation conditions.

\end{abstract}


\section{Introduction}

Stars form due to the gravitational collapse of dense condensations within
molecular clouds and the star formation process is what ultimately leads
to the formation of solar systems and potentially life itself, as has transpired
in our Solar System. Therefore, the formation of the proto-planetary disk is 
a key step in the process of forming a planetary system. Moreover, disks may also play
a role in the formation of binary and multiple star systems via
gravitational instability \citep[e.g.,][]{kratter2010}; about one-third of all star systems and 
50\% of sun-like star systems are found in binary or multiple 
systems with a typical separation of $\sim$50 AU \citep{raghavan2010}. 

The formation of disks is thought to begin early in the star formation process, during
the Class 0 phase of protostellar evolution \citep{andre1993}. 
This is the earliest recognizable phase of the star formation process,
characterized by a protostar surrounded by a dense envelope of gas and dust.
The early formation of disks is inferred from the ubiquity of
outflows and jets during the protostellar phase \citep[e.g.][and references therein]{frank2014}.
Disks form during the collapse of the star forming cloud via conservation
of angular momentum \citep[e.g.][]{ulrich1976, cassen1981,tsc1984}; the angular momentum may derive from bulk
cloud rotation or the net angular momentum of residual turbulent motions.

Disk formation, however, may not be as simple if the protostellar
cloud is magnetized (even weakly) \citep{allen2003, mellon2008,li2013}.
The magnetic field will be dragged inward with the collapsing
material, increasing the local field strength.
The magnetic field then slows the rotation of the inner envelope by
transporting angular momentum to the larger-scale envelope \citep{allen2003,mellon2008}.
This `magnetic-braking' can slow the rotation efficiently enough to suppress disk formation 
and this result has become known as the `magnetic-braking catastrophe.' More recently,
a number of ways around the catastrophe have been found: reduction of magnetic field
strengths via Ohmic dissipation \citep{dapp2010}, misaligned rotation axes and magnetic field
directions \citep{joos2012}, turbulence \citep{seifried2013}, reconnection diffusion \citep{lazarian2012},
and/or reduced ionization (hence less field coupling) \citep{padovani2013}.

While the theoretical difficulty of disk formation is resolved somehow by nature, Class 0
disks remain poorly-characterized observationally. There have been efforts to detect
and resolve Class 0 disks with millimeter interferometers for the past 20 years. Possible
signatures of disks have been detected via dust continuum emission 
\citep[e.g.,][]{chandler1995,brown2000,looney2000,harvey2003,jorgensen2009}; however, these
pioneering efforts were not able to uniquely identify disks due to lack of
sensitivity and/or spatial resolution. Further muddying
the waters, \citet{maury2010} did not resolve structure consistent with disks having R $>$ 100 AU, using 
data from the Plateau de Bure Interferometer (PdBI) with $\sim$0\farcs5 resolution. However, the 
\citet{maury2010} sample was comprised  of two Very Low Luminosity Objects (VeLLOs) \citep[IRAM 04191 and L1521F, ][]{dunham2006,bourke2006} and
three typical Class 0 sources (L1527 IRS, L1448C, NGC 1333 IRAS2A).
Another high-resolution study (0\farcs3) with the Combined Array for Research in Millimeter-wave Astronomy
(CARMA) toward the Class 0 protostellar system L1157-mm also failed to resolve a disk \citep{chiang2012}. 

Following these studies, the Class 0 protostar L1527 IRS (also in the Maury et al. sample)
was found to have evidence for an edge-on disk in 3.8 \micron\ scattered light \citep{tobin2010b}.
Subsequent follow-up observations of this source led to the disk being resolved in the dust continuum 
as well as confirmation of rotational support \citep{tobin2012, tobin2013a}. 
The mass of the protostar was measured to be $\sim$0.2 $M_{\sun}$,
about 20\% of the surrounding envelope mass. Observations of L1527 with the 
Atacama Large Millimeter/submillimeter Array (ALMA) now show
that the mass of the protostar is likely 0.3 $M_{\sun}$, with a rotationally-supported
radius of 54 AU \citep{ohashi2014}. \citet{sakai2014a,sakai2014b} also found evidence for
a chemical change at the interface between the envelope and disk using new data from ALMA.
Evidence of rotationally-supported
Class 0 disks in other systems have now been found with ALMA \citep{murillo2013,lindberg2014,codella2014}.
Thus, the combined results of previous
studies point toward a diversity of disk properties in the Class 0 phase and that
large disks may form in some systems before the end of the Class 0 phase.

Despite these previous efforts, the number of Class 0 protostars observed with
spatial resolutions better than 100 AU is only $\sim$10 at millimeter and 
submillimeter wavelengths. To further
the characterization of Class 0 disks, we have observed a sample 
of 9 Class 0 protostars and 2 Class I protostars in the Perseus 
molecular cloud \citep[D $\sim$ 230 pc;][]{hirota2008,hirota2011} using the CARMA array at 1.3 mm
with a resolution of $\sim$0\farcs3 (70 AU).
The Perseus molecular cloud was chosen for this study
because it is the nearest star forming region with a large number of Class 0 protostars;
the more nearby Taurus and Ophiuchus clouds do not have sufficiently large numbers
of Class 0 protostars from which to draw a meaningful sample.

In this paper, we present the results from our survey, examining the
dust continuum structures, visibility amplitudes, and
compare to radiative transfer models.
 We will also present molecular line follow-up obtained toward some
sources using CARMA and the Submillimeter Array (SMA) as well as the
CO ($J=2\rightarrow1$) outflow emission when detected.
We present the sample, observations, and data reduction in Section 2,
the continuum and molecular line results are described in Section 3, and 
a comparison to radiative transfer models 
is presented in Section 4. The results are discussed in Section 5 and 
we present our summary and conclusions in Section 6.

\section{Observations}

\subsection{Sample Selection}

We have drawn our sample of nine Class 0 sources (Table 1) from the known protostars in the Perseus molecular
clouds based on the various infrared and millimeter studies that have been 
conducted \citep[e.g.,][]{hirano1999, looney2000, enoch2009, 
chen2010, pineda2011,schnee2012,sadavoy2014}. In addition, 
two Class I protostars were located adjacent to 
Class 0 sources and within the CARMA field of view.
\citet{enoch2009} classifies 27 sources
as Class 0 protostars; however, additional sources are now known from more 
recent millimeter observations (references above) and one source
 in the \citet{enoch2009} survey was incorrectly associated with IRAC emission (Per-emb-37). Including,
these additional sources brings the total number of Class 0 protostars in Perseus to $\sim$38.

We chose sources that were not known to be close binaries ($<$ 1000 AU) and had clearly
defined outflow axes from CO observations or \textit{Spitzer} IRAC 
imaging \citep[e.g.][]{jorgensen2006}.
We also focused on sources that had not been previously observed with sensitive,
high-resolution observations, aiming to increase the number of sources
with high resolution observations. Good knowledge of the outflow axis is important 
because this is a proxy for the current angular momentum vector of the system. 
If a sufficiently large disk is present, it is expected to be elongated 
perpendicular to the outflow direction.

The Class 0 sources range in luminosity from 0.7 L$_{\sun}$ to 9.2 L$_{\sun}$ \citep{enoch2009, sadavoy2014}
and the candidate FHSC L1451-MMS has a luminosity less than 0.05 L$_{\sun}$ \citep{pineda2011}. 
This range of luminosity is representative of the distribution of 
protostellar luminosities in Perseus, with only a few systems having luminosities in excess of 10 L$_{\sun}$.
The sources also sample a range of environments: wide binary systems, isolated cores within the cloud,
members of the NGC1333 cluster, the outskirts of the older IC348 cluster,
and the moderately active star forming cloud L1448. This sample enables us to begin more thoroughly
exploring the presence of disks from their dust continuum emission.

\subsection{CARMA Observations and Data Reduction}

CARMA is a heterogeneous interferometer array located in the Inyo mountains of California. It is
 comprised of 6 $\times$ 10.4 m antennas, 9 $\times$ 6.1 m antennas, making up the main array, and 
8 $\times$ 3.5 m antennas that are operated as an independent array 
or as part of the CARMA23 array. 
The observations were conducted in B and C configurations of the main array
between December 2012 and November 2013, see Table 2. B configuration has baseline lengths up to $\sim$1 km
and C configuration has baseline lengths up to $\sim$ 0.35 km. The respective
angular resolutions of the configurations are $\sim$0\farcs3 and $\sim$0\farcs9. Three science
targets were observed in each track, including a test source 0326+277 
to verify the millimeter seeing in B configuration.

The observations were conducted in a standard loop, bracketing the three science pointings (and test source in B-configuration)
with an observation of the phase calibrator (0237+288). Flux (Uranus) and bandpass (3C84) 
calibrators were observed at the beginning of each track. The C-PACS system had been 
used in previous extended configurations for correction of rapid phase variations \citep{perez2010}, but
the system was not available in the 2012 and 2013 observing seasons. However, the
data were taken during exceptionally good weather conditions and standard calibration
methods produced excellent data. The optical depth at 1.3 mm was typically $\sim$ 0.1
during B configuration and the root-mean-squared (RMS) path length (phase noise) was less than 100 \micron\ 
as measured by the phase monitor (see Table 2); 100 \micron\ RMS path length corresponds to $\sim$0\farcs34 seeing, 
approximately the angular resolution provided by B-configuration at 1.3 mm.

The central frequency of our observations was 225.0491 GHz (1.3 mm) and 
the correlator was configured for single-polarization observations with six
500 MHz sub-bands set for continuum observation, yielding 6 GHz 
of continuum bandwidth (dual-side band). The remaining two spectral windows were configured for
spectral line observation with 31 MHz bandwidth and were set to observe
the ($J=2\rightarrow1$) transitions of $^{12}$CO, $^{13}$CO, and C$^{18}$O; the individual channels
were 97 kHz corresponding to $\sim$0.13 \kms\ velocity resolution.
Follow-up observations of Per-emb-14 in November 2013 used dual-polarization mode in order to
have higher spectral line sensitivity.

We reduced and edited the visibility data using standard methods within 
the MIRIAD software package \citep{sault1995}; see \citet{tobin2013a} or \citet{chiang2012} for more details. 
The absolute flux calibration error is estimated to be $\sim$20\%.
The data were imaged using the CLEAN algorithm using
\textit{robust} weighting to optimize the angular resolution and sensitivity.
In order to ensure that our data reduction results in reliable images, we imaged
the test source 0326+277 for each B-configuration data set. Gaussian fits to the source 
demonstrate that it is consistent with a point source, verifying the good millimeter seeing.

\subsection{SMA Observations and Data Reduction}

The SMA is an 8-element interferometer array comprised of 6.1 m antennas and is 
located on Mauna Kea in Hawaii. We observed the source L1448 IRS2 from the CARMA sample with the SMA in the 
Extended configuration on two dates in September 2013 (Table 3). We observed in
two receiver mode with the low frequency receiver tuned to 225.434 GHz (1.3 mm) and
the high frequency receiver tuned to 351.135 GHz (850 \micron). In this mode, the correlator
provides 4 GHz of bandwidth (2 GHz in each of the upper and lower sidebands)
from each receiver. Each 2 GHz band is comprised of 24 sub-bands with a bandwidth of 104 MHz and our
setup had 64 channels per sub-band by default and we used high-resolution settings
to observe C$^{18}$O ($J=2\rightarrow1$) (1024 channels), 
$^{13}$CO ($J=2\rightarrow1$) (512 channels) and $^{12}$CO ($J=2\rightarrow1$) (512 channels)
with the low frequency receiver. With the high frequency receiver, we 
observed the H$^{13}$CO$^+$ and HCO$^+$ ($J=4\rightarrow3$) transitions with 512 channels.

The visibility data were edited and calibrated using the MIR software package, an IDL-based
software package originally developed for the Owens Valley Radio
Observatory and adapted by the SMA group. The error in the absolute flux calibration is
estimated to be $\sim$10\%. The calibrated visibilities were exported to
MIRIAD format for imaging. We detected all the CO ($J=2\rightarrow1$) isotopologues and HCO$^+$ ($J=4\rightarrow3$),
but we will only discuss the CO isotopologues and continuum results in this
paper.

\subsection{Spatial Scale Sensitivity}

The incomplete sampling of the uv-plane limits the maximum size of structures that can be
fully recovered with an interferometer. Moreover, the inclusion of only B and C or Extended and Compact
configurations of CARMA and the SMA, respectively, limits the sensitivity to spatially extended objects.
In order to characterize the sensitivity to different spatial scales, we generated a series of uniform brightness model images
of symmetric disks having a variety of diameters and sampled them with the uv-coverage of the same configurations
as our observations. We find that 50\% of the model flux density can be recovered at 1.3 mm from a 4\arcsec\
diameter disk in CARMA C-configuration, a 1\farcs75 disk for CARMA B-configuration, a 9\arcsec\ disk for SMA Compact configuration, and a 4\arcsec\ disk
for SMA Extended configuration. The SMA Extended configuration observations at 850 \micron\ can recover a 2\farcs7 disk.
Note that these sizes are optimistic because the uv-coverage in the simulations is not split between sources as in the observations
and some observed data are flagged. Furthermore, a uniform brightness disk is not representative of the structure of protostellar sources, but this
example is meant to present a limiting case.

\section{Results}

\subsection{Continuum Emission Structure}
We detect 1.3 mm dust continuum emission toward all observed sources in 
both B and C configurations\footnotetext{Per-emb-24 was only observed in C-array and was not observed in
B-array due to only having a $\sim$3$\sigma$ (3.2 mJy/beam) detection.}. The detected continuum 
structures broadly fit into four categories: disk-like structures (Section 3.1.1), 
candidate multiples (Section 3.1.2), featureless or unresolved (Section 3.1.3), and
asymmetric resolved structures (Section 3.1.4). Some of the continuum
data are highly structured in the combined B and C-array imaging, and others may only have resolved
structure in the highest resolution data (B-array) alone. 
Images of the Class 0 protostars with sensitivity to structure on 
arcsecond scales with the combined B and C-array datasets are shown in 
Figure \ref{sample-lr}. The highest resolution images,
emphasizing structure on sub-arcsecond scales, are shown in Figure \ref{sample-hr}.
Furthermore, two Class I protostars were within the primary beams of Class 0 sources and their images
are shown in Figure \ref{sample-CI}. The 1.3 mm flux densities, RMS noise, and resolution for each
image are given Table 4. 
In addition, we fit two-dimensional Gaussians to the images shown in Figures 1 and 2. 
This was done with the \textit{imfit} task in CASA\footnotetext{CASA stands for Common Astronomy Software Application \citep{casa}
and can be obtained from http://casa.nrao.edu} to determine the deconvolved source sizes and position angles, these values are
listed in Table 4.

For a source to be classified as `disk-like,' it must have a deconvolved position angle within 30\degr\ of
orthogonal to the outflow axis; the relative position angles of outflow and continuum sources are given 
in Table 4. Disks are expected to be aligned orthogonal to the outflow
directions, given that the protostar and disk together are thought to be responsible for
outflow generation \citep{frank2014}. We have adopted a somewhat loose limit for the relative
position angles for several reasons. First, outflow position angles are generally defined by eye
and asymmetries in the outflow emission can lead to incorrect position angles. Second,
outflows can wander with time \citep[e.g., L1157-mm;][]{gueth1996} and have bends and
kinks \citep[e.g., L1448C;][]{hirano2010}, adding uncertainty to the outflow axis. Lastly,
the continuum emission from the protostellar envelope is entangled with that of the disk, potentially
adding systematic uncertainty to the position angle of the continuum.

\subsubsection{Disk-like Structures}

Two sources in our sample show strong evidence of disk-like structures in the dust continuum: 
L1448 IRS2 (Figures \ref{sample-hr} and \ref{IRS2-SMA}) and Per-emb-14 (Figure \ref{sample-hr}).
They are well-resolved in the dust continuum and have deconvolved position angles within 30\degr\ of
being orthogonal to the outflow. 
The outflow directions of these sources are listed in Table 1 and known from both \textit{Spitzer}
IRAC imaging and/or CO outflow mapping. 

\begin{figure}
\begin{center}
\includegraphics[scale=0.9,trim=0.5cm 0.5cm 0.5cm 9cm, clip=true]{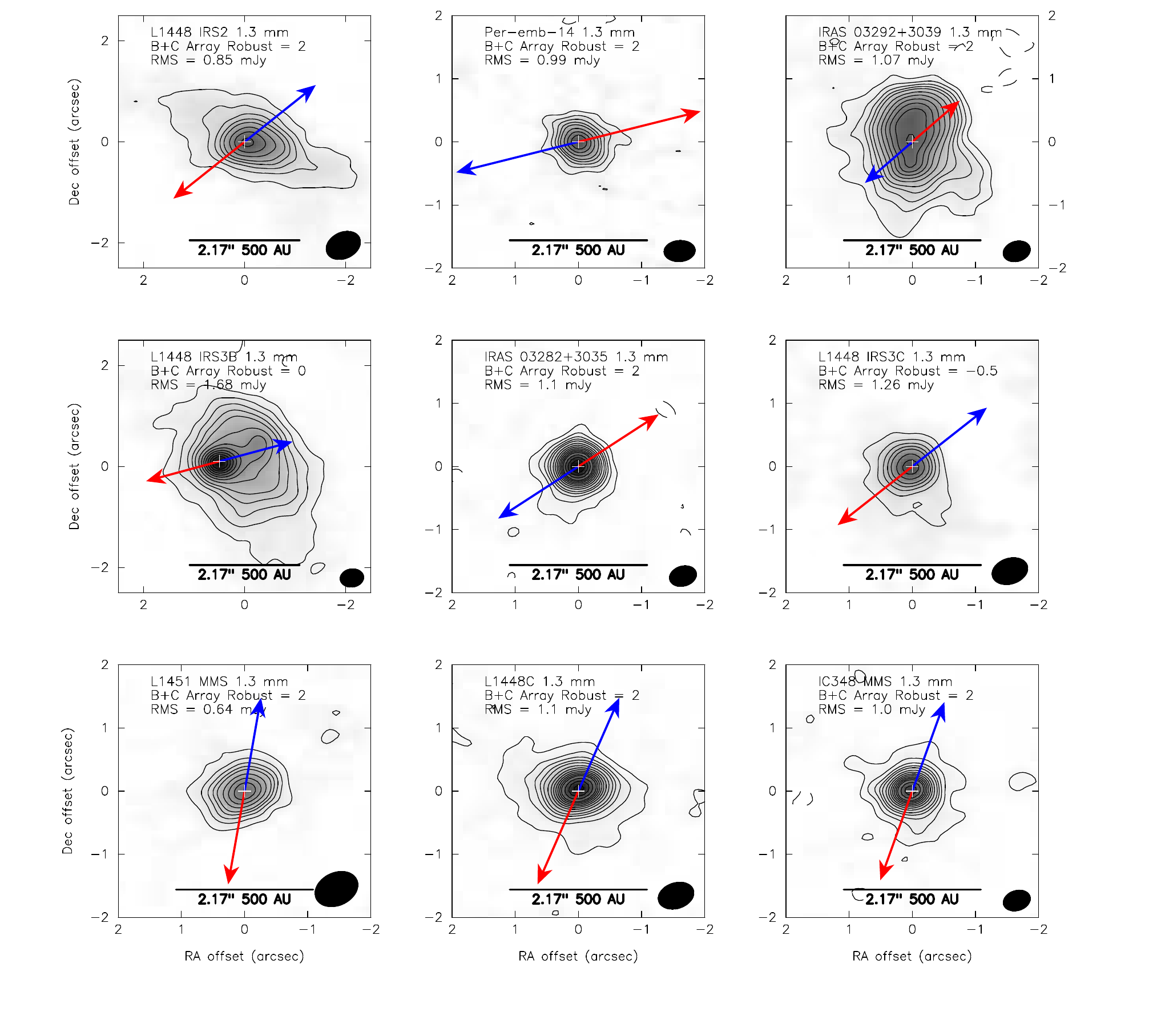}
\end{center}
\caption{Continuum images at 1.3 mm of the full sample, emphasizing the structure of
the sources on the scale of a few arcseconds. Images are ordered from top to bottom 
by their amount of resolved structure; the most resolved structures are at the top. 
The contours in all images are [-3, 3, 6, 9, 12, 15, 20, 25, 30, 35, 40, 45, 50, 60, ...] $\times$ $\sigma$, where
$\sigma$ is denoted in each panel and in Table 4. The beam size is shown in the lower right corner
of each image and exact beam sizes are given in Table 4. The blue and red arrows denote the direction of the blueshifted
and redshifted outflows, respectively.}
\label{sample-lr}
\end{figure}

\begin{figure}
\begin{center}
\includegraphics[scale=0.9,trim=0.5cm 0.5cm 0.5cm 9cm, clip=true]{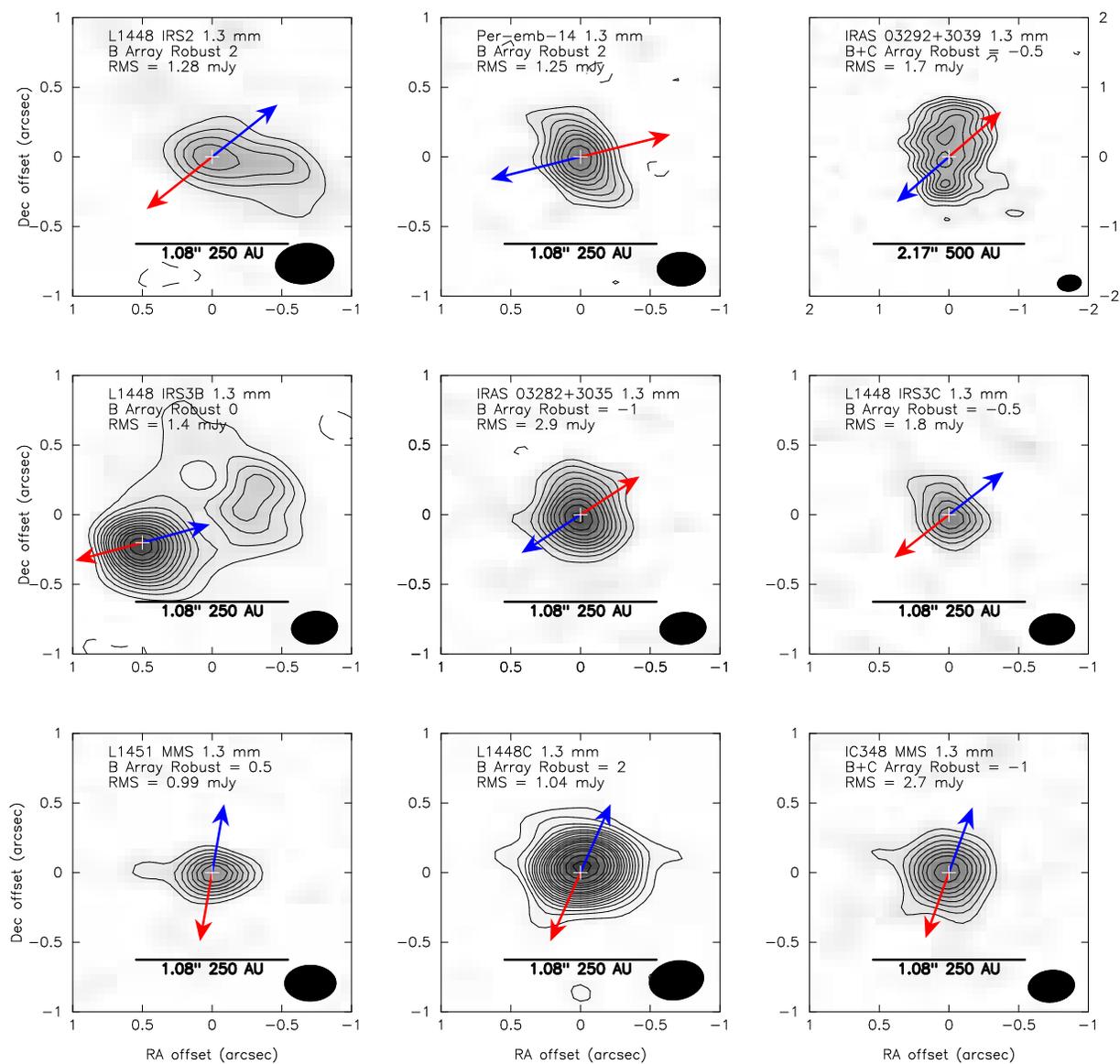}
\end{center}
\caption{Same as Figure 1, but with emphasis on sub-arcsecond structures. All plots are zoomed-in relative to Figure 1, except IRAS 03292+3039.}
\label{sample-hr}
\end{figure}

\begin{figure}
\begin{center}
\includegraphics[scale=0.9,trim=0.5cm 5cm 7cm 9cm, clip=true]{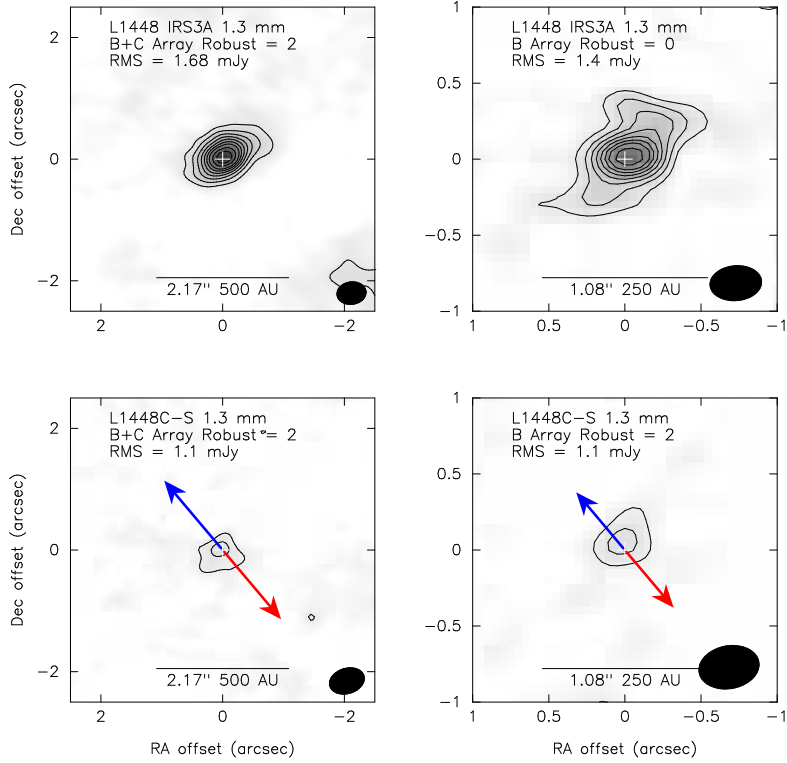}
\end{center}
\caption{Continuum images at 1.3 mm of the neighboring Class I sources that were within the primary beam:
 L1448 IRS3A (top panels) and L1448C-S (bottom panels). 
The contours in all images are [-3, 3, 6, 9, 12, 15, 20, 25, 30, 35, 40, 45, 50, 60, ...] 
$\times$ $\sigma$, where $\sigma$ is denoted in each panel and in Table 4. The beam 
size is shown in the lower right corner of each image and exact dimensions are given in Table 4.
The blue and red arrows denote the direction of the blueshifted
and redshifted outflows, respectively.}
\label{sample-CI}
\end{figure}

\begin{figure}
\begin{center}
\includegraphics[scale=0.7]{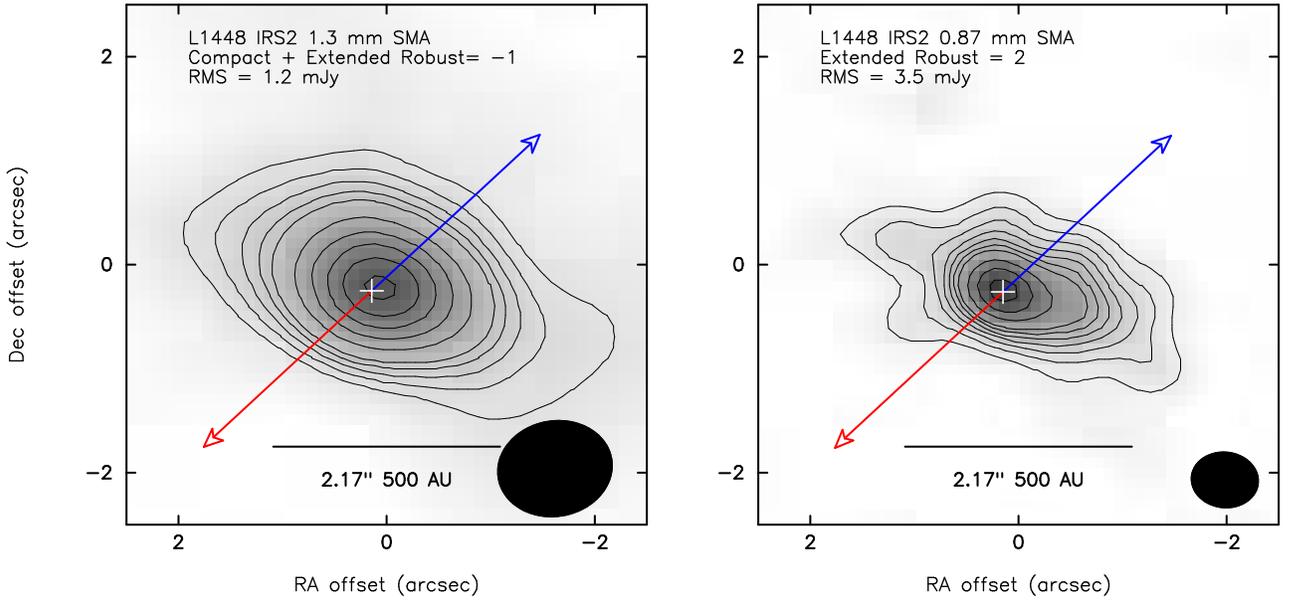}
\end{center}
\caption{L1448 IRS2 images at 1.3 mm (Extended \& Compact) and 
850 $\micron$ (Extended-only) from the SMA. The 1.3 mm image traces similar structure
as the CARMA 1.3 mm images in Figures 1 and 2. The 850 \micron\
image at higher resolution still traces larger-scale structure, but does show the asymmetry
evident in the higher resolution 1.3 mm image. The contours in the 1.3 mm images start at 7$\sigma$, increasing 
by 2$\sigma$ until 15$\sigma$ at which point they increase by 5$\sigma$; the 850 \micron\
contours increase in the same sense, except that they start at 3$\sigma$. At 1.3 mm  and 850 \micron\
$\sigma$ = 1.2 mJy and 3.5 mJy, respectively. The beam sizes are 1.11\arcsec\ $\times$ 0.92\arcsec\ 
and 0\farcs65 $\times$ 0\farcs54 at 1.3 mm and 850 \micron, respectively. 
The blue and red arrows denote the direction of the blueshifted
and redshifted outflows, respectively.}
\label{IRS2-SMA}
\end{figure}

The disk-like structure toward L1448 IRS2 has a
radius of $\sim$300 AU 
and is at an angle of 71\degr\ with respect to the outflow 
in the B and C-array combined image (Figure \ref{sample-lr}).
The SMA Extended and Compact combined image (Figure \ref{IRS2-SMA}) also shows this structure. 
The B-array data alone do not appear
symmetric and the emission is only extended toward the southwest 
at an angle of 68\degr\ with respect to the outflow. 
This extension does not appear to trace a 
discrete source and the SMA Extended image at 850 \micron\ also exhibits this structure 
(Figure \ref{IRS2-SMA}).

Per-emb-14, also known as NGC 1333 IRAS 4C \citep{smith2000}, appears rather featureless in the C and B-array combined
image (Figure \ref{sample-lr}), but the B-array data alone reveal a compact extended structure 
at an angle of 70\degr\ with respect
to the outflow direction (Figure \ref{sample-hr}). The radius of this 
structure is $\sim$100 AU and resembles the expected structure of
an edge-on protostellar disk. An outflow has not been detected toward this source in CO \citep{curtis2010, plunkett2013}.
However, a fan-shaped scattered light nebula is obvious in \textit{Spitzer} maps as well as H$_2$ emission
knots extending to the east of the source position from which the outflow position angle is derived;
the IRAC imaging is shown in the Appendix. An edge-on orientation could
also explain the lack of a CO outflow detection because the projected velocities would be low and 
likely confused with the cloud emission.

In addition to Per-emb-14 and L1448IRS2, the sources L1448C, IRAS 03282+3035, and L1448 IRS3C 
have deconvolved position angles that are within 30\degr\ of orthogonal to the outflow. 
Therefore, these sources could also be disk candidates, but they are not as well-resolved
as Per-emb-14 and L1448IRS2; they will be discussed further in the following sections.

\subsubsection{Candidate Multiple Sources}
Our high-resolution data also reveal several candidate multiple sources with more than 
one continuum peak observed. Toward L1448 IRS3B 
we observe extended structure perpendicular to the
outflow in the combined B and C-array images as well as the presence of another continuum
peak. The B-array imaging alone reveals that there are at least two 
continuum sources present. The secondary source is located to the northwest by 
$\sim$0\farcs9. The secondary also appears to be marginally resolved, with a separation of
$\sim$0\farcs2, slightly less than the minor axis of the beam (0\farcs23).
The nature of the secondary is not immediately clear since it is almost exactly along the
axis of the outflow, but it may be a binary companion.
The deconvolved position angle in Table 4 for the low-resolution image of L1448 IRS3B does not reflect the 
emission that is extended orthogonal to the outflow because
the fitting routine attempts to fit both the main protostar and the companion
along the outflow direction.

IRAS 03292+3039 also has structured continuum emission that is
extended in the north-south
direction at an angle of 49\degr\ with respect to the outflow. 
There are no strong peaks standing out from this structure, 
even when imaging with higher resolution, but there are three weak sub-peaks 
separated from each other by $\sim$0\farcs4. 

\subsubsection{Smooth Continuum Structures}

The sources IC348-MMS, L1448C, L1451-MMS have rather smooth continuum structures. They all have resolved
envelopes extended over several resolution elements in the B and C combined imaging
as shown in Figure \ref{sample-lr}. The higher resolution imaging from the
B-array data alone does not reveal strong evidence of sub-structure and they
are all consistent with smooth source structure at this resolution (Figure \ref{sample-hr}). However,
the deconvolved position angle of L1448C has an angle of 83\degr\ with respect to the outflow, perhaps indicative
of a disk-like structure at the limit of our resolution.

\subsubsection{Asymmetric Resolved Structures}
Several sources show evidence of resolved structure at high-resolution, but these are not clearly
disk-like in appearance and are often only extended toward one side. L1448 IRS3C 
(also called L1448 NW \citep{terebey1997}) appears 
resolved 
at an angle of 99\degr\ with respect to outflow,
but only toward the north-east (Figure \ref{sample-hr}). 
IRAS 03282+3035 is also extended at a slight angle (109\degr) with 
respect to the outflow direction, but only toward the north (Figure \ref{sample-hr}). 
The Class I source L1448 IRS3A shows a roughly symmetric structure in the 
northwest and southeast directions (Figure \ref{sample-CI}). The outflow direction from this
source is uncertain \citep[see Appendix and ][]{kwon2006}, so this could be a disk-like 
structure, but it is uncertain without clear knowledge of the outflow direction.

The strongest asymmetric structure is found toward IRAS 03292+3039 (Figures \ref{sample-lr} and \ref{sample-hr}). 
This source was also identified as a candidate multiple, due to its multiple peaks. However, its extended source
structure stands out relative the other sources in the sample. We will
discuss the nature of this source further in Section 5.2.

\subsection{Known Multiple Systems}

Several systems in our sample were previously known to have wide components separated by more than 1\arcsec.
L1448C is known to have a companion denoted L1448C-S by \citet{jorgensen2006},
also detected by \citet{tobin2007} and \citet{chen2013}. 
IC348-MMS and IRAS 03282+3035 were found to have wide binaries detected
at millimeter wavelengths by \citet{chen2013}. Finally, L1448 IRS3 is comprised of three components 
\citep[IRS 3B, IRS 3A, IRS 3C or NW;][]{looney2000}.

We have confidently detected all components of the L1448 IRS3 system, and
we also find that L1448 IRS3B has a closer companion separated by only 0\farcs9.
We detect the Class I companion toward L1448C
with a flux density of $\sim$ 7 mJy, separated by 8\farcs1 ($\sim$1900AU).
However, we do not detect the companion toward IRAS 03282+3035, reported by \citet{chen2013}
with a separation of 1\farcs5. We should
have detected this source with a signal-to-noise ratio of 7.6. The non-detection
of the companion may indicate that it is not real and it could be a feature 
extended along the outflow cavity.

The wide binary ($\sim$15\arcsec) known as IC348-MMS2 is not detected because
it lies near the edge of the CARMA primary beam, but it does have further
detections at both submillimeter and far-infrared wavelengths \citep{palau2014}. In addition,
 \citet{rodriguez2014} detected IC348-MMS2 and
another source at 2~cm and 3.3~cm located $\sim$ 3\arcsec\ southwest of 
IC348-MMS, denoted JVLA~3a and IC348-MMS is JVLA~3b. JVLA~3a is undetected in our data and
if it were emitting as a typical protostar, it should produce significant 
millimeter-wave emission; JVLA~3a source is also brighter than IC348-MMS at cm wavelengths. 
The Figures shown by \citet{palau2014} 
indicate that JVLA~3a may be coincident with the origin of the outflow emission at
4.5 \micron\ as well as the \textit{Spitzer} 24 \micron\ source, while IC348-MMS (JVLA~3b) appears slightly
offset.

\subsection{Molecular Line Observations}
We attempted to observe molecular line emission in the course of all the continuum observations, but
in nearly all cases the $^{13}$CO and C$^{18}$O emission was not strongly detected
due to observing 3 sources per track.
Therefore, we obtained follow-up observations of the strong disk candidates L1448 IRS2
and Per-emb-14 with longer integration time in order to characterize their molecular line kinematics in the 
$^{13}$CO and C$^{18}$O ($J=2\rightarrow1$) transitions.
The sources in L1448 IRS2, L1448C, L1448 IRS2, L1451-MMS, and the three sources in L1448 IRS3 did have 
detections of their outflows in $^{12}$CO emission (in addition to $^{13}$CO for L1448 IRS2)
and we discuss these observations in the Appendix. 

\subsubsection{L1448 IRS2}

\begin{figure}
\begin{center}
\includegraphics[scale=0.65,trim=0.5cm 5cm 3cm 5cm, clip=true]{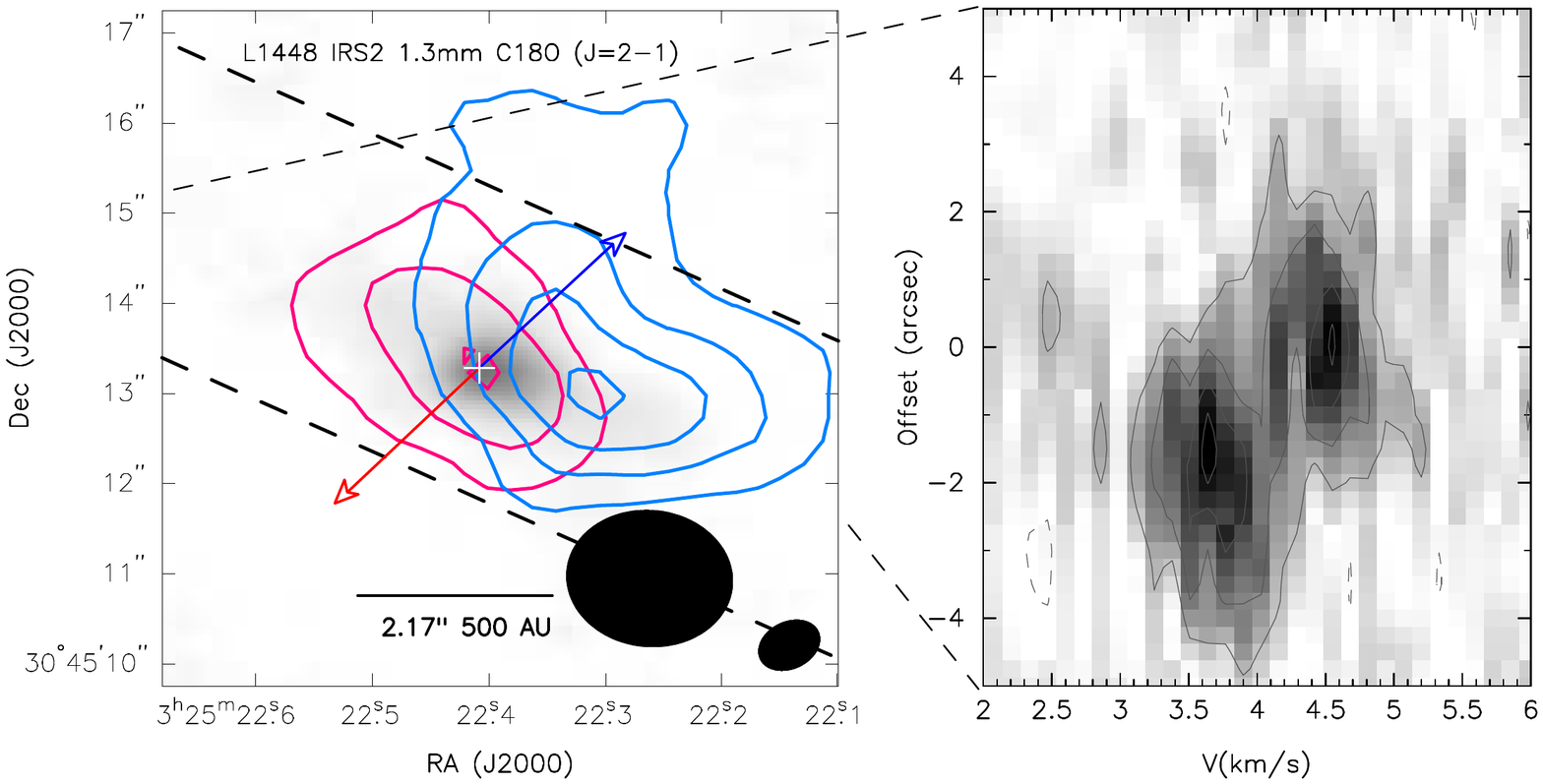}
\includegraphics[scale=0.65,trim=0.5cm 5cm 3cm 5cm, clip=true]{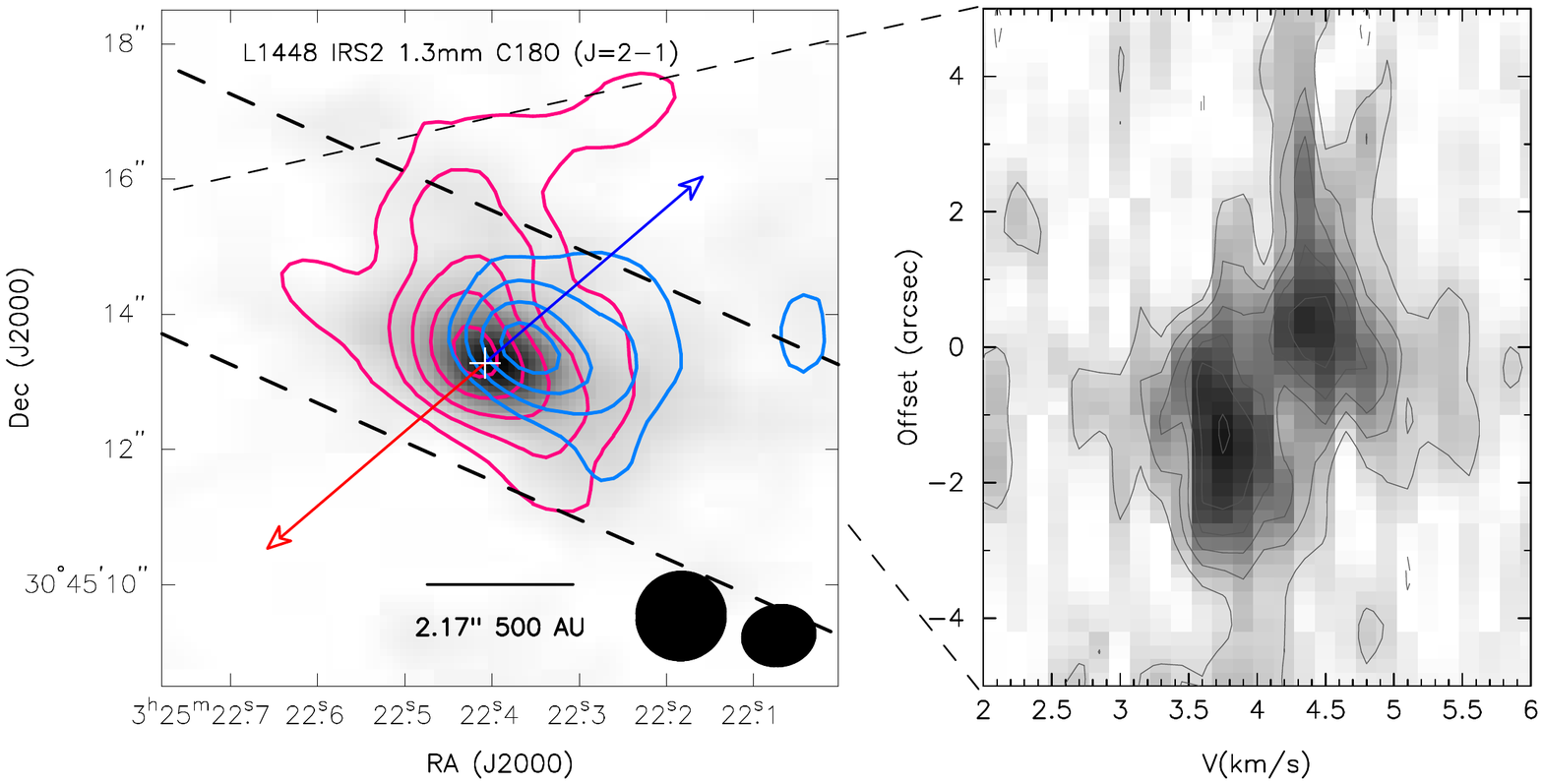}
\end{center}
\caption{Integrated intensity images of red and blue-shifted C$^{18}$O 
($J=2\rightarrow1$) emission overlaid on 1.3 mm dust continuum imaging and
position-velocity diagrams taken from the region enclosed by the parallel dashed lines (right). 
The top panels are from CARMA C and B-array and SMA C$^{18}$O Compact 
data, with 0\farcs5 tapering applied; the lower resolution data show a
clear velocity gradient on 5\arcsec ($\sim$1200 AU scales) at an angle of 70\degr\ with respect 
to the outflow position angle. The bottom
panels show the higher resolution SMA Compact and Extended array observations of 
C$^{18}$O; the direction of the velocity gradient changes to have an angle of only 30\degr\ with
respect to the outflow position angle on scales less than 2\arcsec\ ($\sim$500 AU).
The contours in the CARMA and SMA imaging (top) are [-6, 6, 9, ...]$\times$ $\sigma$
where sigma is 0.46 K and 0.48 K for the red and blue shifted emission respectively. 
The contours for the SMA-only imaging are [-5, 9, 13, 17]$\times$ $\sigma$
where sigma is 0.68 K and 0.62 K for the red and blue shifted emission respectively.
The larger beams are for the line data and are 1\farcs82 $\times$ 1\farcs48 
and 1\farcs31 $\times$ 1\farcs29 for the low and high resolution images 
respectively; the smaller beams are for the continuum data, the same as 
in Figures 1 and \ref{IRS2-SMA}. The blue and red arrows denote the direction of the blueshifted
and redshifted outflows, respectively.
}
\label{IRS2-kinematics}
\end{figure}

The presence of the large continuum structure toward L1448 IRS2 
is suggestive of
a disk, but molecular line observations are necessary to characterize its kinematics
and determine its nature. 
The C$^{18}$O ($J=2\rightarrow1$) integrated intensity maps and 
position-velocity diagrams are shown in Figure \ref{IRS2-kinematics}.
The lower resolution image (top panels) is a combination of SMA Compact and CARMA C-array observations,
showing a velocity gradient
 on $\sim$5\arcsec\ (1200 AU) scales in both the blueshifted and redshifted
integrated intensity and position-velocity plot. 
The position angle between the peak blueshifted and redshifted
emission is 70\degr\ east of north, making a 68\degr\ angle with respect to the outflow. Therefore, this velocity
gradient could be interpreted as rotation.

The SMA Compact and Extended
map is shown in the bottom panel of Figure \ref{IRS2-kinematics} 
with the respective position-velocity plot. The addition of the sensitive
Extended array data shows that the velocity gradient direction changes 
to a position angle of 100\degr\ east of north and
is now more in the direction of the outflow on $\sim$2\arcsec\ scales. There
are several possibilities that could produce a velocity gradient along the outflow
that we will discuss further in Section 5.2. Therefore, 
we cannot confirm rotation on scales less than $\sim$300 AU toward L1448 IRS2.

\subsubsection{Per-emb-14}
\begin{figure}
\begin{center}
\includegraphics[scale=0.75]{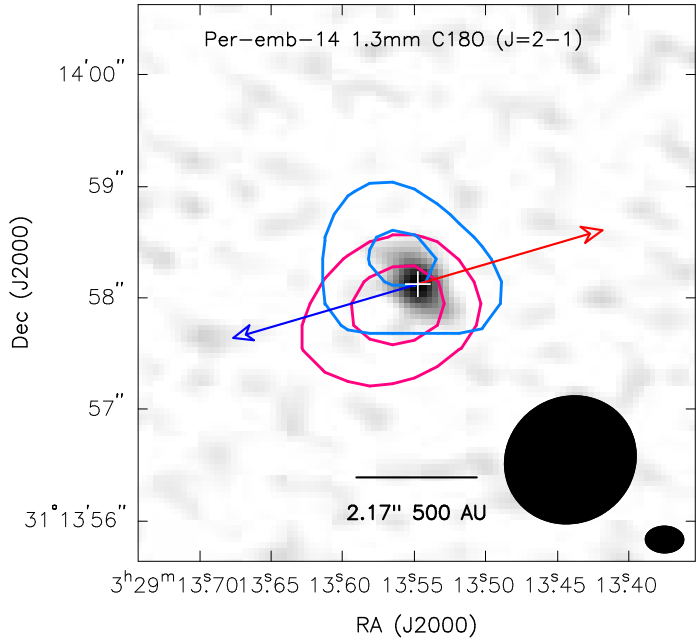}
\end{center}
\caption{Per-emb-14 (NGC 1333 IRAS4C) with CARMA C$^{18}$O overlaid on the B and C-array dust continuum
image. The integrated intensity of the
red and blue-shifted emission is suggestive of rotation on the scale of the disk candidate,
but the signal-to-noise is far too low for us to determine if the rotation is consistent with 
Keplerian. The contours are $\pm$3$\sigma$ and 5$\sigma$, where $\sigma$ = 1.74 K and 1.35 K
for the red and blue-shifted maps respectively. The larger beam drawn in the lower right 
corresponds to the C$^{18}$O data and is 1\farcs19 $\times$ 1\farcs11; the smaller beam is for the continuum
data and is the same as in Figure 2. The blue and red arrows denote the direction of the blueshifted
and redshifted outflows, respectively.
}
\label{peremb14-kinematics}
\end{figure}

We have also obtained higher sensitivity molecular line imaging toward 
Per-emb-14 (NGC 1333 IRAS4C) in the C$^{18}$O ($J=2\rightarrow1$) transition
with CARMA in C-array; $^{13}$CO ($J=2\rightarrow1$) was not detected.
The C$^{18}$O emission is weak, but we have made 
integrated intensity maps of the red and blue-shifted emission toward the source that are shown in  
Figure \ref{peremb14-kinematics}. The blue and red-shifted emission are offset from each other,
with the velocity gradient being perpendicular to the outflow and
consistent with the axis of the
resolved continuum source. The offset is only $\sim$0\farcs5, which is about
40\% of the beam width; however, the velocity centroid
accuracy is approximately equivalent to the beam width divided by the signal-to-noise
ratio \citep{reid1988}, 0\farcs25 in this case. 
While the gradient is most likely real,
the signal-to-noise is not high enough to determine if it is indicative of rotational support.

\subsection{Visibility Amplitude Profiles}

\begin{figure}
\begin{center}
\includegraphics[scale=0.28]{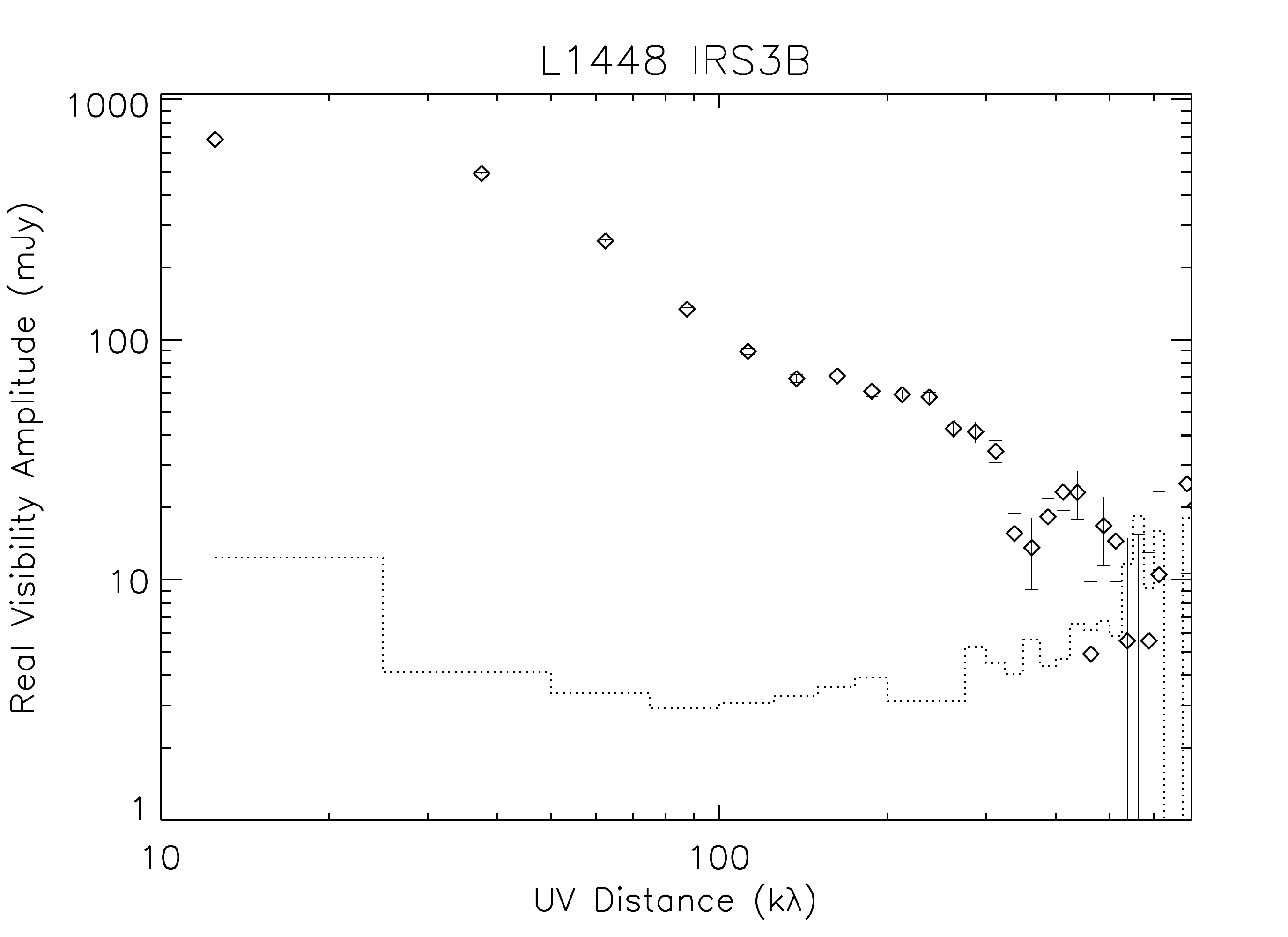}
\includegraphics[scale=0.28]{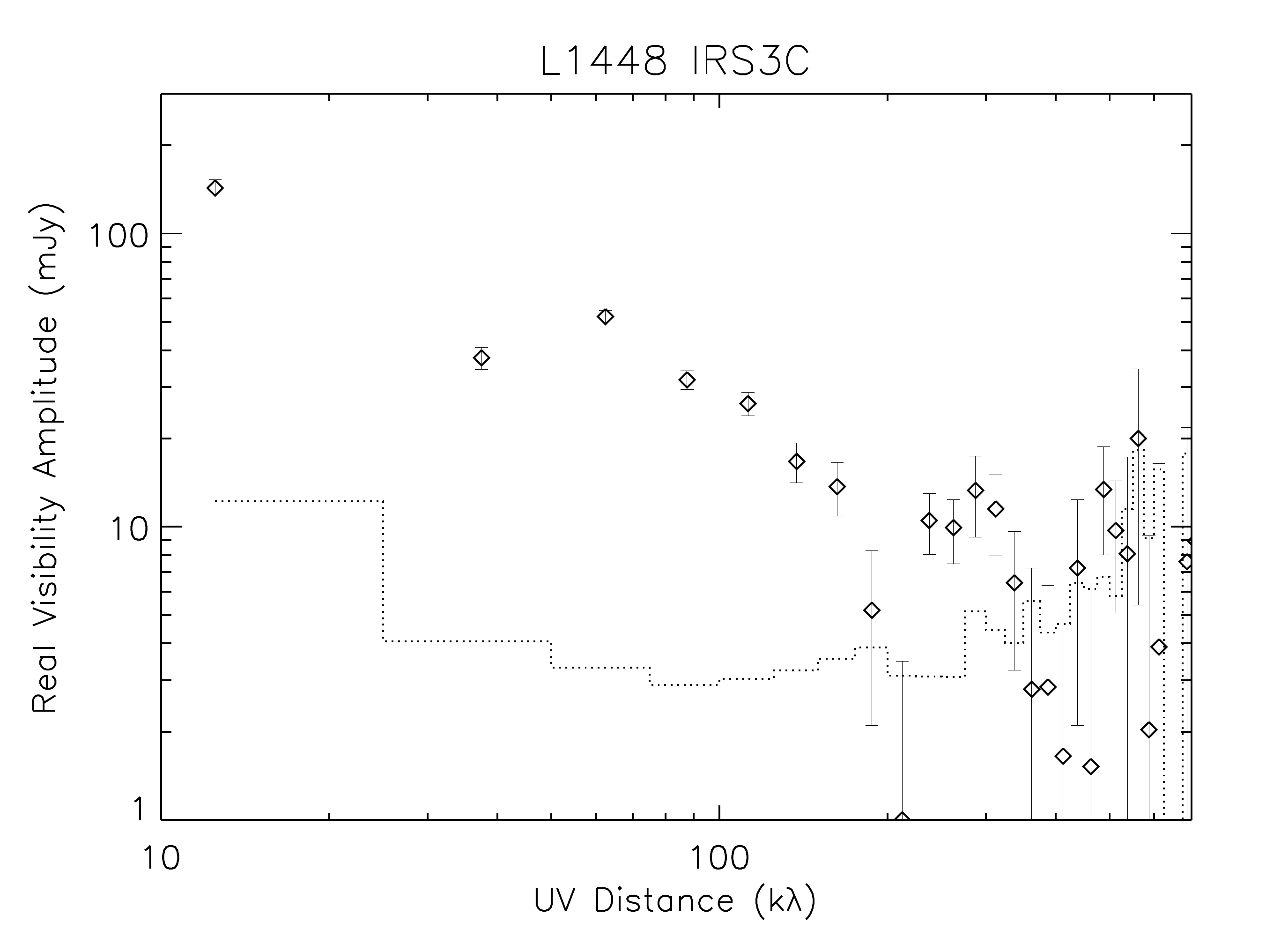}
\includegraphics[scale=0.28]{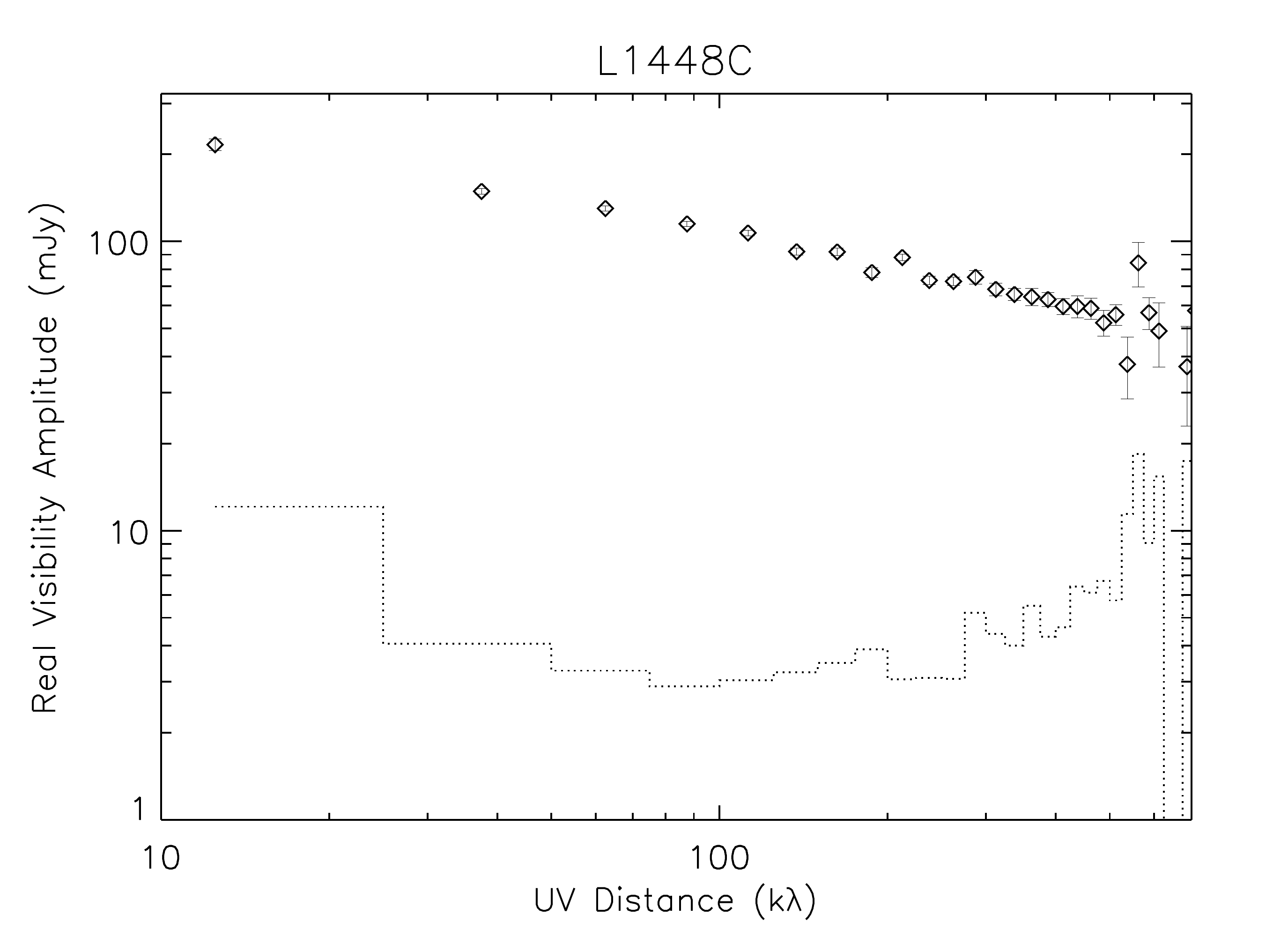}
\includegraphics[scale=0.28]{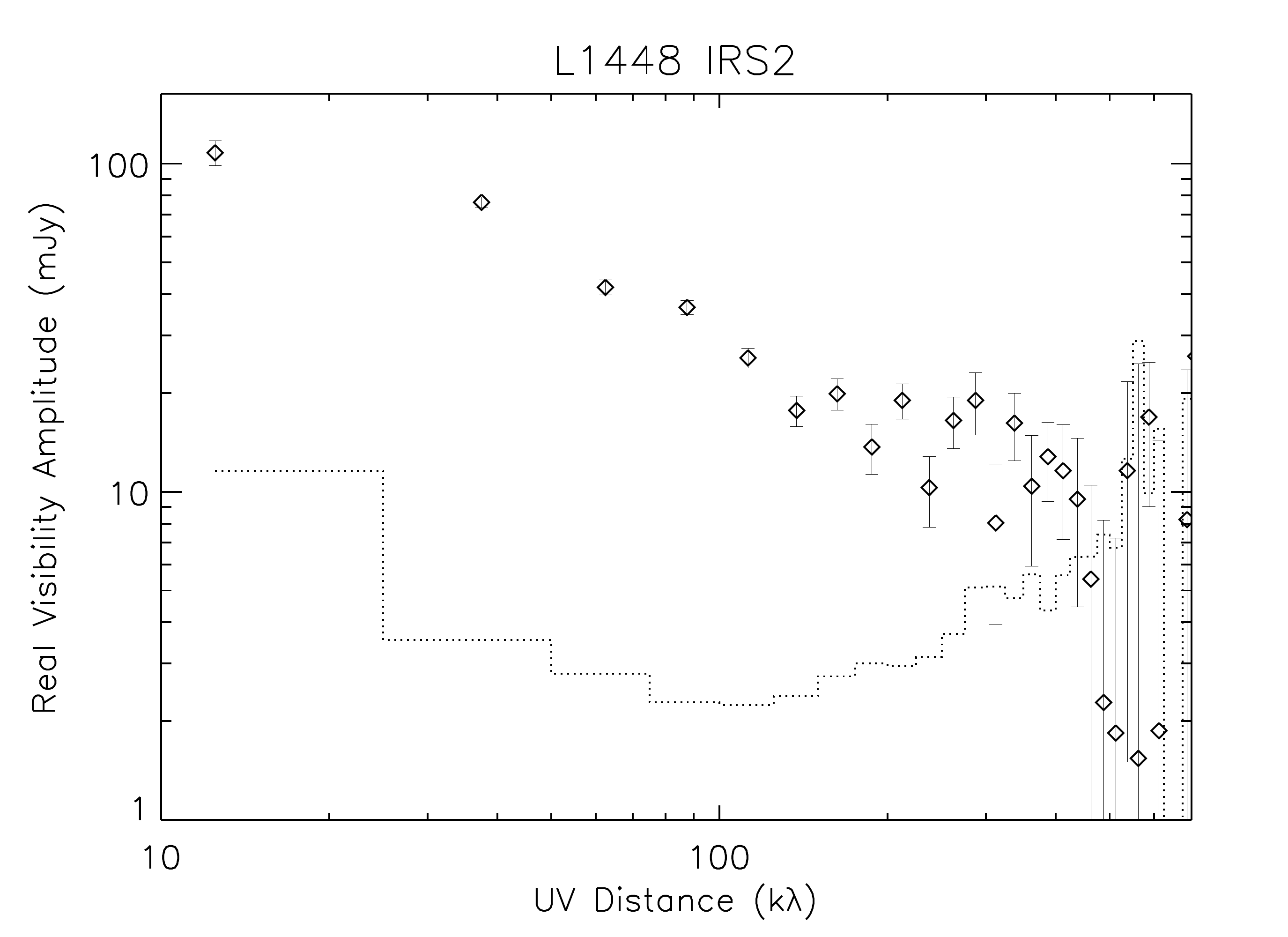}
\end{center}
\caption{Binned visibility amplitude plots versus projected uv-distance at 1.3 mm. The visibility data
are averaged within 25 k$\lambda$ bins. The sources show a variety of structures 
from flat visibility amplitudes toward L1451-MMS, power-law decline in L1448C,
high scatter for L1448 IRS2 indicating resolved structure, steep declines at a variety
of uv-distances (Per-emb-14, IRAS 03292+3039, IRAS 03282+3035), and evidence of multiplicity
L1448 IRS3B. The dotted line is the expected visibility amplitude from noise alone.
}
\label{uvamps}
\end{figure}

The well-sampled uv-plane provided by CARMA produces deconvolved images that 
reliably recover complex structure; however, the visibility amplitude data themselves still
provide details of source structure that are not immediately apparent in the
deconvolved images. The circularly averaged visibility amplitudes are plotted against projected uv-distance for each
source and are shown in Figure \ref{uvamps}, with uv-coverage out to $\sim$600 k$\lambda$.
These data show a variety of structures from roughly constant amplitude out to 600 k$\lambda$ (0\farcs33)
to very little emission on baselines longer than $\sim$200 k$\lambda$ (scales smaller than 1\arcsec).

L1448C and L1451-MMS exhibit the flattest visibility amplitudes. L1448C shows
a linear decline in the log-log plots; the linearity is evident in L1448C out to the longest baselines. 
L1451-MMS shows the least amount of visibility amplitude decline of all the observed sources.

Per-emb-14, IRAS 03282+3035, IC348-MMS, and IRAS 03292+3039 all decrease slowly until a scale specific
to each source and then drop quickly. The scales are consistent with the size of the resolved
structures apparent in the deconvolved images. L1448 IRS2 also had $\sim$4\arcsec\ diameter
flattened structure, but the visibility amplitudes decrease more slowly than IRAS 03292+3039
likely due to the structure being smaller in one dimension. L1448 IRS2 also shows variations 
at uv-distances $>$ 100 k$\lambda$, likely due to the resolved 
structure seen in the images on $>$1\arcsec\ scales.

L1448 IRS3B showed clear evidence of multiplicity in its deconvolved image (Figure \ref{sample-hr}).
The visibility amplitudes flatten at $>$100 k$\lambda$ and then
begin to drop at $\sim$250 k$\lambda$ (0\farcs8), the approximate separation of 
the candidate companion source. The visibility amplitude structures of L1448 IRS3C and L1448 IRS3A
are difficult to interpret due to their proximity to L1448 IRS3B; moreover, both sources show abrupt drops
in visibility amplitude at $\sim$200 k$\lambda$ and $\sim$ 400 k$\lambda$. The visibility data will be further interpreted
in conjunction with radiative transfer models in Section 4.

\subsection{Dust Continuum Masses}

The dust emission observed toward the protostars is directly proportional
to the mass of the emitting dust, provided that the dust opacity, the temperature
of the emitting dust, and the optical depth are known. At millimeter wavelengths
the dust opacities are low enough such that optically thin emission is a reasonable
assumption. However, the dust opacity is only known to within a factor of 3 to 5 
from models \citep[e.g.,][]{ossenkopf1994, ormel2011} and empirical 
estimates \citep[e.g.,][]{beckwith1990}. Moreover, there can be opacity variations from source
to source \citep{kwon2009, chiang2012} and radial opacity 
variations in envelopes \citep{kwon2009} and disks \citep{perez2012}. Since these sources all have protostars,
there will also be a temperature gradient in the envelope. However for simplicity, we will assume
a constant, average temperature of 30 K, appropriate for the environs within a few $\times$ 100 AU 
from the protostar.

The mass of the emitting dust can be calculated with the relationship
\begin{equation}
M = \frac{D^2 F_{\lambda} }{\kappa_{\lambda} B_{\lambda}(T_{dust}) };
\end{equation}
where D is the distance, $\kappa$ is the dust opacity at the observed wavelength,
$B(T_{dust}$) is the Planck function, and $T_{dust}$ is the assumed temperature of the emitting material.
We are adopting the dust opacities of \citet{ossenkopf1994} for this calculation
($\kappa_{1.3mm}$ = 0.899 cm$^2$ g$^{-1}$, dust-only opacity),
which are appropriate for the conditions of protostellar clouds. Finally, the total mass can be
calculated by assuming a dust to gas mass ratio, the standard value of which is 1:100. The systematic
uncertainty in the derived masses can be factors of several due to uncertainties in the dust opacity 
and dust temperature.

We will use this method to estimate the mass of the emitting material from
the observed flux density at 1.3 mm, assuming that we are only detecting
thermal dust emission. To measure the flux densities, we will employ two methods. First, we will
measure the flux density in the deconvolved images at both low and high resolution and use these
flux measurements to calculate the mass. Second, we will measure the flux densities directly from 
the visibility amplitudes on 50 k$\lambda$ and 100 k$\lambda$ scales. On these scales the envelope contribution
is small and the residual contribution can be estimated from single-dish flux density measurements.

\subsubsection{Masses from Deconvolved Images}

To measure the continuum flux
densities directly from the deconvolved images, we sum the flux above the 3$\sigma$ level 
within a 2.5\arcsec\ aperture around the protostar position. Thus, we only include
the flux density bounded by our 3$\sigma$ image contours shown in Figures \ref{sample-lr} and \ref{sample-hr}. 
We perform these measurements on both the lower resolution images and the high resolution
images. The integrated flux densities, peak flux densities, RMS noise, and inferred masses
are given in Table 4. 

The lowest mass measured in the low-resolution imaging is 0.017 $M_{\sun}$ toward L1451-MMS and the
largest mass is 0.48 $M_{\sun}$ toward L1448 IRS3B. These mass measurements would include
contributions from both the disk and envelope. The median and average low resolution masses
are 0.1 $M_{\sun}$ and 0.13 $M_{\sun}$, respectively, with a standard deviation of 0.14 $M_{\sun}$.
The masses from the high resolution data, 
on the other hand, have filtered more of the large-scale
emission and may better probe the mass from the compact inner envelope and/or disk. 
The largest high-resolution mass is 0.19 $M_{\sun}$
toward IRAS 03292+3039 and the lowest is 0.011 $M_{\sun}$ toward L1451-MMS. 
The median and average high resolution masses are 0.067 $M_{\sun}$ and 
0.064 $M_{\sun}$, respectively, with a standard deviation of 0.054 $M_{\sun}$.

The masses at both high and low-resolutions have variations over an order of magnitude, though the standard deviation 
of the high-resolution masses is over a factor of two less than the standard deviation at low-resolution. 
Moreover, there are no trends in mass (or flux density) with other protostar
properties such as bolometric temperature (T$_{bol}$) or bolometric luminosity (L$_{bol}$).

\subsubsection{Masses from Visibility Amplitudes}

We also measured the masses directly from the visibility amplitude data that are shown in Figure \ref{uvamps}.
This enables us to determine better the spatial scales of the emitting material. We measured the 
flux densities
at 50 k$\lambda$ and 100 k$\lambda$, corresponding 
to $\sim$4\arcsec\ (920 AU) and $\sim$2\arcsec\ (420 AU), respectively,
and any emission must be coming from this or smaller scales. The flux densities and
masses are given in Table 5.

The median masses derived from the flux densities at 50 k$\lambda$ and 100 k$\lambda$ are 0.09 $M_{\sun}$ and
0.07 $M_{\sun}$ respectively, with respective average masses of 0.1 $M_{\sun}$ and 0.05 $M_{\sun}$. The
standard deviations of the visibility amplitude masses are lower than those measured from the deconvolved
images, 0.08 $M_{\sun}$ and 0.03 $M_{\sun}$ in units for 50 k$\lambda$ and 100 k$\lambda$, respectively. 

Despite the isolation of emission from small-spatial scales, these
masses still include any envelope flux that originates on these scales. \citet{jorgensen2009}
developed a method to correct for the residual envelope flux density at 50 k$\lambda$. They
 determined that an envelope with a $\rho$ $\propto$ R$^{-1.5}$ 
density profile will contribute at most 4\% of its total emission at 50 k$\lambda$; we verified this
and also determined that at 100 k$\lambda$ the envelope contribution is only 2\% of its total flux density.
If the density profile were steeper, $\rho$ $\propto$ R$^{-1.8}$ for instance, \citet{jorgensen2009} 
found that the envelope could then contribute up to 8\% to the flux density at 50 k$\lambda$. 

The total envelope flux density for the sample in \citet{jorgensen2009} was determined from
SCUBA 850 \micron\ data. For our sources, there were 1.1 mm Bolocam data taken toward all protostars
in the Perseus cloud. We scaled the 1.1 mm flux densities to 1.3 mm, assuming 
$\beta$ = 1.78 from \citet{ossenkopf1994}; the flux densities at 50 k$\lambda$ are typically 10\% to 25\% of
the Bolocam flux densities. We then subtracted the estimated envelope contribution of 
4\% and 2\% of the scaled Bolocam flux density for 50 k$\lambda$ and 100 k$\lambda$ scales, respectively.

As an independent check, we applied this method to the resolved disk of L1527 IRS \citep{tobin2013a}.
The 870 \micron\ and 3.4 mm visibility amplitudes at 50 k$\lambda$ were 314 mJy and 19.7 mJy respectively.
The single-dish flux densities of L1527 at 870 \micron\ and 3.3 mm are 1690 mJy and 33.3 mJy respectively \citep{shirley2011}.
Assuming the single-dish flux densities only contribute 0.04 of the 50 k$\lambda$ flux, the
corrected 50 k$\lambda$ flux densities are 246 mJy and 18.4 mJy at 850 \micron\ and 3.4 mm respectively. These values
are close to the integrated flux densities from the highest resolution images of the resolved disk in \citet{tobin2013a}.

The median envelope-corrected masses at 50 k$\lambda$ and 100 k$\lambda$ are 0.052 $M_{\sun}$ and
0.046 $M_{\sun}$ respectively, with respective average masses of 0.068 $M_{\sun}$ and 0.040 $M_{\sun}$.
The standard deviations of the masses are 0.08 $M_{\sun}$ and 0.03 $M_{\sun}$ at 50 k$\lambda$ and 100 k$\lambda$, respectively, with roughly the same amount
of scatter as in \citet{jorgensen2009}.
A histogram of disk masses at 50 k$\lambda$ and 100 k$\lambda$ is shown in Figure \ref{disk-mass-histo}.
We show the corrected and uncorrected masses at 50 k$\lambda$ and 100 k$\lambda$ versus
T$_{bol}$ in Figure \ref{disk-mass-lbol-tbol}. Our sample shows no sign of 
a correlation between the calculated masses and $T_{bol}$, 
while the results of \citet{jorgensen2009} may have shown a weak correlation. However,
the T$_{bol}$ values we use are calculated with the 
inclusion of longer wavelength \textit{Herschel} data \citep{sadavoy2014} and may be more accurate.
For instance, T$_{bol}$ = 47 K for L1448C in \citet{sadavoy2014}, but \citet{jorgensen2009}
lists T$_{bol}$ = 77 K for this source and \citet{enoch2009} gives T$_{bol}$ = 69 K.
We also only cover a narrow range of T$_{bol}$ since we are focused on Class 0 sources, while
\citet{jorgensen2009} had equal numbers of Class 0 and Class I sources.

%old
%median mass 100    0.0700000
%median mass 50    0.0880000
%average mass 100    0.0527000
%average mass 50    0.0968000
%stdev mass 100    0.0345770
%stdev mass 50    0.0840658
%median corr mass 100    0.0460000
%median corr mass 50    0.0520000
%average corr mass 100    0.0403000
%average corr mass 50    0.0721000
%stdev corr mass 100    0.0319028
%stdevcorr  mass 50    0.0733234

%new
%median mass 100    0.0700000
%median mass 50    0.0880000
%average mass 100    0.0527000
%average mass 50    0.0968000
%stdev mass 100    0.0345770
%stdev mass 50    0.0840658
%median corr mass 100    0.0457000
%median corr mass 50    0.0516000
%average corr mass 100    0.0403800
%average corr mass 50    0.0681200
%stdev corr mass 100    0.0318802
%stdevcorr  mass 50    0.0761868

\section{Model Comparison}

The interpretation of the 1.3 mm continuum emission can be aided by comparison to radiative transfer
models of axisymmetric envelopes, disks, and envelopes with embedded disks. We have run a small
grid of models using the Hyperion code \citep{robitaille2011}. This is a Monte Carlo radiative transfer code 
that calculates
radiative equilibrium and generates high signal-to-noise images of 
dust emission using ray tracing. 
With these models,
we aim to determine if the data are consistent with envelope-only models or if a compact structure
is required. 
We ran three varieties of models:
disk-only models, rotating--collapsing envelopes \citep[CMU envelopes;][]{cassen1981,ulrich1976},
and power-law envelopes; both types of envelope models are run with and without disk 
components. 

The models are run with three variations of disk 
mass 0.0 $M_{\sun}$, 0.01 $M_{\sun}$, and 0.1 $M_{\sun}$ and
with disk radii of 10 AU, 30 AU, 50 AU, 100 AU, 300 AU, and 500 AU. The disks have a surface density
profile $\propto$ R$^{-1}$ and a scale height of 10 AU at a radius of 100 AU; the surface
density profile and scale height are chosen to be consistent with a
viscous accretion disk in vertical hydrostatic equilibrium \citep{ss1973}. 
The envelopes all have masses of 
5.25 $M_{\sun}$, a 1 L$_{\sun}$ central star, an outer radius of 10,000 AU, and an outflow
cavity with a half-opening angle of 10\degr. 
The envelope mass corresponds to a CMU model with
a mass infall rate of 5 $\times$ 10$^{-5}$ $M_{\sun}$ yr$^{-1}$ and this was chosen to be representative
of a young protostellar system dominated by the mass of the envelope.
The inner radius of the envelope and disk are
equivalent to the dust destruction radius, adopted to be where the dust temperature is 1600 K. 
For the CMU envelopes, we assumed that the disk radius
was equivalent to the centrifugal radius (R$_C$), the radius where material is rotationally supported. 
The power-law envelopes have either a $\rho$ $\propto$ R$^{-1.5}$ or R$^{-2}$ spherical
density profile; these density profiles are representative of free-fall 
collapse and the singular isothermal sphere \citep{shu1977} (or alternatively the Larson-Penston solution \citep{larson1969}), respectively.

The models were set up to generate output images at 1.3 mm with 5 AU pixels and we Fourier transformed
the images using the MIRIAD task \textit{fft}. We azimuthally averaged the Fourier transformed images
to compare with the observed visibility amplitudes. The normalized visibility amplitudes from the
envelope-only models (0.0 $M_{\sun}$ disks) and the disk-only models are shown in Figure \ref{disk-models}.
As expected, the CMU models with increasing R$_C$ have more rapidly declining visibility amplitudes
and a power-law envelope with $\rho$ $\propto$ R$^{-2}$ has a more slowly decreasing
visibility amplitude than the R$^{-1.5}$ envelope and CMU envelopes. The disk-only models
have flat amplitudes for disks with small radii, but disks with radii $>$ 100 AU show a factor of 5 decline
at uv distances $>$ 100 k$\lambda$ and a 100 AU disk has a factor of 5 decline at 500 k$\lambda$.
The disk-only models also show variation in visibility amplitudes due to the disks with smaller
radii being optically thick, resulting in less emergent flux.

Disks embedded within envelopes are shown in Figures \ref{cmu-models} and \ref{power-models}.
The addition of the disk components does add some flattening to the visibility amplitude
profiles, but they are always declining. The model with R$_C$ = 500 AU does, however, have a small upward bump.
The disks also change the ratio of the visibility amplitudes at small uv-distances relative to
large uv-distances (large scales to small scales, respectively). For assumed envelope
masses of 5.25 $M_{\sun}$, a 0.01 $M_{\sun}$ disk does not dramatically affect the visibility amplitude
profiles, but a 0.1 $M_{\sun}$ disk has a more profound effect. We also notice that a 100 AU, 0.1 $M_{\sun}$
disk has a flatter visibility amplitude profile between 100 k$\lambda$ and 400 k$\lambda$ than disks
with radii between 10 AU and 50 AU. This is due to the small, massive disks being 
optically thick as mentioned in the previous paragraph.

A qualitative comparison of the model visibility amplitude profiles to the data reveal that many
sources have profiles that decline more slowly than the
models of envelopes with embedded disks. Furthermore, some sources look more like
the disk-only models than disk+envelope models (e.g., Per-emb-14, IRAS 03282+3035, IC348-MMS).

A more quantitative comparison of the models and observations is shown in Figure \ref{uvratios},
where we plot the ratio of visibility amplitude at 25 k$\lambda$ to the 
visibility amplitude at 250 k$\lambda$. This
plot gives a sense of the relative contribution of the envelope 
and compact component (modeled as a disk) 
on spatial scales that are an order of magnitude different,
10\arcsec\ ($\sim$2300 AU) and 1\arcsec\ ($\sim$230 AU) scales. These scales are chosen
for both physical and practical reasons; 25 k$\lambda$ is the minimum scale that
we could measure accurate visibility amplitudes due to our chosen array configurations.
Furthermore, a 10\arcsec\ ($\sim$2300 AU) scale is likely to be directly associated with gravitational collapse
and formation of the protostar, while a disk is expected to form on scales 
of about 1\arcsec\ ($\sim$230 AU) or smaller. Lastly, at uv-distances greater than 250 k$\lambda$,
not all sources had sufficient signal-to-noise to accurately measure their visibility amplitude.

The disk-only models are shown as plus signs in Figure \ref{uvratios} with 
an increasing ratio as the disk radius increases and 
the power-law envelopes without disks are plotted as the large symbols. We also show the models with 
increasing disk radii within a given envelope (triangles and squares). An envelope with a 10 AU, 0.1 $M_{\sun}$ 
embedded disk has large ratios because the disk is optically thick, and
the ratios decrease with disk radius until R = 100 AU and then the ratios 
begin to increase again as the disk itself becomes resolved. 
These tracks with disk radii can be raised or lowered by 
changing the ratio of disk mass to envelope mass. 

The key result shown in Figure \ref{uvratios} is that for a 25 k$\lambda$ to 250 k$\lambda$ ratios
less than $\sim$8 (the ratio of the $\rho$ $\propto$ R$^{-2}$ envelope, the large diamond in Figure \ref{uvratios})
a compact component of some size and mass
is required to explain the visibility amplitudes. The most 
likely form of such a compact component is a protostellar disk. This 
criterion is fulfilled for 6 out of 9 Class 0 protostars: L1451-MMS, Per-emb-14, L1448 IRS2, 
L1448C, IC348-MMS, and IRAS 03282+3035.
Furthermore, Figure \ref{uvratios} also demonstrates that massive ($\sim$ 0.1 $M_{\sun}$),
embedded disks with 
radii $<$ 30 AU may not be apparent toward protostars at 1.3 mm because the 25k$\lambda$ 
to 250 k$\lambda$ visibility amplitude ratio are not significantly lower than the
envelope-only case. This happens because emission from the small, massive disks being
optically thick masking much of the emitting material.

\section{Discussion}

The results presented here are currently the largest collection high-resolution 1.3 mm continuum observations
of Class 0 protostars in a single star forming region. Moreover, our data all have comparable 
resolution and sensitivity, a feature lacking from previous studies.

\subsection{Disk Formation}

The formation of disks during the protostellar phase has been thought to occur as a
consequence of angular momentum conservation during protostellar collapse 
\citep{ulrich1976, cassen1981, tsc1984}. This would enable the rapid formation
of large disks during the protostellar phase, dependent entirely on the angular
momentum inherited from the infalling envelope. Hydrodynamic simulations,
without magnetic fields, readily form large disks during collapse \citep[e.g.,][]{yorke1999}.
These disks are massive enough to be gravitationally unstable, forming spiral arms and 
fragments \citep[e.g.,][]{boley2009,vorobyov2010,kratter2010,zhu2011}.

However, magnetic fields, which had been shown to 
potentially slow rotation during the pre-collapse phase \citep[e.g.,][]{basu1998}, were 
shown by \citet{allen2003} to also significantly slow rotation during collapse and
suppress the formation of a rotationally-supported disk \citep{galli2006}. These studies 
were verified by \citet{mellon2008} and \citet{hennebelle2008}; disk formation 
was prevented even in the presence of very weak magnetic fields. More 
recent studies showed that by including non-ideal MHD effects in simulations,
initially small rotationally-supported disks could form \citep{dapp2010}. The infalling material
would still undergo magnetic braking, but the magnetic fields would be dissipated in the high-density
material close to the protostar, enabling disk formation of a small rotationally supported disk.
The size of the rotationally-supported region will grow with time, but is expected to remain $<$ 10 AU 
throughout the Class 0 phase \citep{dapp2012}. \citet{machida2008}
also simulated  the formation of disks with non-ideal MHD and found that massive 100 AU
disks could form while in the Class 0 phase. A study by \citet{joos2012} explored less ideal initial conditions,
with misaligned magnetic fields and rotation axes. These simulations showed that even in the ideal MHD
limit disks could form if the magnetic fields are misaligned with respect to the rotation axis.

The misaligned magnetic field scenario may be plausible because
the TADPOL survey of magnetic field
morphologies \citep{hull2013, hull2014} found no systematic alignment of magnetic fields and outflow axes (presumed
to reflect the rotation axis). Moreover, two of the four known Class 0 systems with
rotationally-supported disks \citep[L1527 IRS and VLA 1623; ][]{tobin2012, murillo2013}
were found to have magnetic fields perpendicular to the outflow axis on $\sim$1000 AU scales and down
to the scale of the disk in L1527 IRS \citep{seguracox2015}. The sources
L1448C, L1448 IRS3B, and L1448 IRS2 had observations in the TADPOL survey. L1448C only had a few detected 
magnetic field vectors, but they are at a $\sim$45\degr\ angle to the outflow, L1448 IRS3B had vectors that
are at an angle of $\sim$90\degr, and L1448 IRS2 had vectors that are both aligned and 
misaligned with respect to the outflow. 

To summarize, theory predicts two different scenarios for disk formation, that depend on the 
included physics and the initial conditions, specifically the treatment 
and importance of magnetic fields.
However, recent work with misaligned magnetic fields and non-ideal MHD \citep[e.g.,][]{tomida2015}
seems to have reduced the significance of the `magnetic braking catastrophe.'

\subsection{Evidence for Class 0 Disks}

Thus far, rotationally-supported disks have been detected toward four Class 0 protostars: 
L1527, VLA 1623, RCrA IRS7B, and HH212 MMS \citep{tobin2012,murillo2013,codella2014,lindberg2014}. 
L1527, VLA 1623, and RCrA IRS7B all have both extended dust
emission perpendicular to the outflow direction and Keplerian rotation signature, while HH212 MMS is
only resolved kinematically. It is unclear
if the small number of observed Class 0 disks reflects a true paucity due to the physics of
disk formation or if it is simply a lack of observations with high enough resolution and
sensitivity to detect Class 0 protostellar disks (or disk candidates). Moreover, 
 given that
the emission from the disks is entangled with that of the envelope, they are more difficult
to directly resolve than disks around Class I and Class II sources with much less (or no) envelope emission.

To increase the likelihood of detecting disks embedded within infalling envelopes, our sample is slightly
biased toward sources that might have orientations that are within 30\degr\ of edge-on 
and well-defined outflow axes from either infrared scattered light 
or CO emission. This is because edge-on disks should stand out better against the surrounding
envelope due to the higher column density through the disk midplane.
While our data are very sensitive and have a resolution of 
$\sim$0\farcs3 (70 AU), this is barely at the limit where
we would expect to detect evidence for disks around Class 0 protostars. 
Indeed, for a source two-beams across, the
diameter would have to be at least $\sim$140 AU. 
The spatial resolution achieved by our survey is a factor of 
$\sim$1.6 more coarse than the observations
toward L1527 IRS in \citet{tobin2012}. 
If L1527 IRS was at the distance to 
Perseus, it would only be marginally resolved in our observations.

\subsubsection{Sources with Evidence for R $>$ 100 AU Disks}

Despite the difficulties posed by limited spatial resolution, we do detect two strong
candidate disks toward L1448 IRS2 and Per-emb-14 (NGC 1333 IRAS4C). 
The dust continuum structure around L1448 IRS2 is very large, $\sim$300 AU in radius.
Per-emb-14 is more compact and has an apparent radius of $\sim$100 AU.

The new and archival kinematic data for L1448 IRS2 did not enable a clear detection of Keplerian rotation 
around the protostar. Near the protostar, the velocity gradient in 
C$^{18}$O only differs from the outflow position angle by 30\degr. 
The dust continuum emission at higher resolution is extended
toward the southwest and Tobin et al. (in prep.) detects a binary source at the end of this dust extension. 
\citet{takakuwa2014} showed that infall from a circumbinary disk, shepherded by the binary sources could
produce a radial velocity gradient along the outflow direction. Thus, the rotation signature on 
$>$200 AU scales could be that of a circumbinary disk and the shift in velocity gradient direction could reflect material
flowing through the disk toward the binary sources. On the other hand, a velocity gradient along the outflow
direction is suspicious given that the outflow can entrain the ambient envelope material \citep{arce2006}. Thus,
we merely suggest radial transport of material through the disk as a possibility to produce
the velocity gradient along the outflow.

Per-emb-14 also appears to be a promising disk candidate from its resolved dust continuum structure. 
Our C$^{18}$O data detected evidence of
a velocity gradient in the expected direction for rotation.
However, higher sensitivity will be required to
verify this velocity gradient and determine if this rotation is Keplerian.

In addition to these two most promising R $>$ 100 AU disk candidates, IRAS 03292+3039 has one of 
the brightest, largest, and most puzzling structures. While the position angle of this source is not
close to being orthogonal to the outflow, the combined imaging from C and B-configurations shows
a $\sim$2.5\arcsec\ (575 AU) diameter structure. Then at the highest resolution,
the emission is rather constant across the source with a brightness temperature of $\sim$8K, indicating
that the dust may be close to being optically thick, in which case the brightness temperature would reflect
the temperature of the emitting material. In addition to the size of this structure, 
strong evidence of rotation on $\sim$3\arcsec\ scales
was shown by \citet{schnee2012} in C$^{18}$O ($J=2\rightarrow1$) observations. Furthermore,
the velocities of the outflow emission traced by CO ($J=2\rightarrow1$) indicate that this source is viewed
at an intermediate inclination, possibly 40\degr\ \citep{yen2015}.
The C$^{18}$O ($J=2\rightarrow1$) data was further analyzed by \citet{yen2015}, finding that 
a disk as large as 850 AU could form in this system.

The large resolved structure, intermediate inclination, and rotation on scales comparable to the 
continuum size are evidence that this source may also be a good disk candidate, even though the continuum
emission is not extended orthogonal to the outflow direction. The mass of the continuum source is also
large, $\sim$0.2 $M_{\sun}$ at the highest resolution and the protostar mass is 
estimated to be $\sim$0.3 $M_{\sun}$ \citep{yen2015}. If the continuum source is indeed a rotationally-supported,
the large ratio of disk to protostar mass could mean that the disk would 
be gravitationally unstable. However, future observations
with higher resolution will be needed to verify if the continuum 
structure is a rotationally supported disk as well as a more robust kinematic
measurement of the protostar mass.

\subsubsection{Sources with Evidence for R $<$ 100 AU Disks}

While there is evidence for large disk-like structures around a few protostars, the
results are not so clear for the remaining sources. 
The sources L1448C, IRAS 03282+3035 and L1448 IRS3C,
have deconvolved position angles that are 7\degr, 19\degr, and 9\degr\ from being orthogonal to the outflow, respectively.
However, the continuum emission is not obviously disk-like in the images, but only higher-resolution/sensitivity
data will be able to determine this for certain, in addition to molecular line observations.

The analysis of visibility amplitude ratios enables us to shed further light on the evidence for   
disks (or at least compact dust structures) on scales smaller than 100 AU. Figure \ref{uvratios}
shows that sources with 25 k$\lambda$ to 250 k$\lambda$ amplitude ratios less than $\sim$8 require
a contribution from a compact component, possibly a disk. A ratio of 8 is the value for a spherical envelope
with a radial density profile $\rho$ $\propto$ R$^{-2}$ and no compact density structure, the steepest we might expect
in idealized star formation models; the ratio of an R$^{-1.5}$ density profile is
$\sim$30. The ratio expected for a 500 AU disk itself is $\sim$5. Thus, the ratio
between 0 and 8 will depend on both the radius of the disk and the amount of
mass in the disk relative to the envelope.

L1451-MMS, Per-emb-14, L1448C, IC348 MMS,
L1448 IRS2, and IRAS 03282+3035 all have a 25 k$\lambda$ to 250 k$\lambda$ 
ratio less than 8 (6 out of 9 Class 0s in the sample). L1448 IRS3C was discussed
previously as having resolved emission nearly orthogonal to the outflow, but it does not have a ratio
less than 8. This could result from its close proximity to L1448 IRS 3B, given that
its visibility amplitudes have unexplained dips at 200 k$\lambda$ and 40 k$\lambda$. IRAS 03292+3039
(discussed in the previous section) also has a ratio $>$ 8, but this could be caused by 
the emission
from the disk and inner envelope being optically thick out to $>$ 1\arcsec\ scales, leading to the emission being resolved-out.

L1448 IRS3B appears to have extended envelope emission orthogonal
to the outflow, encompassing the two sources. The detected emission
is not significantly larger in angular extent than that of L1448 IRS2 and IRAS 03292+3039,
but the emission is not centered on either of the 
continuum
sources. Kinematic observations of L1448 IRS3B by \citet{yen2015} 
show that there is a velocity gradient consistent with rotation on 3\arcsec\ scales, but
higher resolution kinematic observations will be needed to determine the nature of this structure.

\subsubsection{Outlook}

The results from this small survey have shown that at least two (Per-emb-14 and L1448 IRS2), possibly
three (including IRAS 03292+3039) show evidence for resolved disks with radii $>$ 100 AU. 
The remaining sources have evidence for compact structure within their envelopes that could
be evidence for a disk, but higher resolution continuum and kinematic observations are necessary to
characterize them. Despite the small sample, it is encouraging
for studies of Class 0 disks that 3 of 9 Class 0 sources have evidence
for resolved disk-like structure and many of the others have  
compact emission extended perpendicular to their outflows or visibility amplitude
ratios that suggest a disk-like component. While we cannot
prove that the disk candidates are rotationally supported, it is important 
to highlight that we are detecting an abundance of structure on scales less than 500 AU toward 
Class 0 protostars, possibly a signature of protostellar disks.

\subsection{Class 0 Disk Masses}
Assuming that we are probing emission from the protostellar disks, we have been able to calculate their masses
to compare with measurements of other Class 0 sources and more evolved Class I and II objects.
The median masses of the compact components are $\sim$0.05 $M_{\sun}$ (values corrected for estimated envelope emission) for 
measurements taken at both 50 k$\lambda$ and 100 k$\lambda$. This mass is an order of magnitude larger
than the characteristic mass of Class II disks \citep[$\sim$0.005 $M_{\sun}$,][]{andrews2005}
and 5$\times$ larger than the median Class I disk mass measured in \citet{jorgensen2009}. The Class 0 
disk masses measured in \citet{jorgensen2009} did have some sources with even larger masses than we find, which may
support a scenario of typically higher disk masses in the Class 0 phase. A caveat of this comparison
is that while \citet{jorgensen2009} assumed the same dust opacities as our study, \citet{andrews2005} 
assumed $\kappa_{850\mu m}$ = 0.035 (dust plus gas opacity), about a factor of two larger than the
\citet{ossenkopf1994} model at 850 \micron\ (a factor of two larger opacity 
decreases the calculated mass by the same factor). 
However, a larger opacity might be more appropriate for
Class II disks, but more importantly the gas to dust mass ratio in Class II disks might be significantly
less than 100 \citep{williams2014}. The gas to dust ratio uncertainty is potentially more significant than
the opacity uncertainty in the case of Class II disks, while the gas and dust may still be well-mixed
in the Class 0 phase.

The main uncertainty in masses derived in this
analysis is the true envelope density structure.
Numerous studies have shown that
Class 0 envelopes may have a variety of density profiles; 
the most typical profiles are
steeper than $\rho$ $\propto$ R$^{-1.5}$ \citep{looney2003,chiang2008,kwon2009,
chiang2012,tobin2014}. This is important, because steeper envelope density profiles result in
greater contributions of envelope emission at 50 k$\lambda$ and 100 k$\lambda$, leading
to over estimated disk masses.

\subsection{Multiplicity}

Our sample is not large enough to make strong revisions to the current 
multiplicity results toward Class 0 protostars, which was recently
found to be 0.64 by \citet{chen2013}.
We have discovered one new close companion 
(L1448 IRS3B separated by $\sim$210 AU) and we believe that
the 300 AU companion toward IRAS 03282+3035 is spurious. Thus, the multiplicity
fraction as a whole does not change. 
However, we did sensitively sample smaller spatial scales
than previous studies and only confidently find 1 new clear candidate companion (L1448 IRS 3B). 
IRAS 03292+3035 may be another candidate, but its multiplicity is uncertain.

\citet{maury2010} had suggested a lack of multiplicity on 
scales between 
150 AU and 430 AU (corrected for the updated distance to Perseus), but our new binary
detection is separated by $\sim$210 AU and the sample 
of \citet{chen2013} survey was quite incomplete at those scales.
A larger, less-biased sample will better illuminate
the multiplicity frequency as a function of separation; 
such a project is being carried out now with 
the VLA (Tobin et al. 2015 in prep.). 

\section{Summary}

We have presented a CARMA 1.3 mm survey of 9 Class 0 protostars and 2 Class I protostars 
in the Perseus molecular cloud. This is one of the largest high-resolution 
($\sim$0\farcs3, 70 AU) samples of 1.3 mm data taken toward Class 0 protostars (thus far). We also
include kinematic follow-up for two sources (Per-emb-14 and L1448 IRS2). 
The main results from the survey can be summarized as follows.

1) We detect three strong Class 0 disk candidates toward L1448 IRS2, Per-emb-14, and IRAS 03292+3039. These systems
are not yet kinematically confirmed to be Keplerian disks. L1448 IRS2 shows evidence of rotation on scales
$>$ 200 AU, with a change in velocity gradient direction on scales $<$ 200 AU. Per-emb-14 has an
indication of rotation on the scale of the detected disk-like structure, but the spectral line
data have low signal-to-noise and we cannot determine if the rotation is Keplerian. IRAS 03292+3039 does
show evidence for rotation on 500 AU scales, but smaller scale measurements are not yet available.

2) A variety of resolved structures are detected within our continuum data, aside from the disk candidates
listed in 1), ranging from apparent massive inner envelopes or disks
(L1448 IRS 3B and IRAS 03292+3039) to marginally resolved
structures on $<$ 200 AU scales (IRAS 03282+3035, L1448 IRS3C). Finally, three sources do not
appear to have resolved structure on scales $<$ 200 AU (L1448C, IC348 MMS, L1451-MMS).
Moreover, 5 out of 9 sources in the sample have deconvolved position angles that are within 30\degr\ of
orthogonal to the outflow, possibly an indication of compact structure consistent with a disk.

3) Comparison of the observed visibility amplitude ratios to radiative 
transfer models of disks and envelopes enables us to infer that at least 
6 of 9 sources in our sample require a compact component,
possibly represented by a disk (L1451-MMS, Per-emb-14, L1448C, IC348 MMS,
L1448 IRS2 and IRAS 03282+3035). The radiative transfer models also 
show that emission from embedded disks with masses $\sim$0.1 $M_{\sun}$ and radii 
$<$ 30 AU will be optically thick and have lower levels of emission at 1.3 mm relative to larger disks
with the same mass.

4) A candidate companion to L1448 IRS3B is detected with a separation of 0\farcs9 ($\sim$210 AU) and the companion
itself may be comprised of two sources separated by $\sim$0\farcs2. IRAS 03292+3039 may have
multiple components, but the multiple peaks detected do not significantly stand
out from the resolved structure. We do not detect a companion source toward IRAS 03282+3035
that had been detected by \citet{chen2013}.

We conclude that there is evidence for the formation of large disks in the Class 0 phase, 
but Class 0 disks likely have a range of radii and masses that depend on the structure, 
kinematics, and possibly magnetic field properties of their parent cores. Sub-arcsecond resolution
imaging is crucial to characterizing the structure of Class 0 disks and observations with 
ALMA and the VLA at wavelengths longer than 1.3 mm will likely be necessary to characterize emission from
small, massive disks.

We wish to thank the anonymous referee for helpful comments that have
improved the clarity of this paper.
We also wish to thank Mike Dunham and Jes Jorgensen for useful discussions. 
J.J.T. acknowledges support provided by NASA through Hubble Fellowship 
grant \#HST-HF-51300.01-A awarded by the Space Telescope Science Institute, which is 
operated by the Association of Universities for Research in Astronomy, 
Inc., for NASA, under contract NAS 5-26555. 
This work is supported by grant 639.041.439 from the Netherlands
Organisation for Scientific Research (NWO).
This work is supported by the European Union A-ERC grant 291141
CHEMPLAN. L. L. acknowledges the support of DGAPA, UNAM, CONACyT (M\'exico), for financial support.
Support for CARMA construction was derived from the states of Illinois, California, and Maryland, 
the James S. McDonnell Foundation, the Gordon and Betty Moore Foundation, the Kenneth T. and 
Eileen L. Norris Foundation, the University of Chicago, the Associates of the California 
Institute of Technology, and the National Science Foundation. Ongoing CARMA development 
and operations are supported by the National Science Foundation under a cooperative 
agreement, and by the CARMA partner universities. The Submillimeter Array is a joint 
project between the Smithsonian Astrophysical Observatory and the Academia Sinica 
Institute of Astronomy and Astrophysics and is funded by the Smithsonian 
Institution and the Academia Sinica. The National Radio Astronomy 
Observatory is a facility of the National Science Foundation 
operated under cooperative agreement by Associated Universities, Inc.
This research made use of APLpy, an open-source plotting package 
for Python hosted at http://aplpy.github.com. This research has made use 
of NASA's Astrophysics Data System Bibliographic Services.

The authors wish to recognize and acknowledge the very significant 
cultural role and reverence that the summit of Mauna Kea has always 
had within the indigenous Hawaiian community.  We are most fortunate 
to have the opportunity to conduct observations from this mountain. 

{\it Facilities:}  \facility{CARMA}, \facility{SMA}, \facility{Spitzer}

\appendix

\section{Outflow Maps}

Our observations included the $^{12}$CO and $^{13}$CO  ($J=2\rightarrow1$) transitions that, when detected,
enabled us to examine the
outflow structure toward some of our protostellar targets at resolutions
that are higher than typically achieved for outflow emission.

\subsection{L1448 IRS3}
The L1448 IRS3 region
is of particular interest because of the complex, overlapping outflows
in the region \citep{kwon2006} and there have not been higher angular resolution
observations toward these sources. We show the outflow channel maps in Figure \ref{IRS3-outflows}
from a combination of our higher-resolution data and the CO data observed in CARMA D-array
by \citet{hull2014}, gaining better sensitivity to extended structures.
The sources are marked with crosses and we can clearly identify the outflow of L1448 IRS3B, the
southern most source. There appears to be a contribution from the L1448C outflow in the southern
parts of the map. We also can identify, for the first time, an 
outflow associated with the northern-most source, L1448 IRS3C. 
For this source, we mainly detect the redshifted side
of the outflow and the blueshifted side is observed in only a few velocity channels. Nevertheless,
both the blue and red-shifted sides of the outflow have a cone-like morphology that can be traced back to
an origin at the position of L1448 IRS3C. The asymmetric brightness of the blue 
and red-shifted outflow components may result from L1448 IRS3C being located near 
the edge of the L1448 cloud and having less entrained material
on the blue-shifted side.

\citet{kwon2006} had associated what we now believe to be the
 red-shifted side of the L1448 IRS3C outflow with L1448 IRS3A, the more evolved
Class I protostar in the region. However, the bipolar pattern of the CO emission seems to clearly originate at
the position of IRS3C. We do see some redshifted CO emission near L1448 IRS3A that seems to be extended toward
the northeast; however, we cannot clearly identify a position angle for this outflow. Scattered light morphology
shown by \citet{tobin2007} may suggest that the outflow has a slight north-west position angle, but the complexities
of the emission from the three protostars within 30\arcsec\ of each other make such an identification difficult.

\subsection{L1451-MMS}
We detected CO ($J=2\rightarrow1$) toward L1451-MMS, the candidate first hydrostatic core 
identified by \citet{pineda2011}, who also found a compact CO outflow
in their lower resolution data. We confirm the presence of the CO outflow with our 
higher-resolution data as shown in the channel maps in Figure \ref{L1451MMS-outflow}.
Moreover, we also find that the continuum emission is slightly offset from the
line connecting the blue and red-shifted outflow; this slight offset was also 
seen in the data from \citet{pineda2011}. The offset may be related to 
the outflow being driven into an inhomogeneous medium. This offset is also in
the direction of the 3$\sigma$ extension seen in the continuum emission. 
The detection of the outflow with our high-resolution data 
demonstrates the compactness of the outflow, otherwise it would
have been resolved-out.

\subsection{L1448 IRS2}

We detect the outflow toward L1448 IRS2 in CO ($J=2\rightarrow1$), as shown in Figure \ref{L1448IRS2-12CO},
with excellent sensitivity. The high resolution and sensitivity is achieved with a combination of SMA Compact and 
Extended configurations as well as CARMA C and B configurations. The outflow is clearly offset from the
1.3 mm continuum peak by about 1\arcsec\ and the outflow cavities appear asymmetric, opening wider to the
southwest than the north east. This asymmetry is evident in both the blue and red-shifted lobes of the outflow.
The asymmetry in the opening angle could reflect the structure of the ambient medium in the direct vicinity
of the protostars, with the envelope being less dense to the southwest and enabling the 
cavities to open more widely. Furthermore, the outflow would appear more 
symmetric if the driving source was shifted southwest a few arcseconds, along
the extended dust continuum emission.

The combined sensitivity of the CARMA and SMA maps also enabled $^{13}$CO ($J=2\rightarrow1$) to be
resolved in the outflow toward L1448 IRS2 (Figure \ref{L1448IRS2-13CO}). 
The $^{13}$CO emission traces more compact structure located
near the continuum source, but the asymmetry and spatial offset evident in the $^{12}$CO emission
are still present. Moreover, the location of the blue-shifted intensity peak is coincident with the 
blue-shifted intensity peak in C$^{18}$O, evidence that the compact C$^{18}$O may be
affected by the outflow.

\subsection{L1448C}

Lastly, we were able to detect the compact inner outflow toward L1448C in CO ($J=2\rightarrow1$) with
only the CARMA B and C-array data. We were able to obtain a good detection with such little time on source
due to L1448C having extremely bright CO emission. Maps with superior sensitivity 
and comparable resolution do exist \citep[e.g.,][]{hirano2010}, but 
we include the map for the sake of completeness.

\subsection{Per-emb-14}

Per-emb-14 does not currently have a CO outflow detection, possibly due to the source being edge-on. However, the
\textit{Spitzer} IRAC images of the source reveal a scattered light nebula at the position of the source, along
with shock features extended along the inferred blueshifted side of the outflow.
These data were taken as part of \textit{Spitzer} GO program 30516 and were reduced with the same methods described in 
\citet{tobin2010a} and are more sensitive than the data presented by \citet{gutermuth2008}. We infer that the eastern side
of the outflow is blueshifted because it is brighter and it is known that the geometrical effects will cause
the redshifted side to be more extincted and have a fainter nebula \citep{whitney2003a}.

\begin{small}
\bibliographystyle{apj}
\bibliography{ms}
\end{small}

\clearpage

\begin{figure}
\figurenum{7b}
\begin{center}
\includegraphics[scale=0.28]{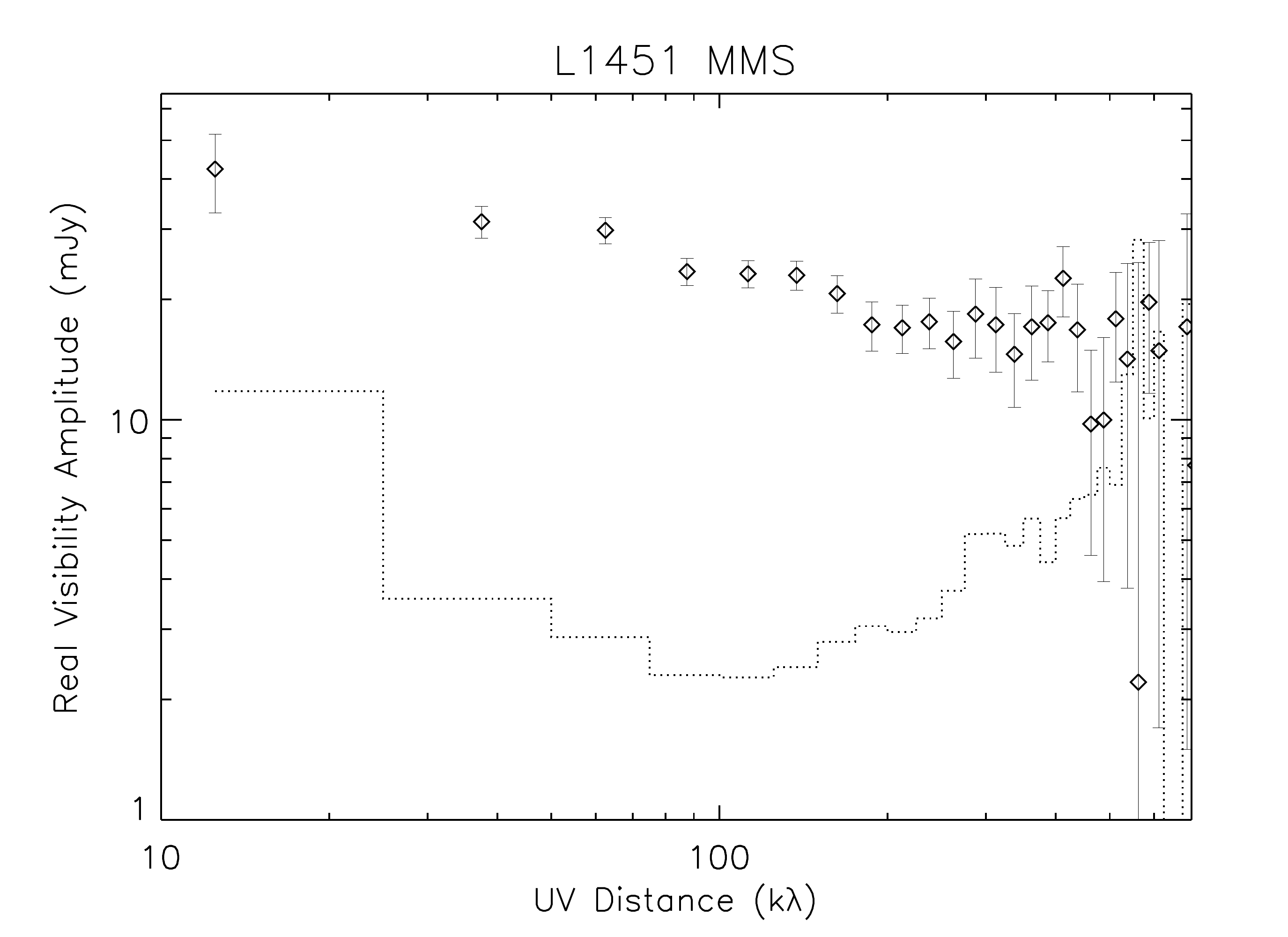}
\includegraphics[scale=0.28]{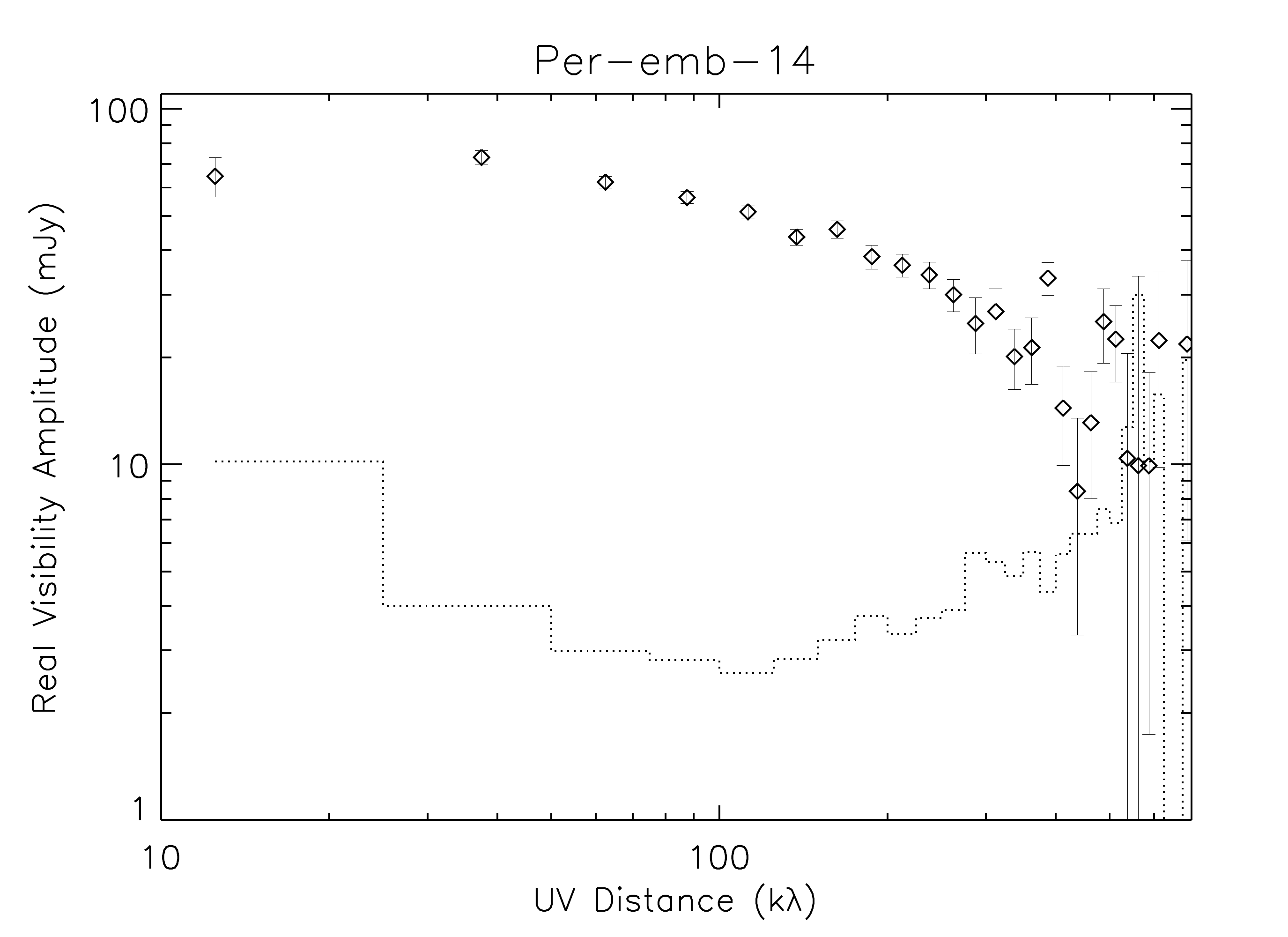}
\includegraphics[scale=0.28]{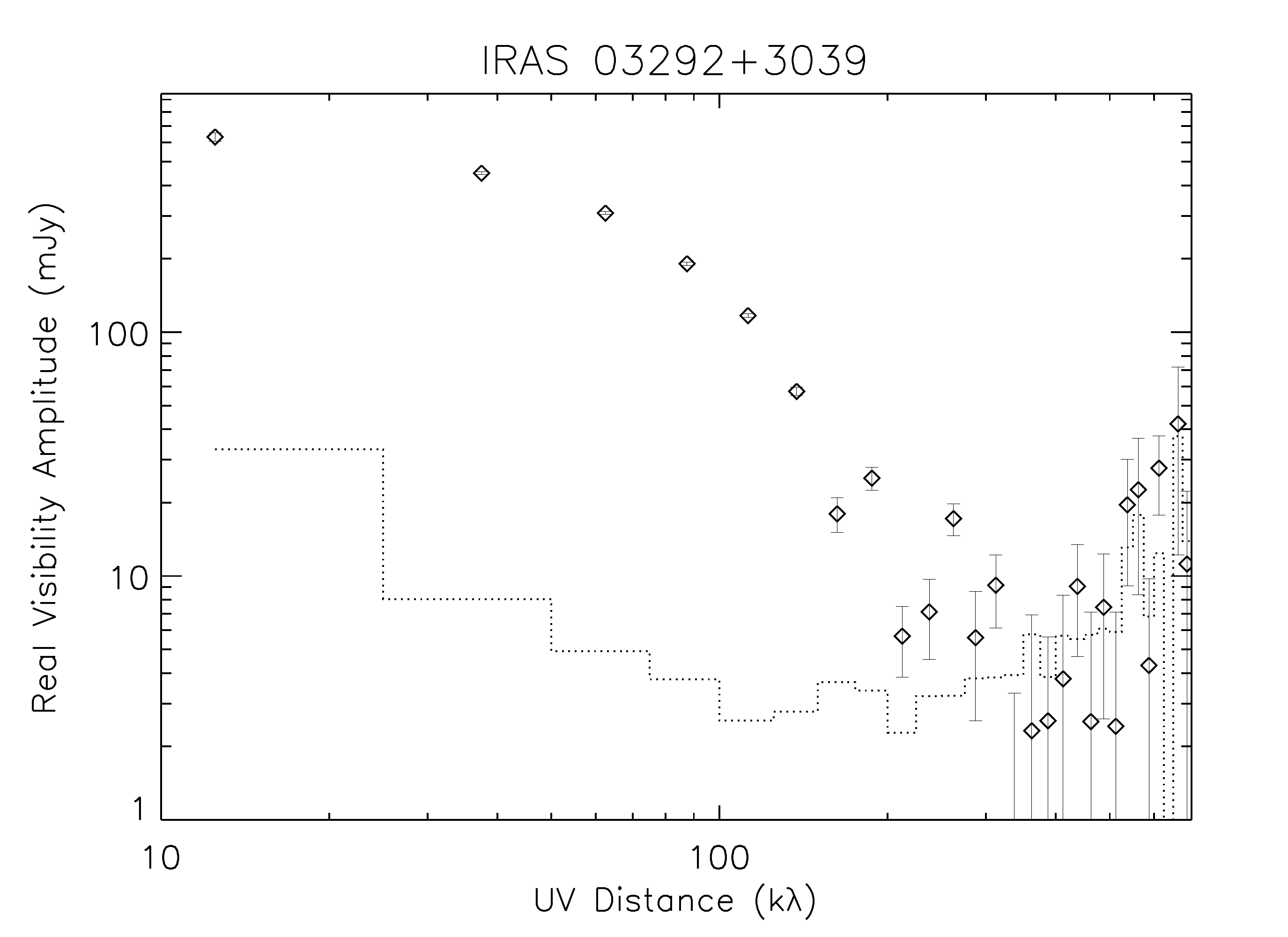}
\includegraphics[scale=0.28]{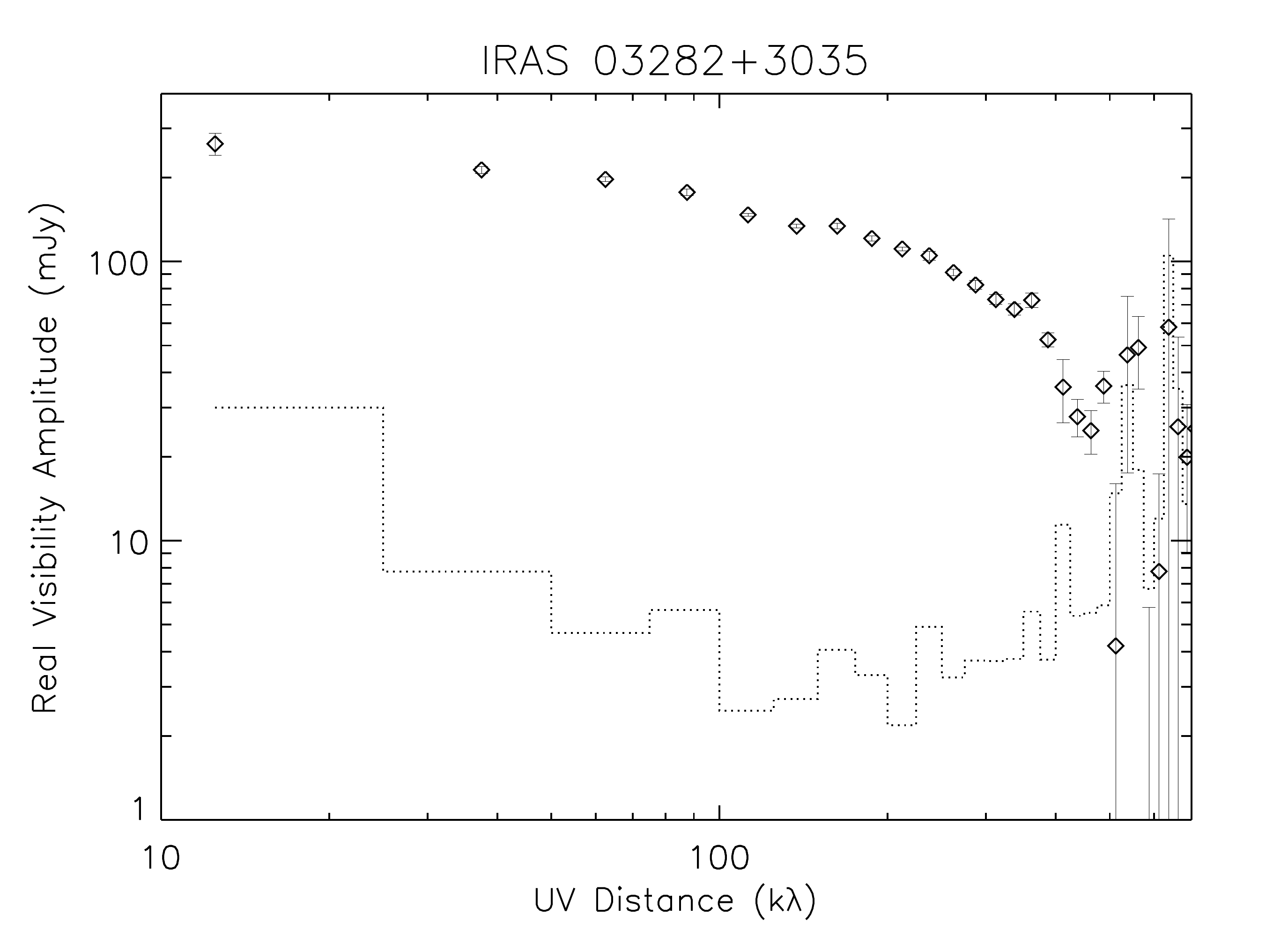}
\includegraphics[scale=0.28]{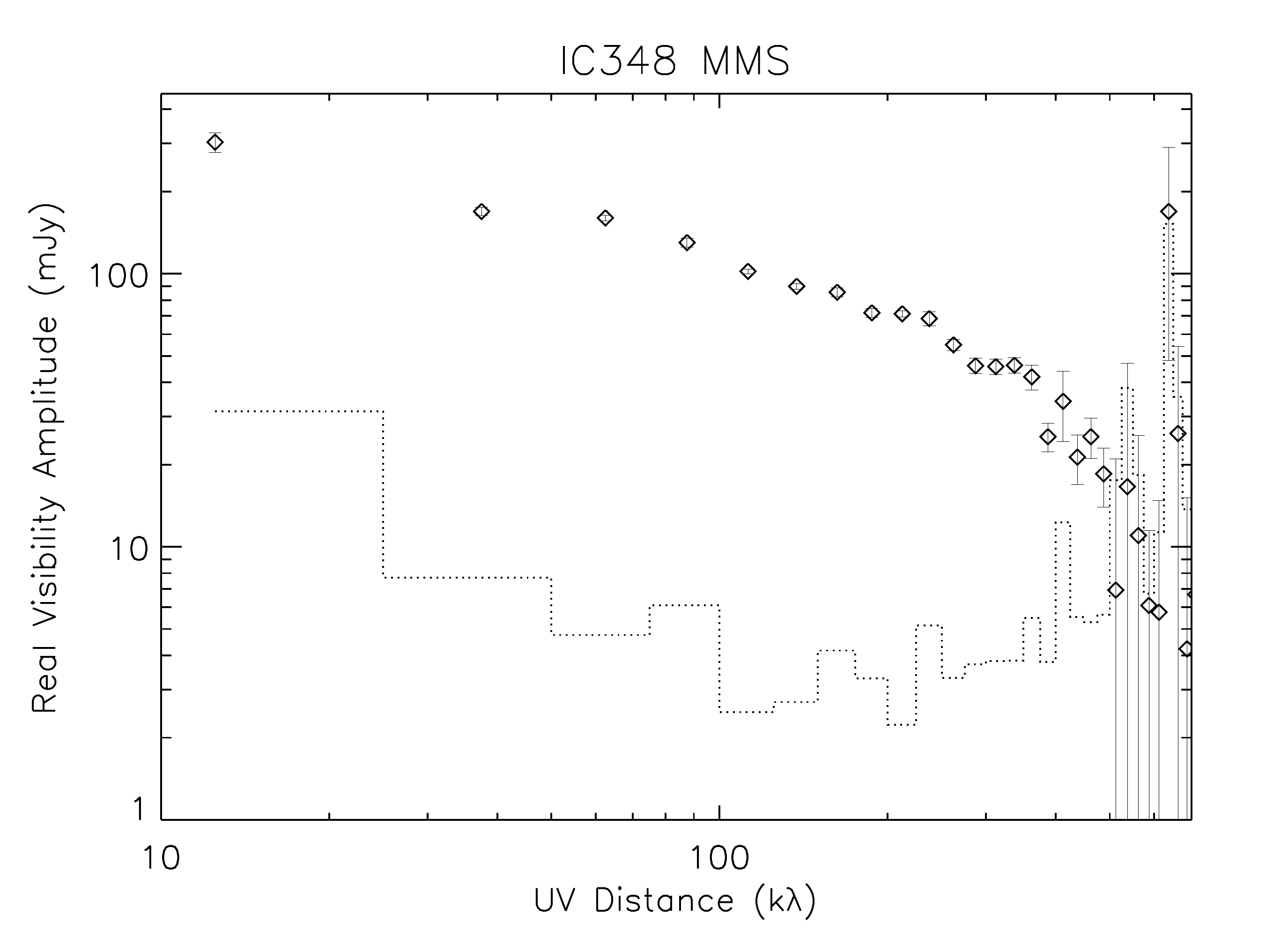}

\end{center}
\caption{}
\end{figure}

\begin{figure}
\figurenum{7c}
\begin{center}

\includegraphics[scale=0.28]{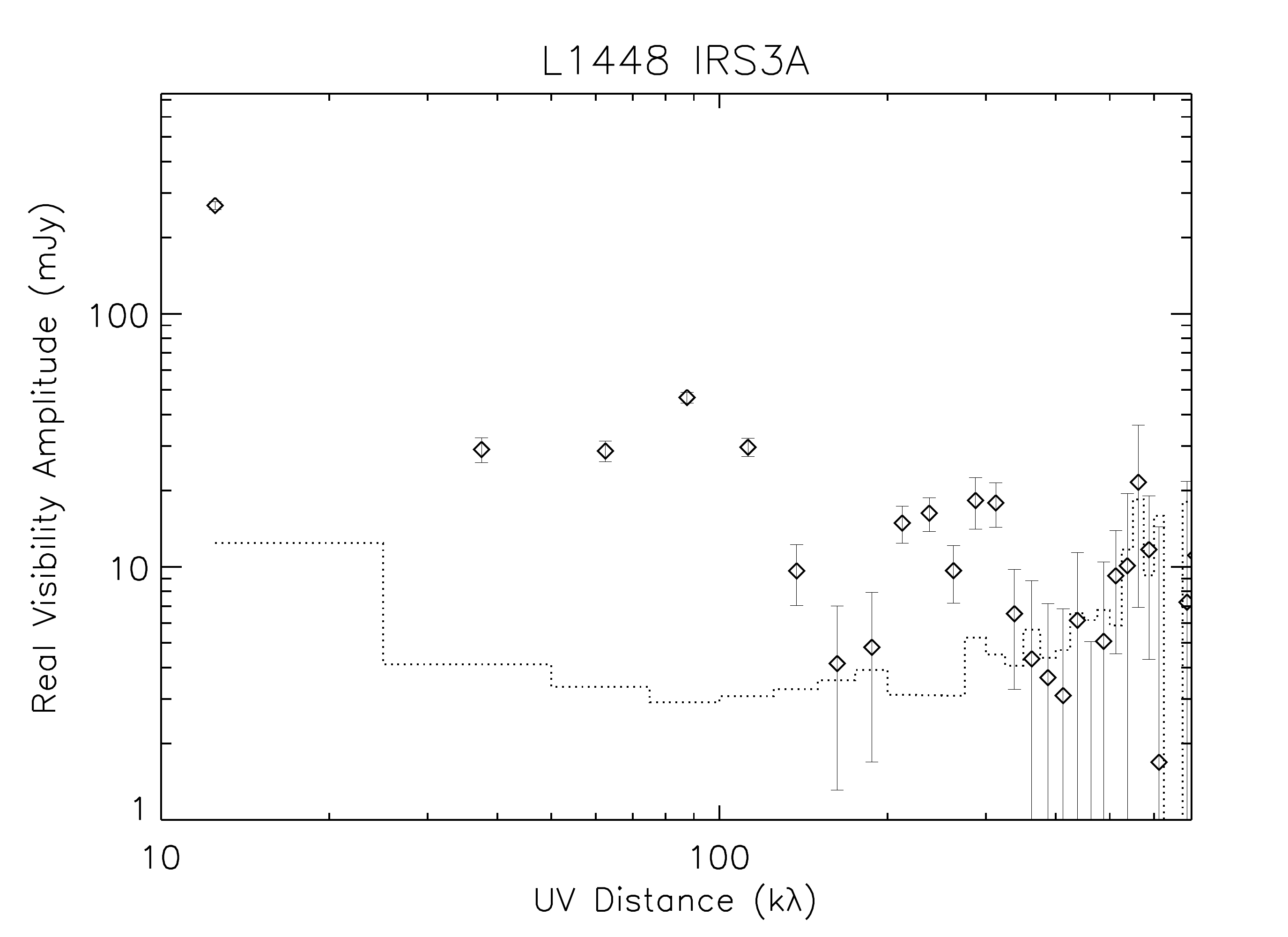}
\includegraphics[scale=0.28]{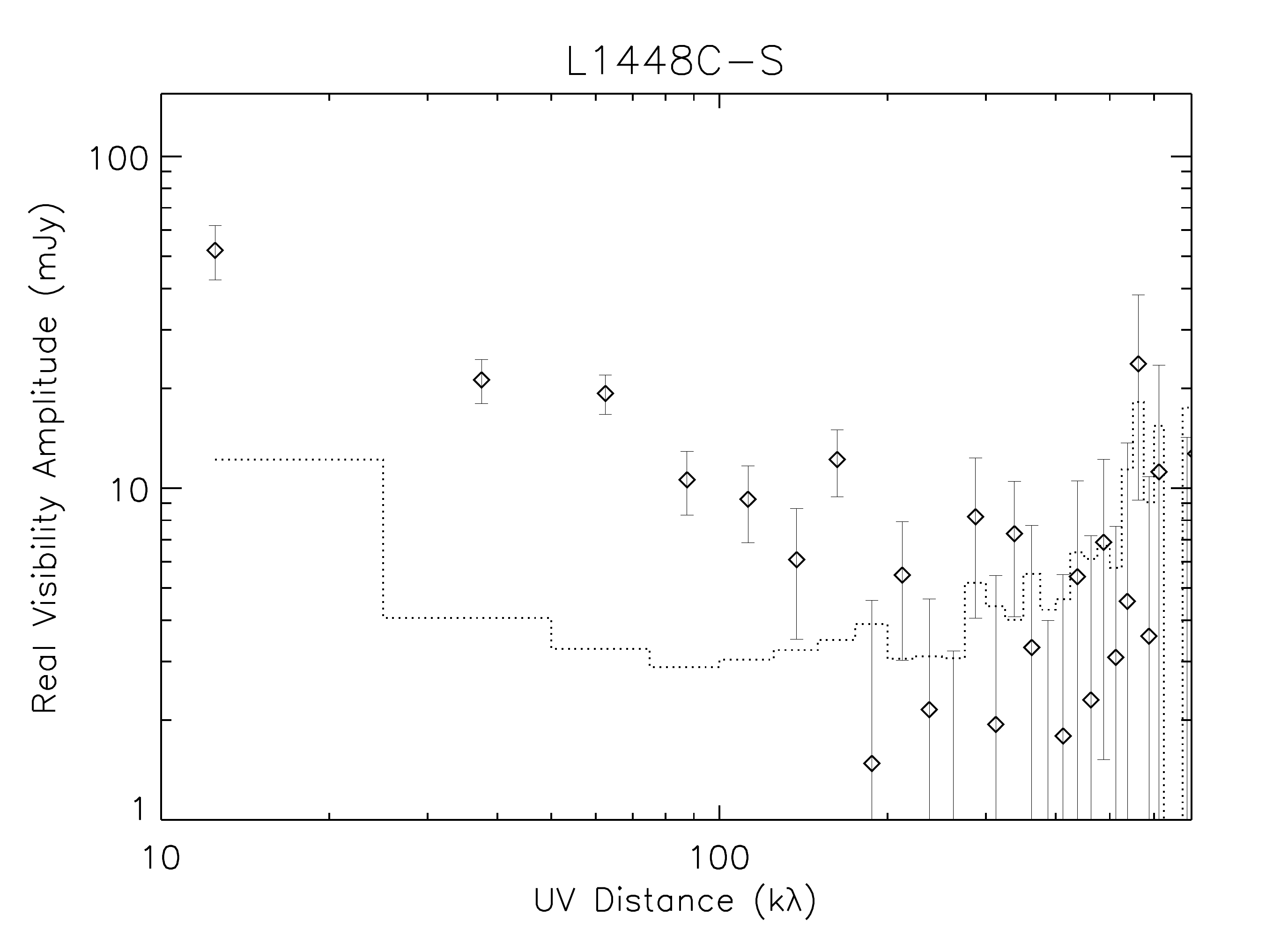}
\end{center}
\caption{}
\end{figure}

\begin{figure}
\begin{center}
\includegraphics[scale=0.3]{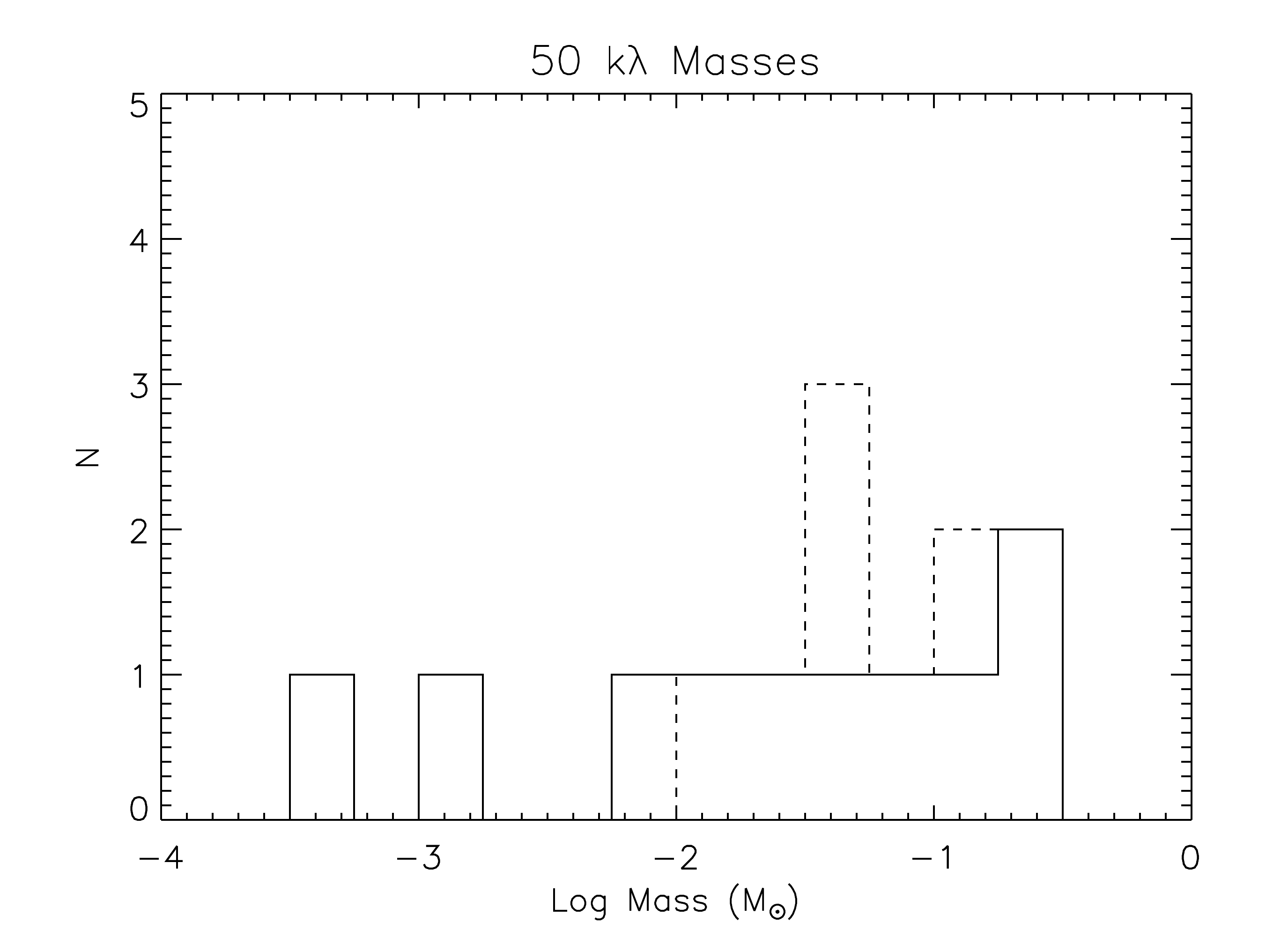}
\includegraphics[scale=0.3]{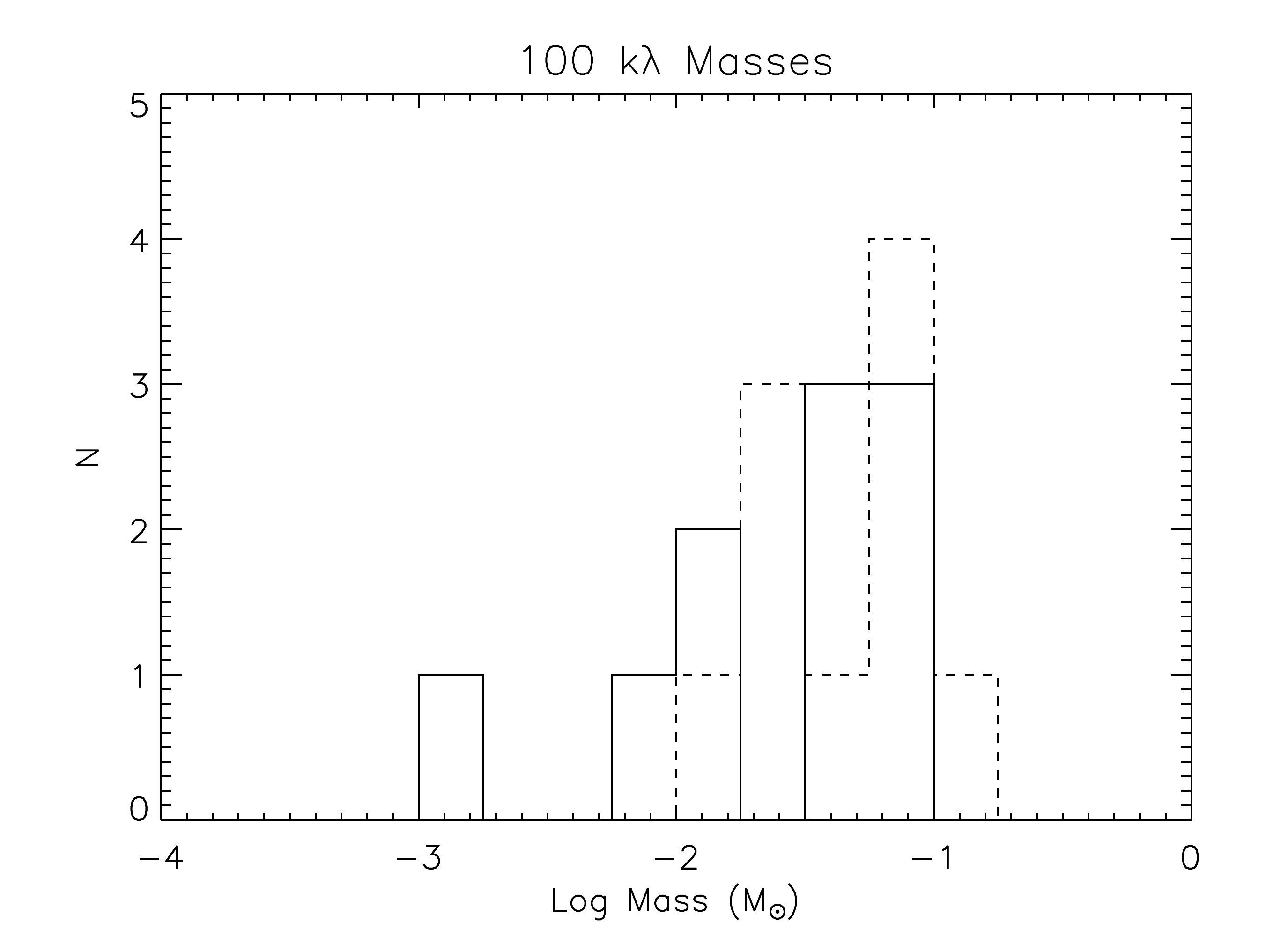}
\end{center}
\caption{Histogram plot of disk masses inferred from the flux densities at 50 k$\lambda$ (top) and
100 k$\lambda$ (bottom). Masses that have been corrected for estimated envelope 
contribution (solid line) and with no correction (dashed line) are shown. The median disk masses 
corrected for envelope emission are 0.052 $M_{\sun}$ and 0.046 $M_{\sun}$ at  50 k$\lambda$ flux and 100 k$\lambda$, respectively.
The respective median disk masses that are not corrected for envelope emission are 0.09 $M_{\sun}$ and 0.07 $M_{\sun}$, respectively.}
\label{disk-mass-histo}
\end{figure}

%new
%median mass 100    0.0700000
%median mass 50    0.0880000
%average mass 100    0.0527000
%average mass 50    0.0968000
%stdev mass 100    0.0345770
%stdev mass 50    0.0840658
%median corr mass 100    0.0457000
%median corr mass 50    0.0516000
%average corr mass 100    0.0403800
%average corr mass 50    0.0681200
%stdev corr mass 100    0.0318802
%stdevcorr  mass 50    0.0761868

\begin{figure}
\begin{center}
\includegraphics[scale=0.275]{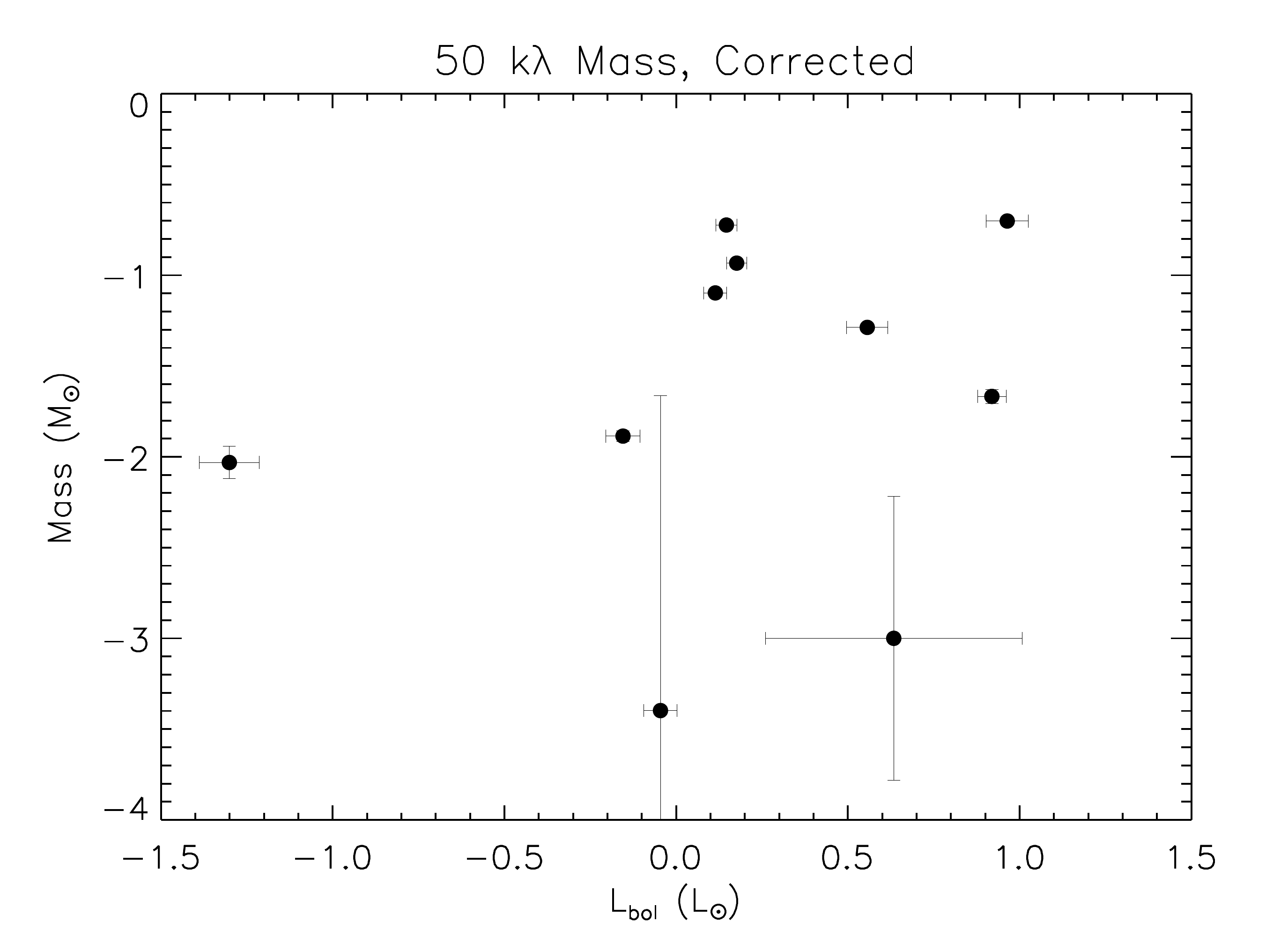}
\includegraphics[scale=0.275]{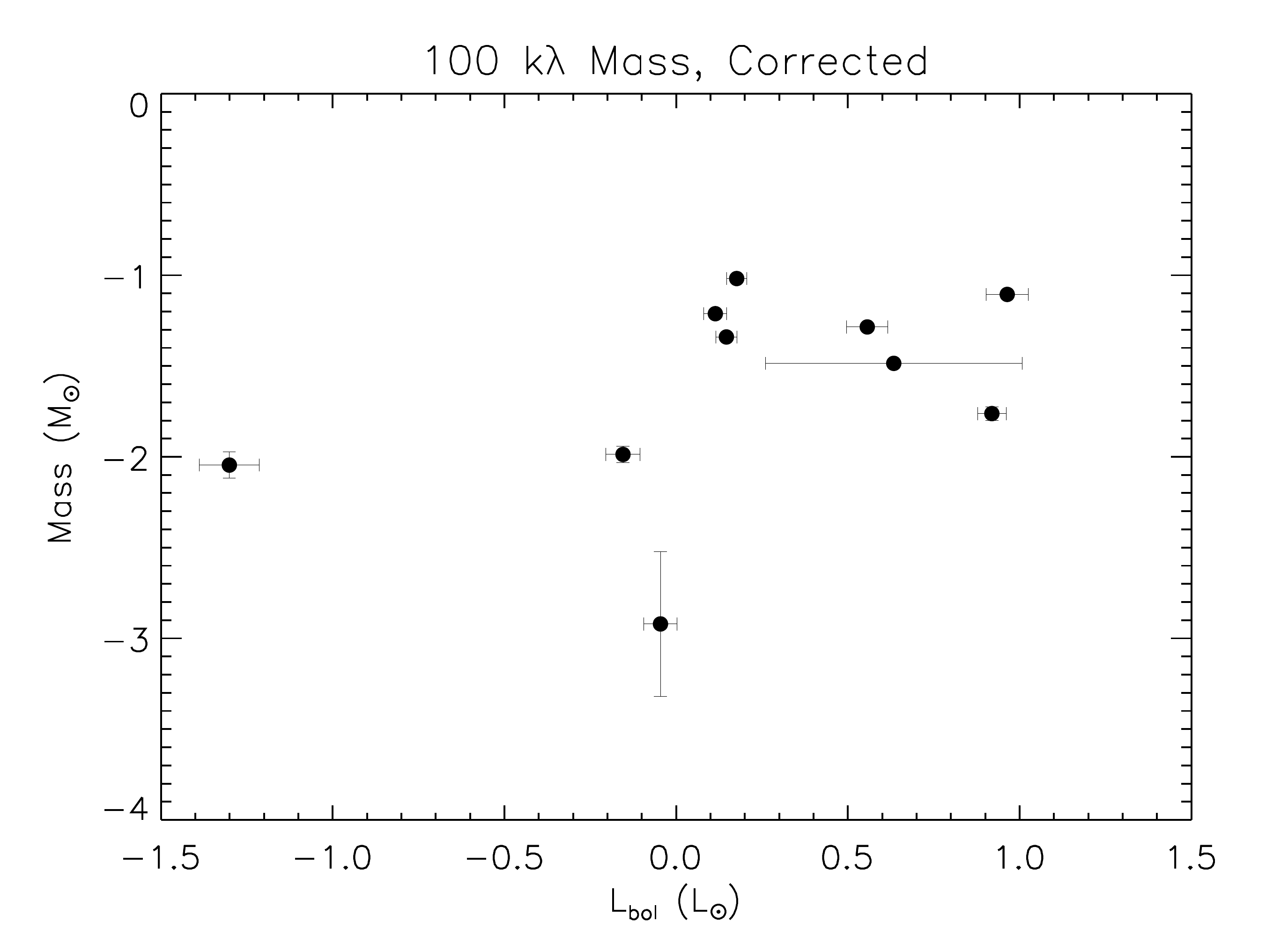}
\includegraphics[scale=0.275]{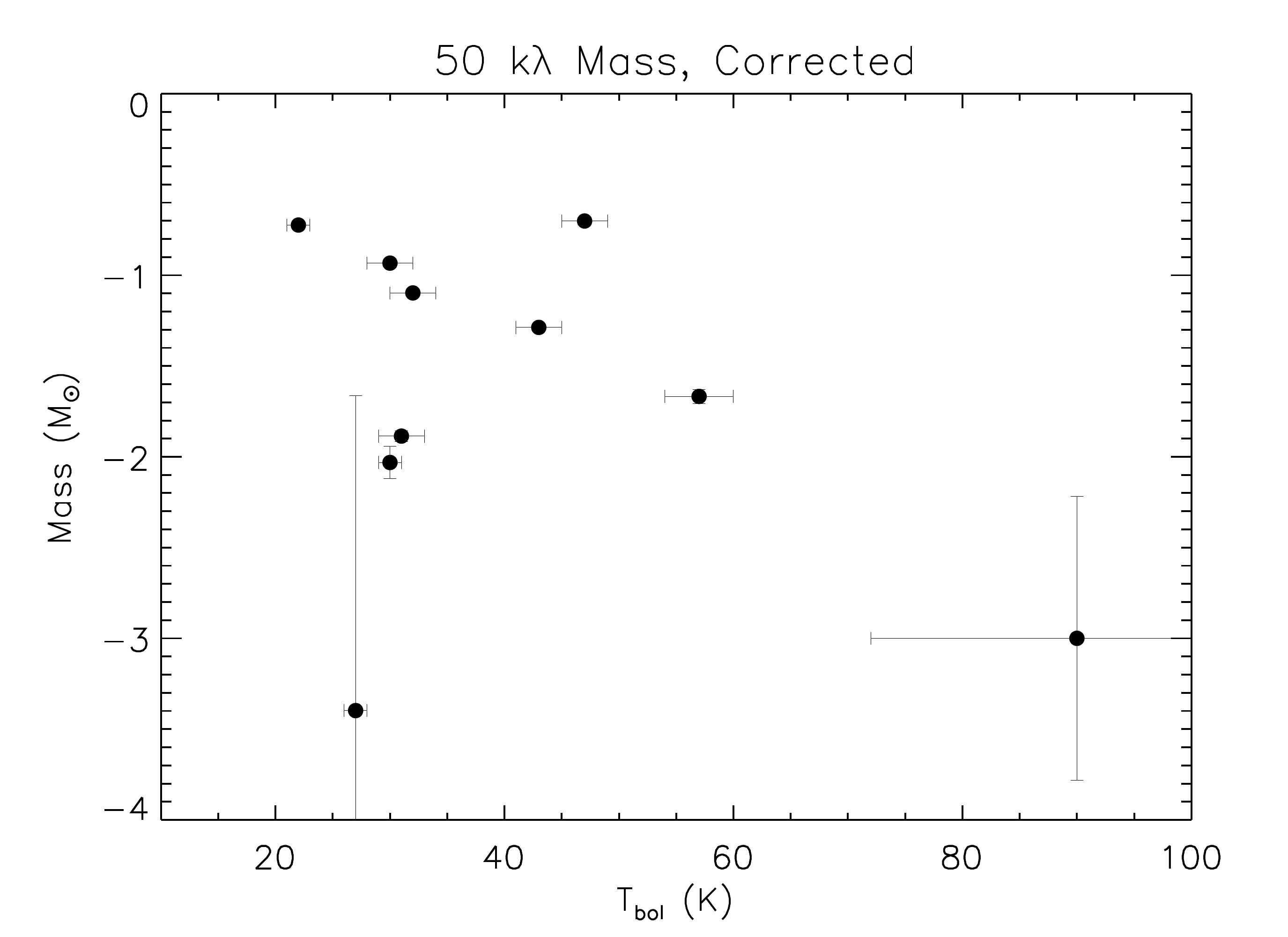}
\includegraphics[scale=0.275]{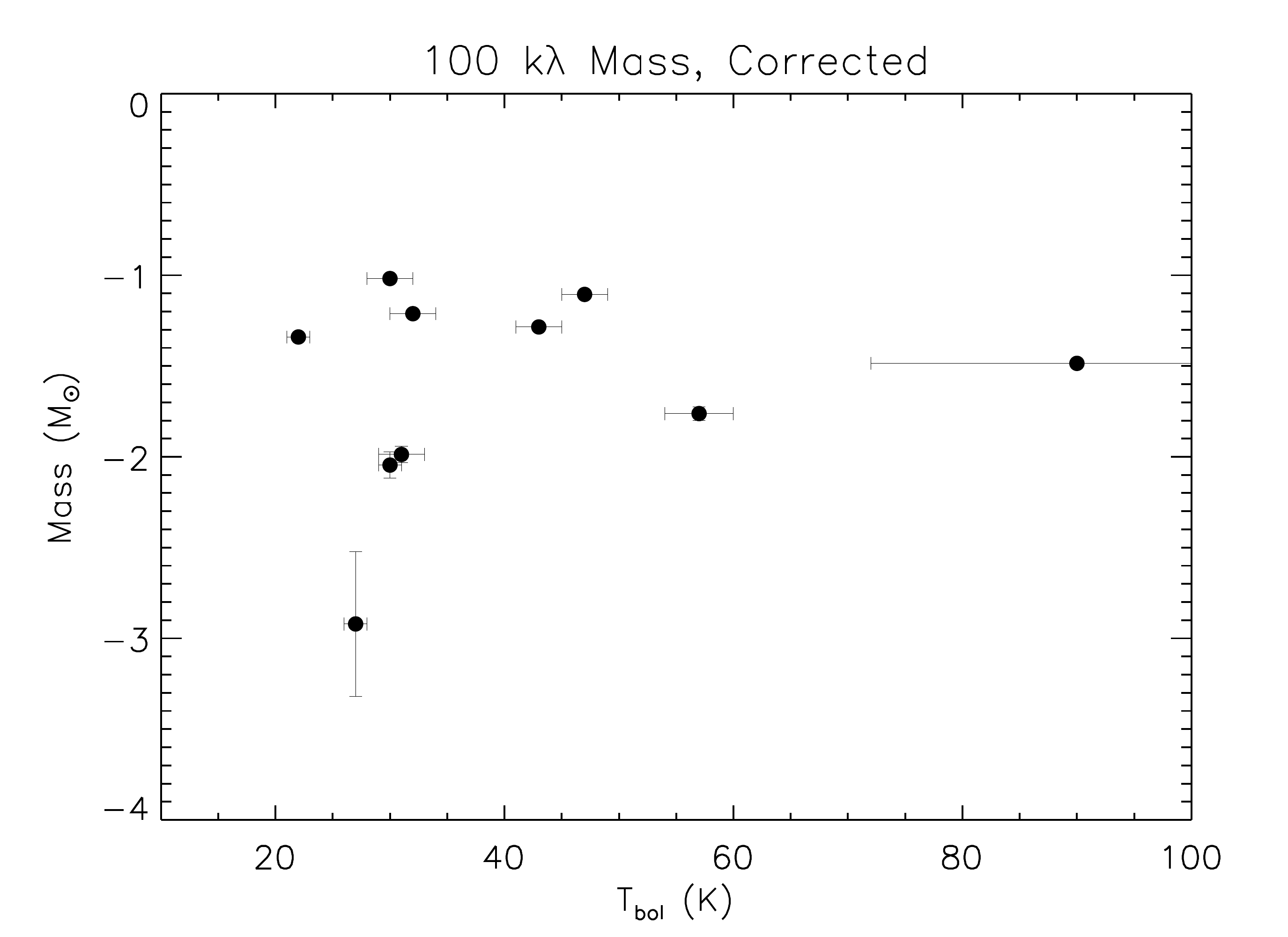}
\end{center}
\caption{Plot of (corrected) 50 k$\lambda$ disk masses versus L$_{bol}$ (top left)
and T$_{bol}$ (bottom left); the same plots but
for 100 k$\lambda$ masses are plotted on the right. 
There is no apparent trend in the inferred disk
masses as a function of L$_{bol}$ or T$_{bol}$. \citet{jorgensen2009} also
found inconclusive evidence for Class 0 disk masses to trend with these parameters.
The lack of correlation indicates that the disk properties (at least mass)
may not depend specifically on evolutionary state and luminosity, but rather initial 
conditions.}
\label{disk-mass-lbol-tbol}
\end{figure}

\begin{figure}
\begin{center}
\includegraphics[scale=0.29]{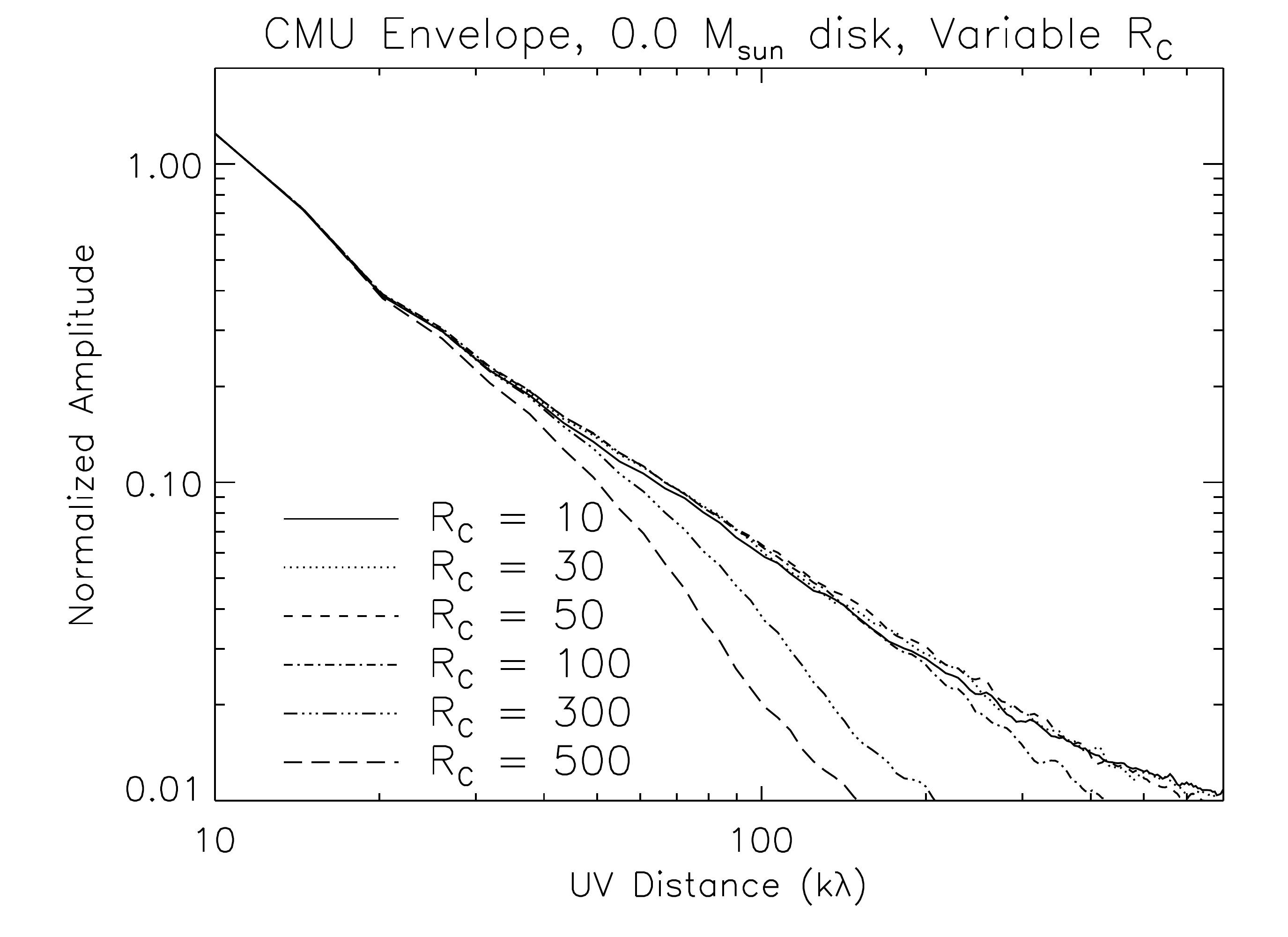}
\includegraphics[scale=0.29]{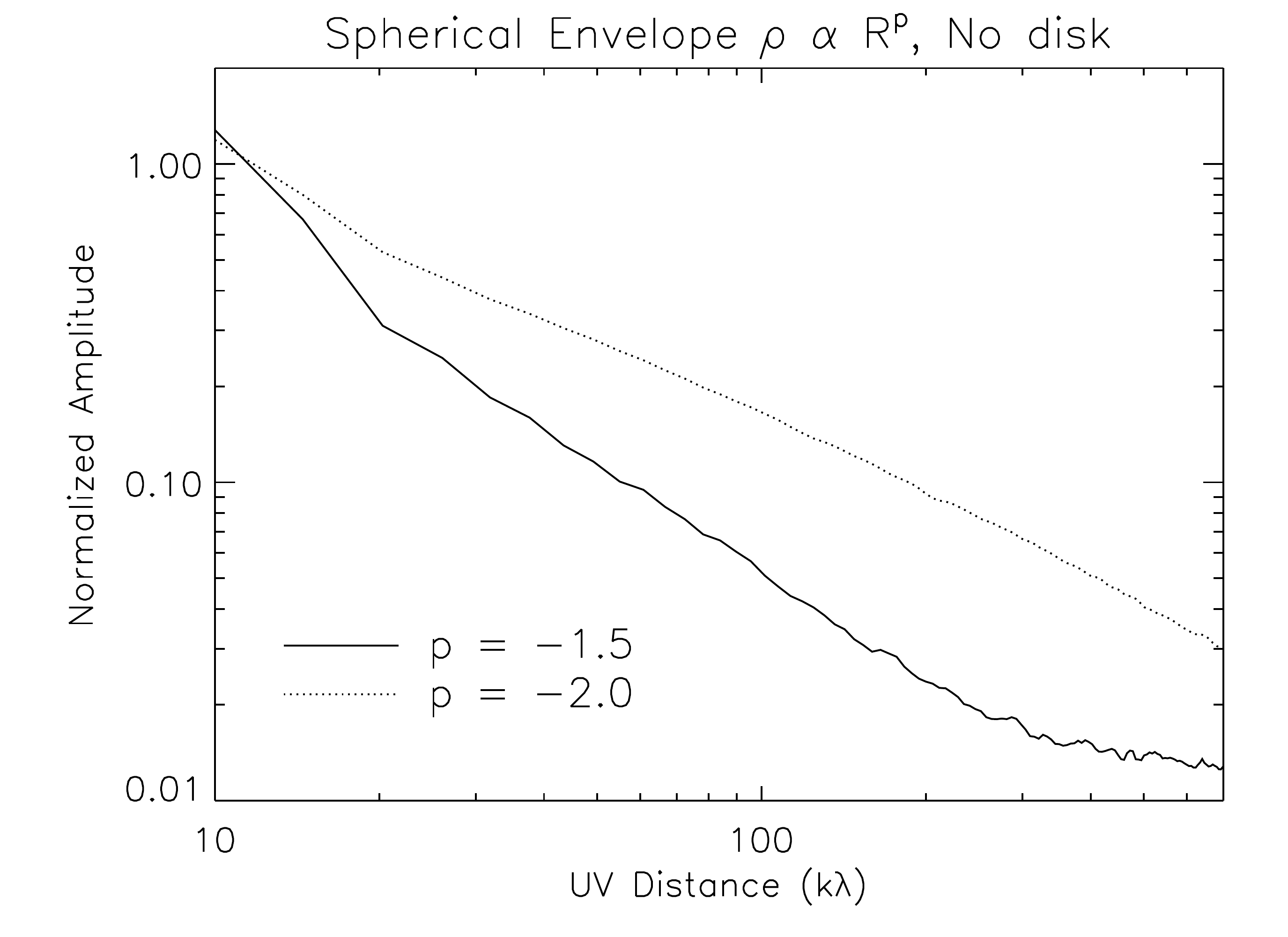}
\includegraphics[scale=0.29]{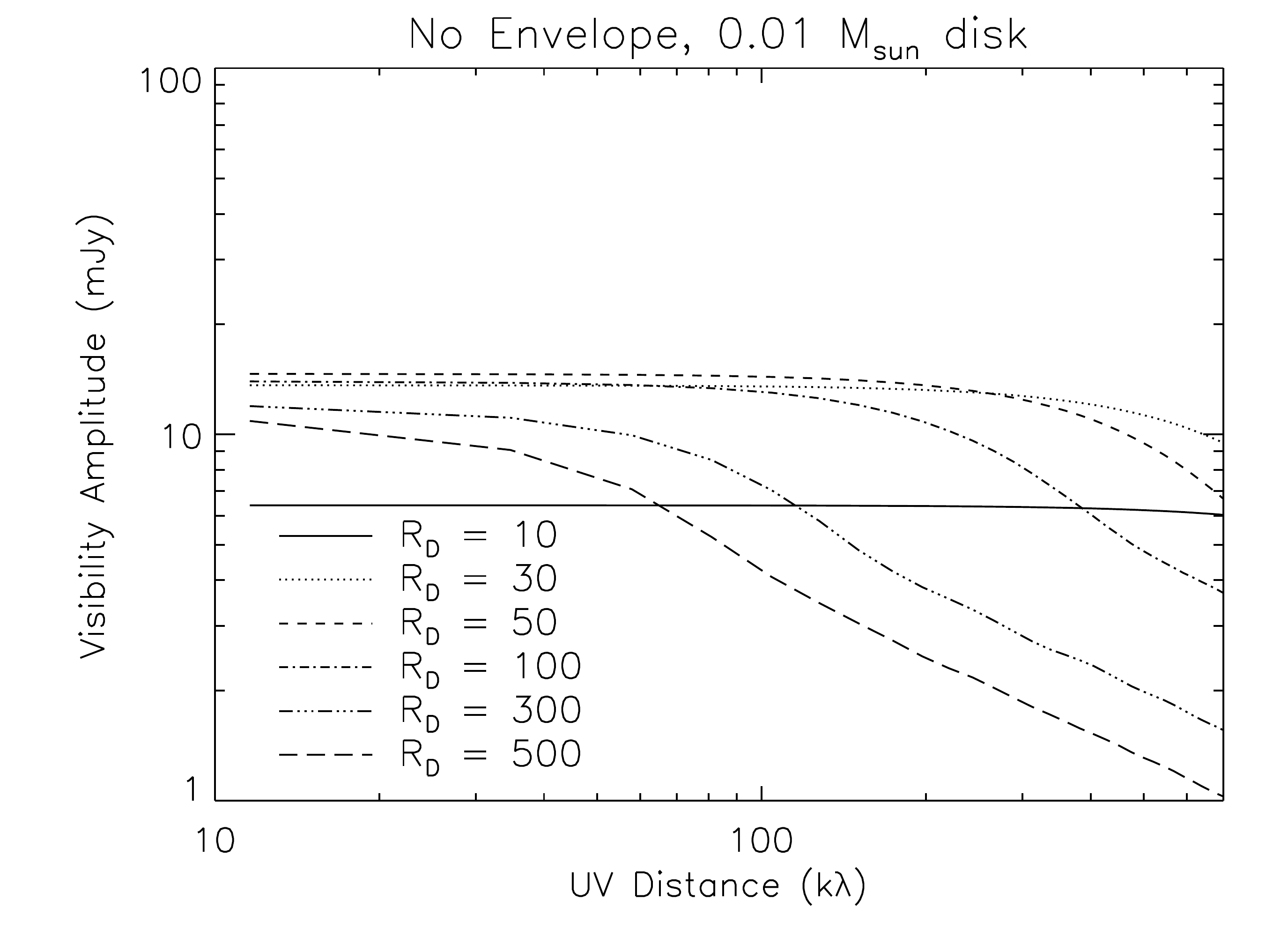}
\includegraphics[scale=0.29]{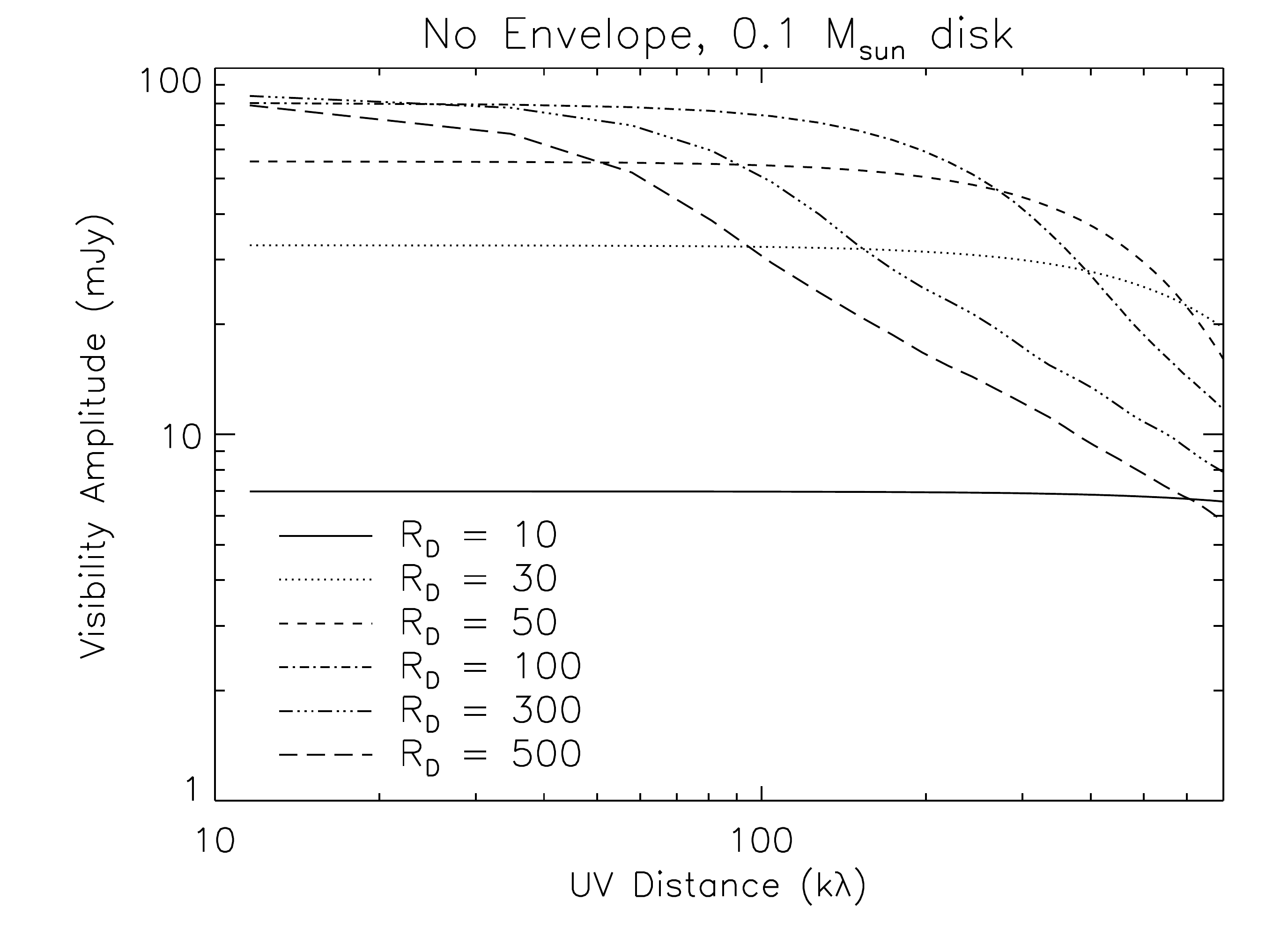}
\end{center}
\caption{Visibility amplitudes for envelope models with and without disk components. The upper left
shows a standard rotating collapse model (CMU) without a disk component. The upper right
shows power-law envelopes with density profiles proportional to R$^{-1.5}$ and R$^{-2.0}$ with
no embedded disk. The lower left shows disk-only models with sizes between 10 AU and 500 AU with 
a total mass of 0.01 $M_{\sun}$. The lower right shows disk-only models having a total 
mass of 0.1 $M_{\sun}$. The visibility amplitudes of the disk-only models are not normalized in order to show
that for fixed mass, the disks with small radii have high optical depths thereby reducing the emergent
1.3 mm flux.}
\label{disk-models}
\end{figure}

\begin{figure}
\begin{center}
\includegraphics[scale=0.29]{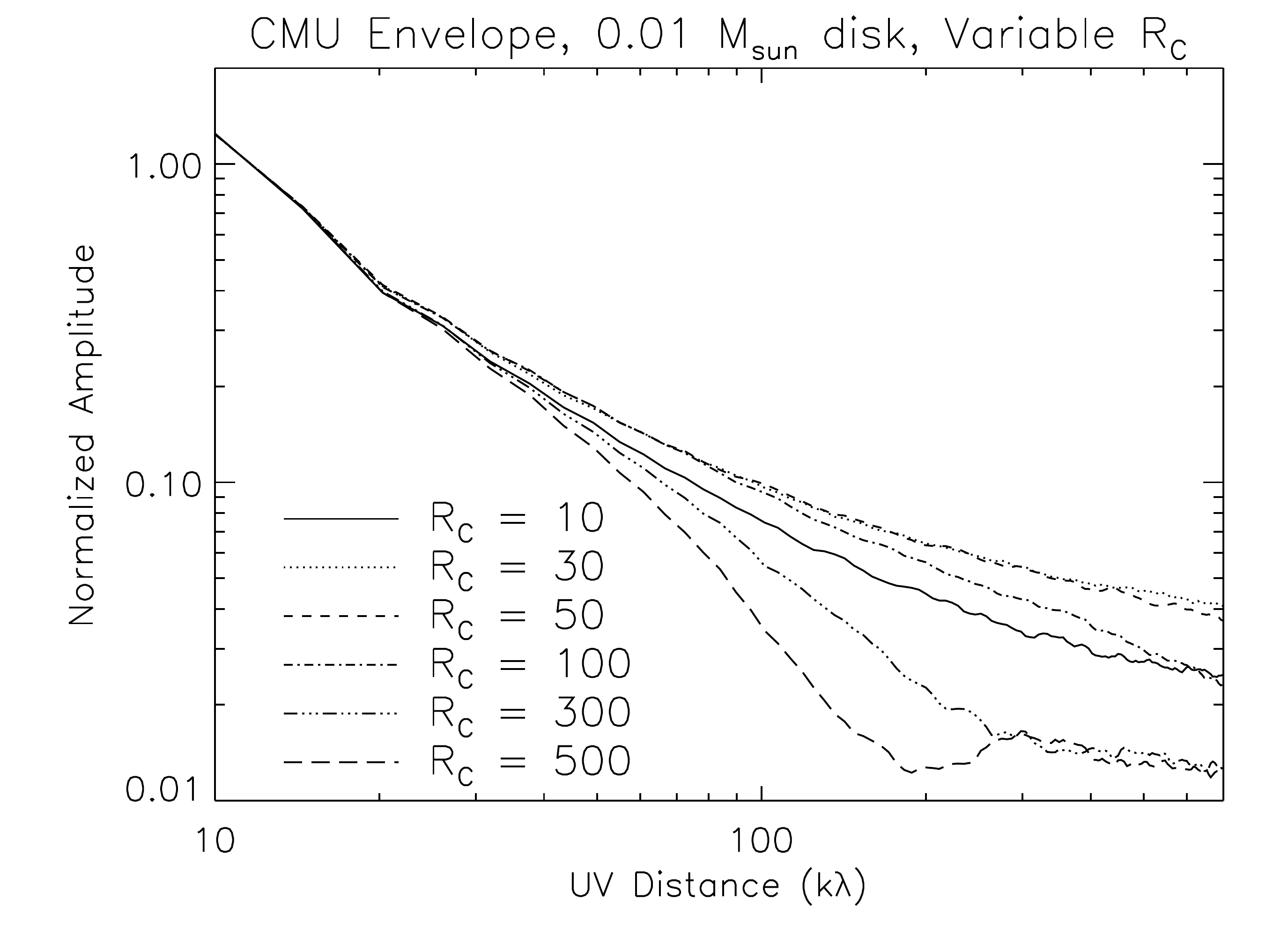}
\includegraphics[scale=0.29]{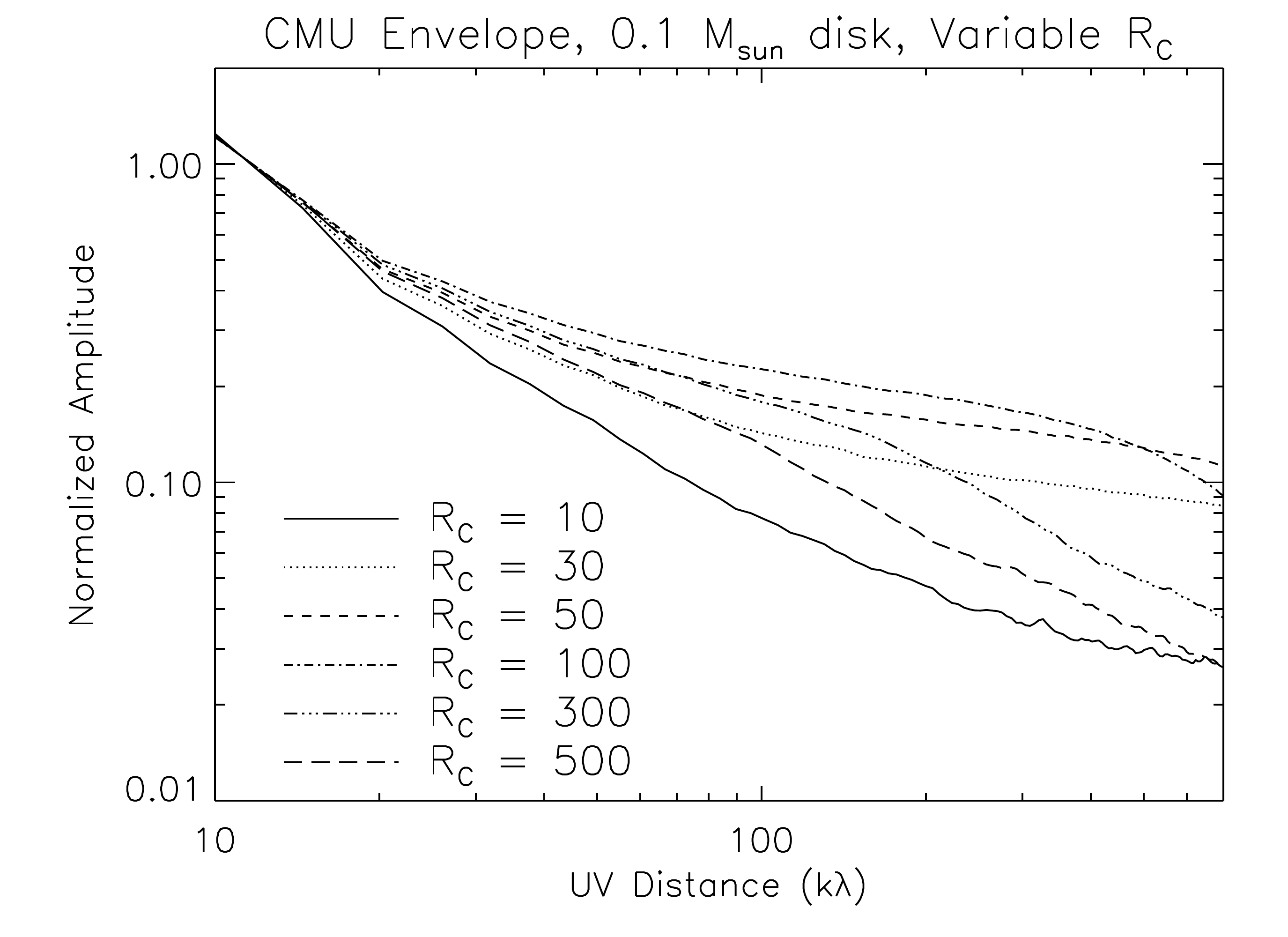}
\end{center}
\caption{Visibility amplitudes for CMU Envelope models with disk components. 
The left panel shows CMU envelope models with an embedded disk of sizes between 10 AU and 500 AU with 
a total mass of 0.01 $M_{\sun}$. The right panel shows CMU envelope models with disks having a total 
mass of 0.1 $M_{\sun}$. The envelopes have a total mass of 5.25 $M_{\sun}$.}
\label{cmu-models}
\end{figure}

\begin{figure}
\begin{center}
\includegraphics[scale=0.29]{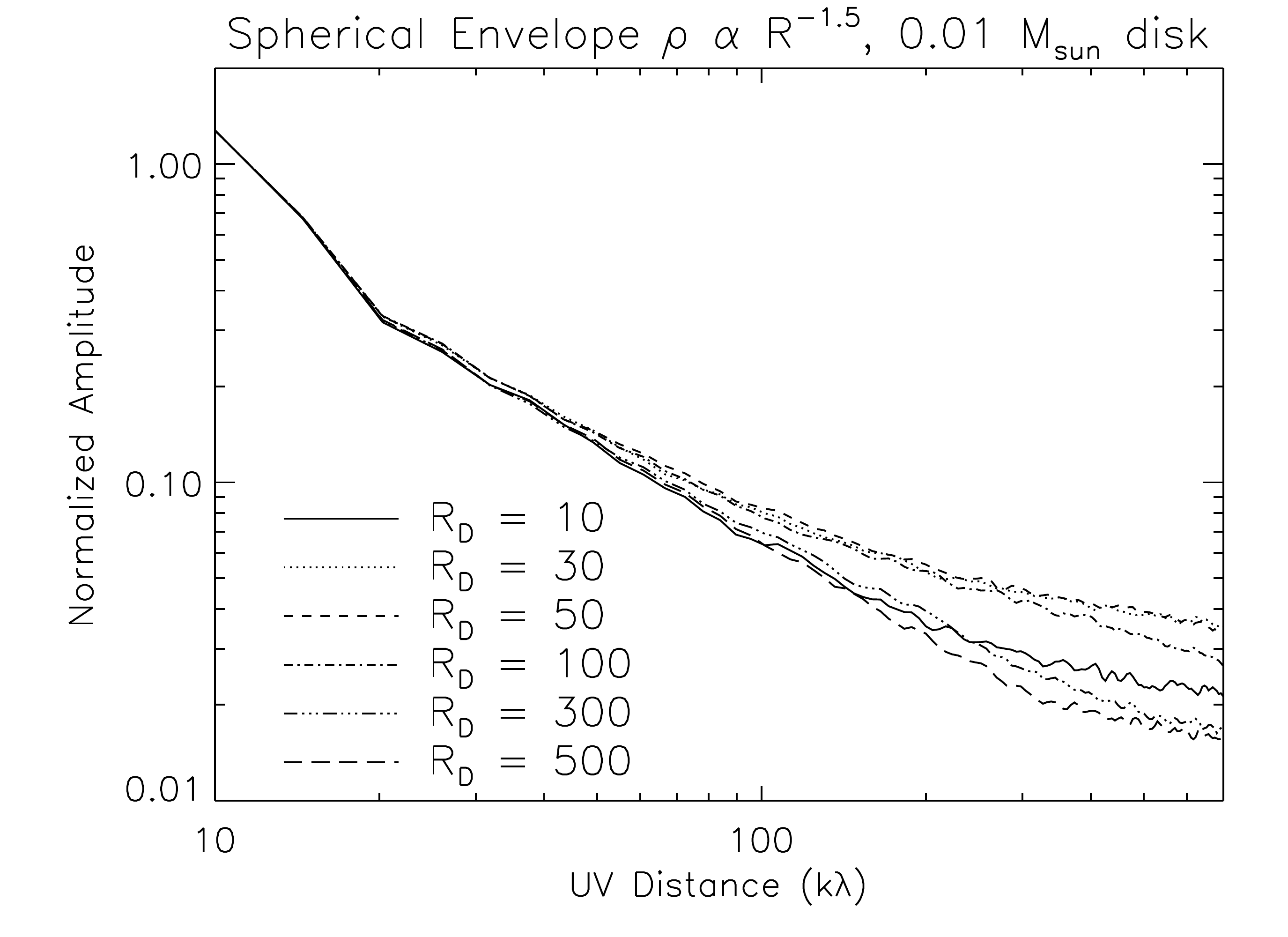}
\includegraphics[scale=0.29]{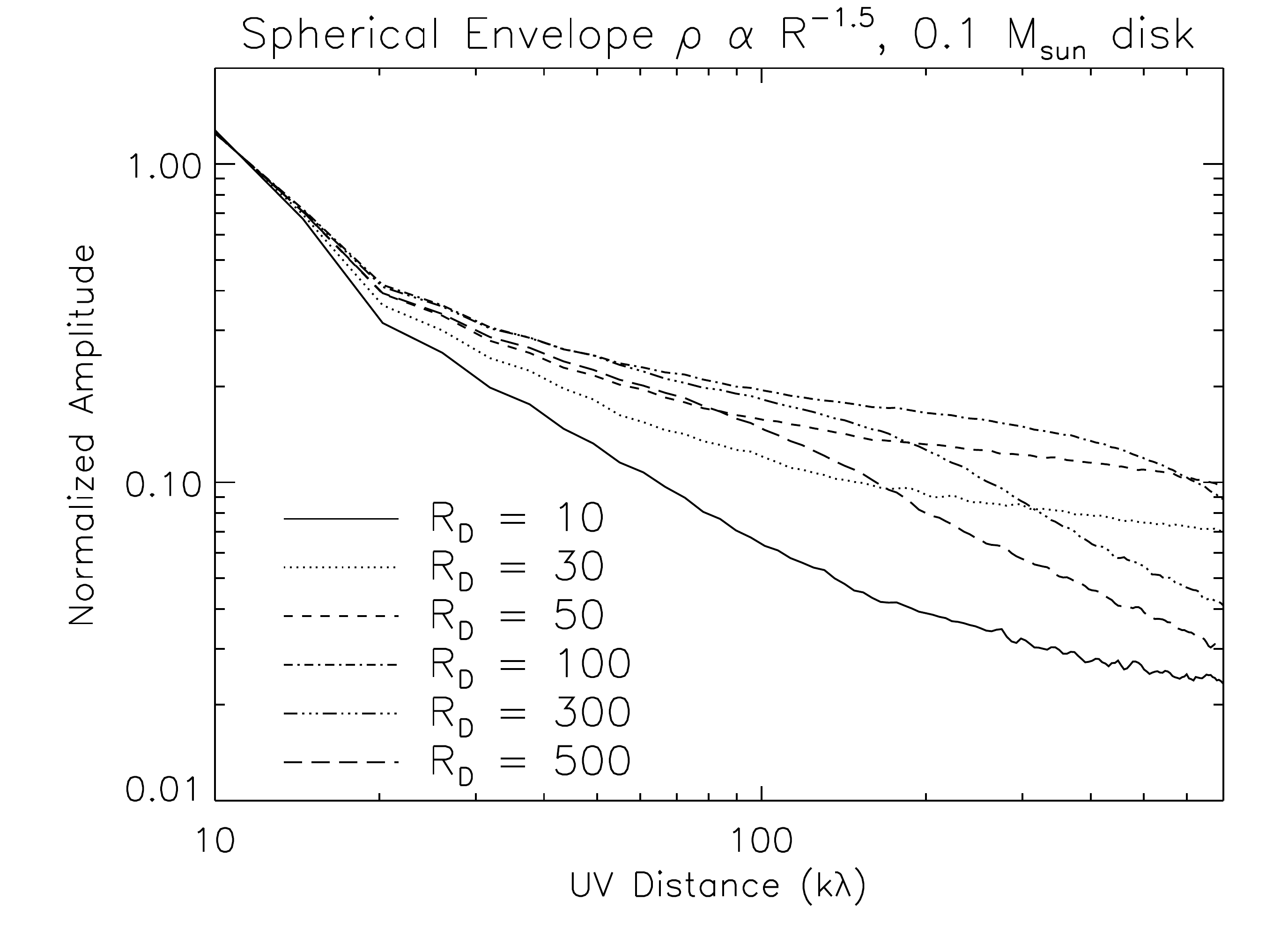}
\includegraphics[scale=0.29]{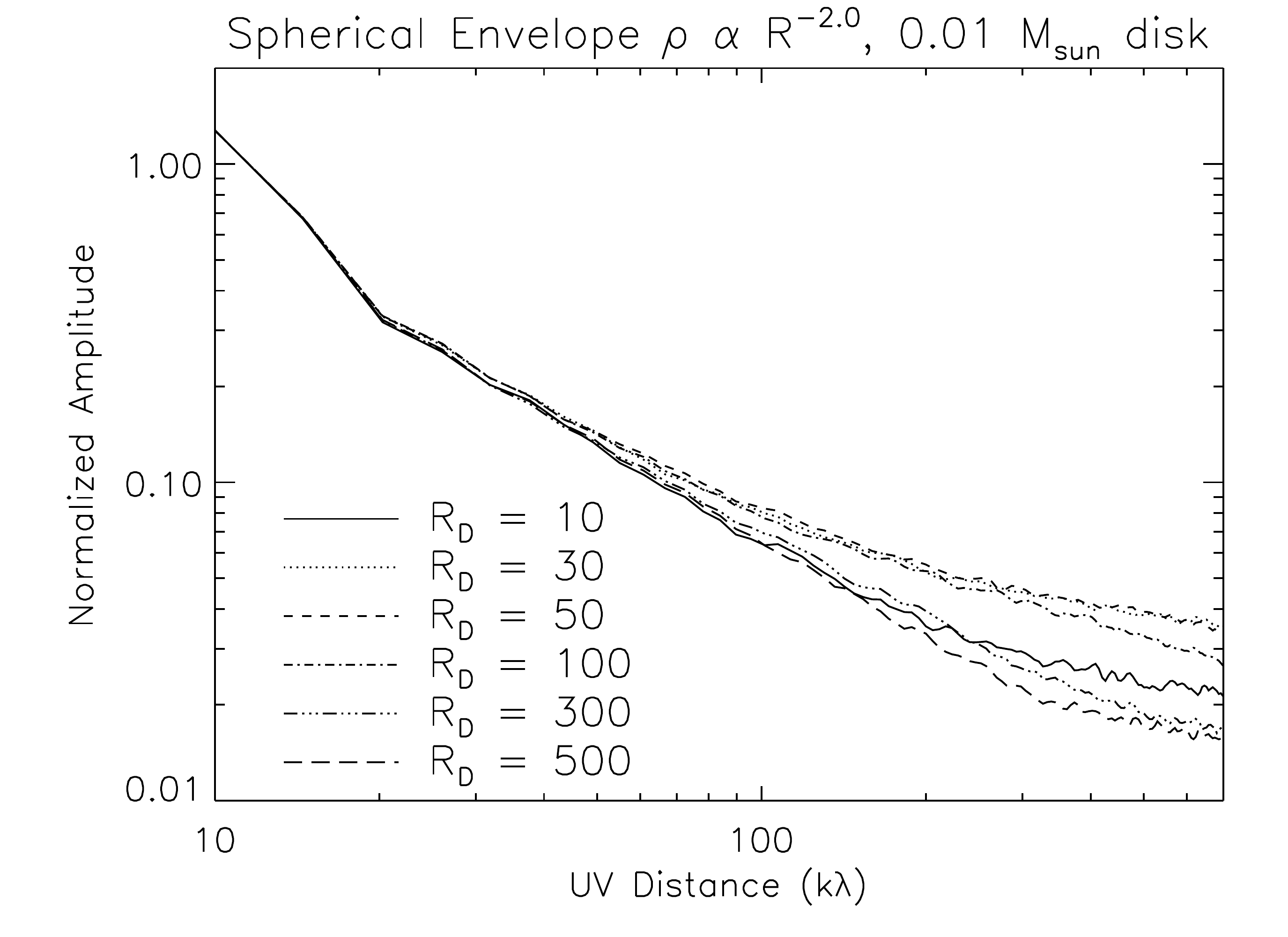}
\includegraphics[scale=0.29]{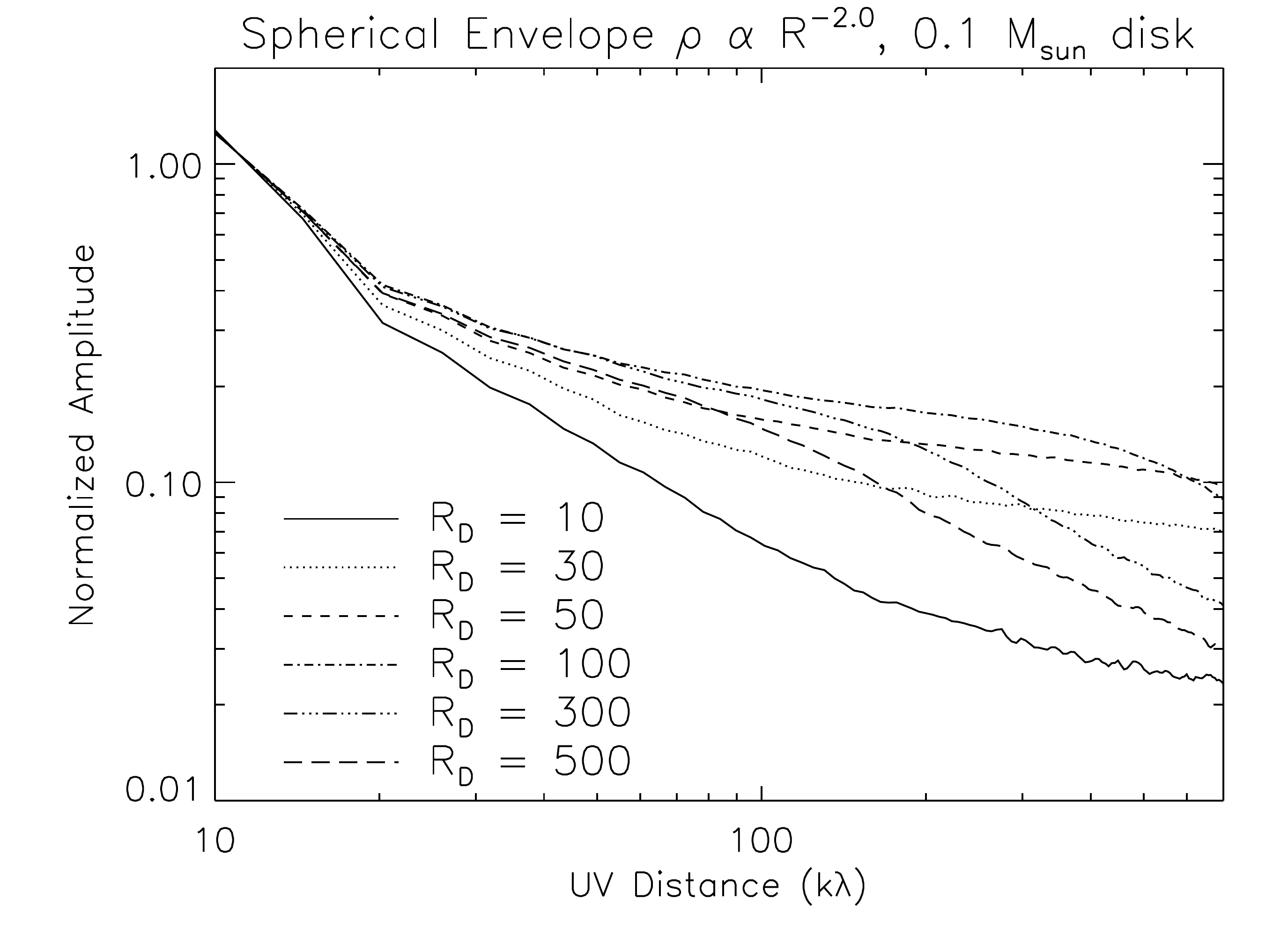}
\end{center}
\caption{Visibility amplitudes for Power-law Envelope models with disk components. 
The upper left panel shows $\rho$ $\propto$ $R^{-1.5}$ envelope models with embedded disks of sizes between 10 AU and 500 AU with 
a total disk mass of 0.01 $M_{\sun}$. The upper right panel is the same as the upper left, except that the disks have
a total mass of 0.1 $M_{\sun}$. The lower left panel shows $\rho$ $\propto$ $R^{-2.0}$ 
envelope models with embedded disks having a total mass of 0.01 $M_{\sun}$. 
The lower right panel is the same as the left, except that the disks have
a total mass of 0.1 $M_{\sun}$. The envelopes have a total mass of 5.25 $M_{\sun}$.}
\label{power-models}
\end{figure}

\begin{figure}
\begin{center}
\includegraphics[scale=0.75]{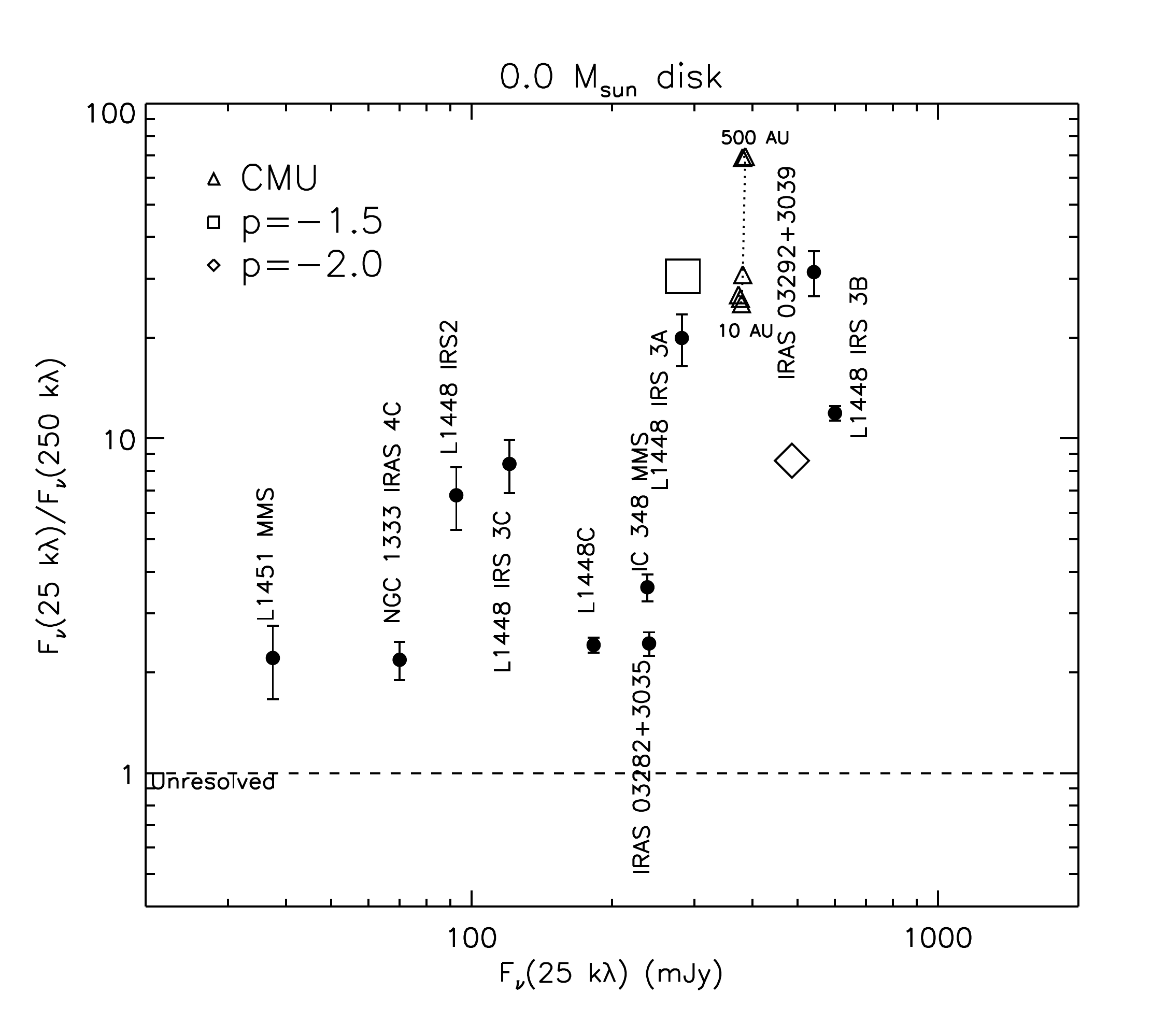}
\end{center}
\caption{Visibility amplitude ratio plots for the visibility amplitudes at 
25 k$\lambda$ and 250 k$\lambda$; corresponding to 10\arcsec\ 
(2300 AU) and 1\arcsec\ (230 AU) scales, respectively. The sources are plotted as the solid circles
and the models are plotted as the crosses (disk only), triangles (CMU models), squares ($\rho$ $\propto$ $R^{-1.5}$), and
diamonds ($\rho$ $\propto$ $R^{-2.0}$). This figure (a) shows models without 
a disk component, but still a rotationally-flattened region for the CMU case.  
Figure 13b shows models with a 0.01 $M_{\sun}$ disk component, and Figure 13c
shows models with a 0.1 $M_{\sun}$ disk component. These plots imply that
sources with a 25 k$\lambda$ to 250 k$\lambda$ ratio less than $\sim$8 require a disk or another
compact component, or a density profile steeper than $\rho$ $\propto$ $R^{-2.0}$. Envelope models
with an embedded disk can be made to have lower ratios by decreasing the envelope mass, emphasizing
the disk component. The centrifugal/disk radii are labeled at 10 AU, 100 AU, and 500 AU for the CMU and disk models
such that the variation of the visibility amplitude ratios can be traced. The F$_{\nu}$(25k$\lambda$) values for the
disk-only models have been scaled by the factors noted in each plot such that the points for all radii are visible.
}
\label{uvratios}
\end{figure}

\begin{figure}
\figurenum{13b}
\begin{center}
\includegraphics[scale=0.75]{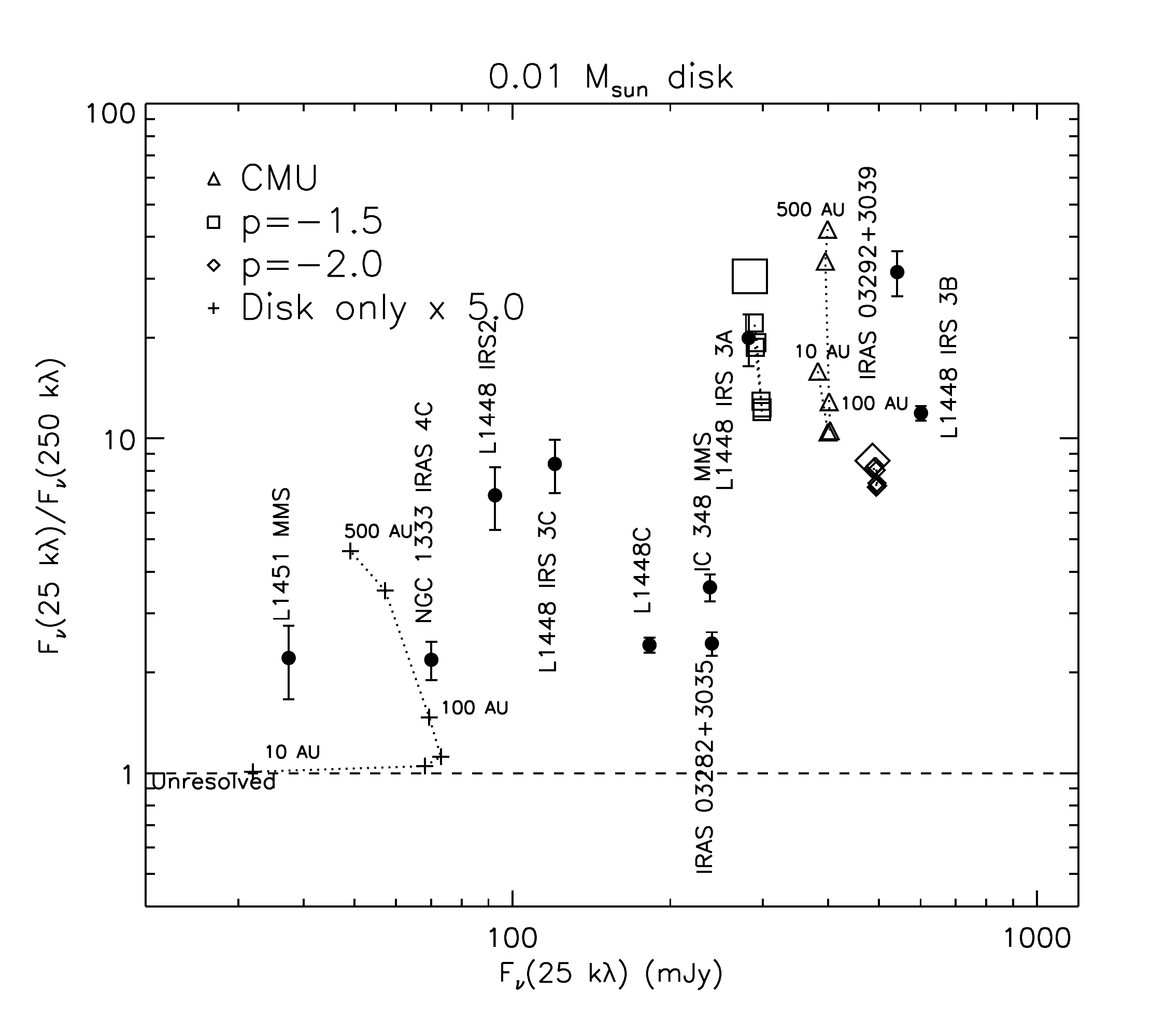}

\end{center}
\caption{}

\end{figure}

\begin{figure}
\figurenum{13c}
\begin{center}
\includegraphics[scale=0.75]{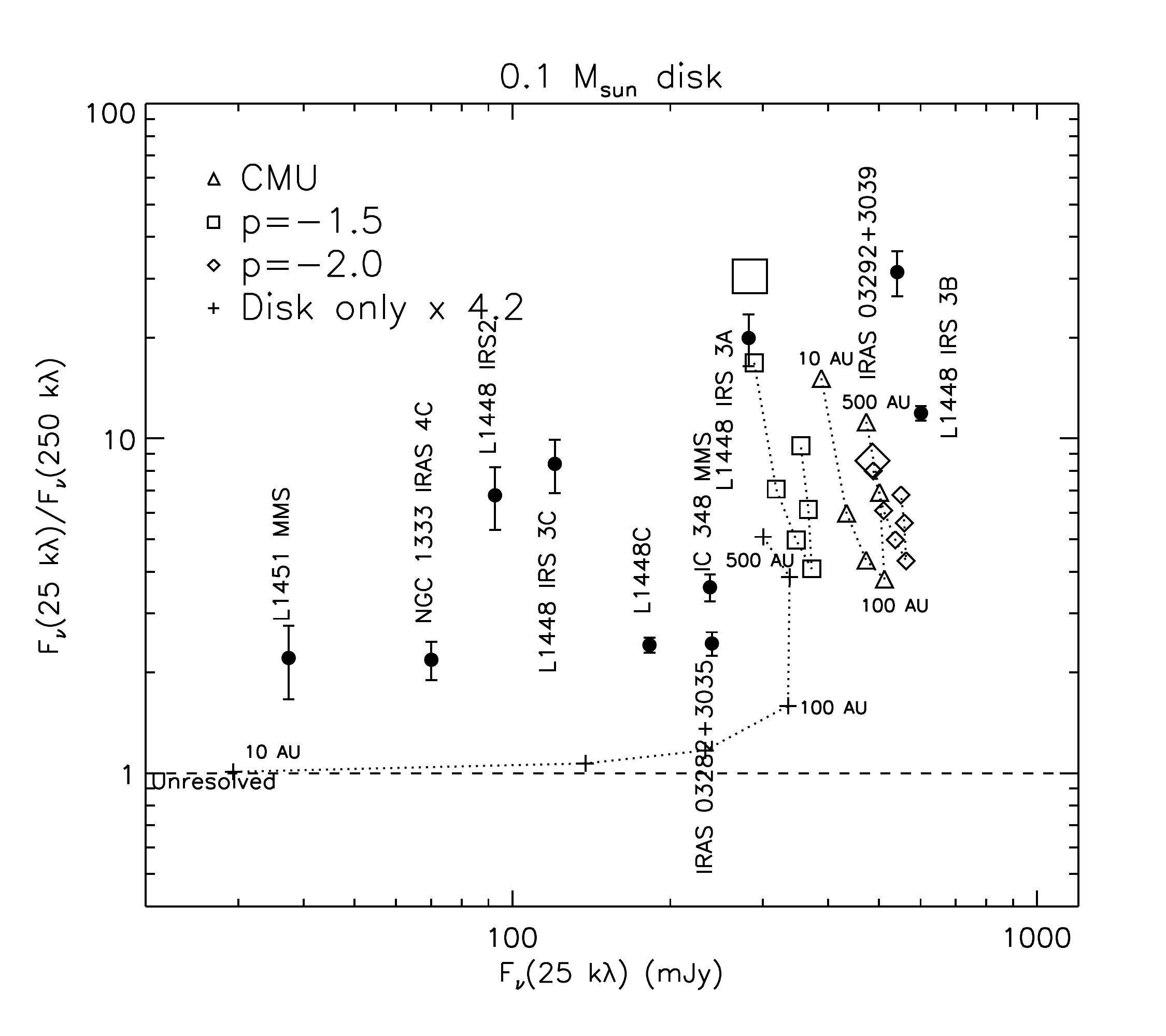}

\end{center}
\caption{}

\end{figure}

\begin{figure}

\begin{center}
\includegraphics[scale=0.75]{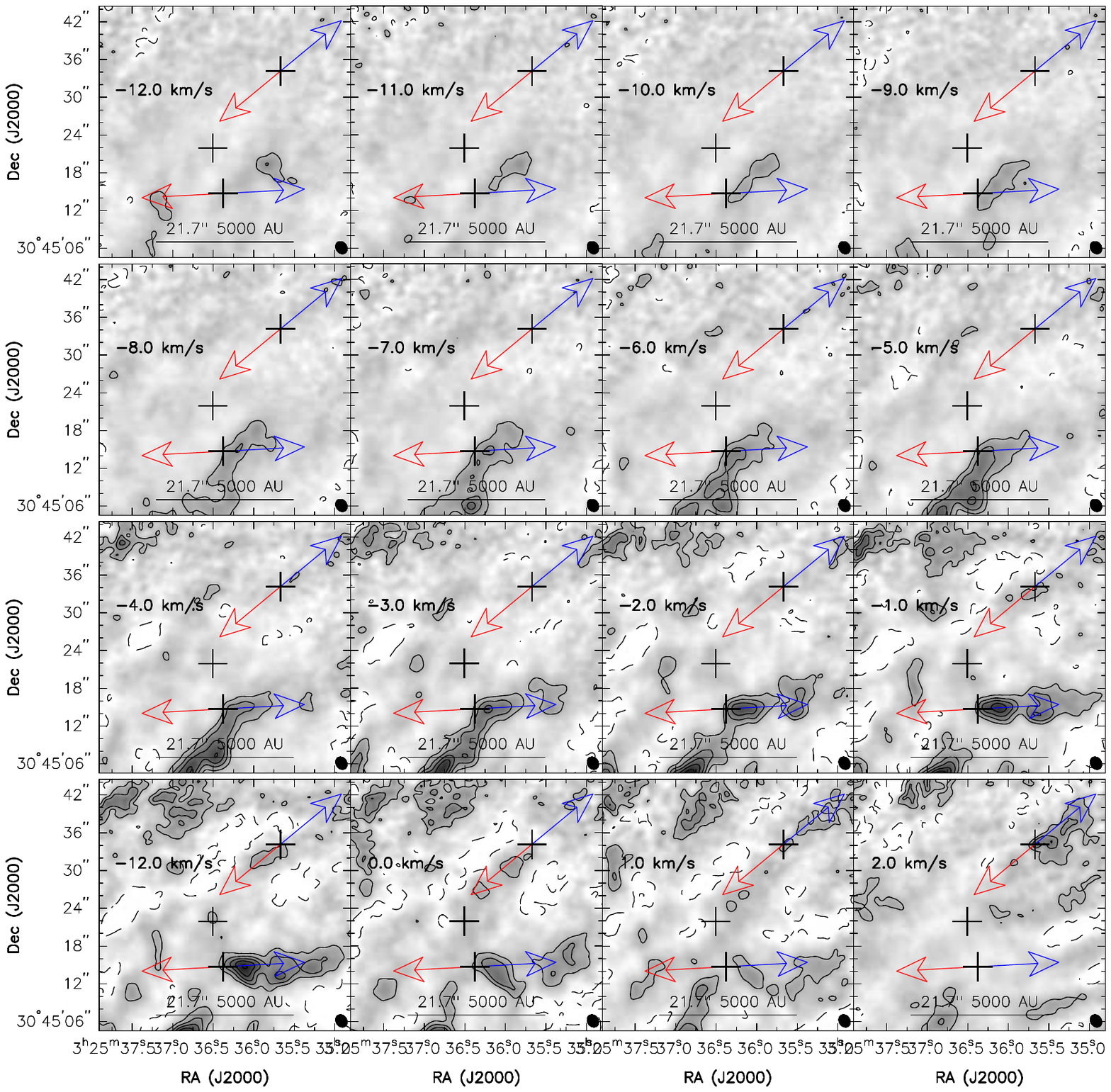}
\end{center}
\caption{Outflows in the L1448 IRS3 region. The sources are marked with crosses;
the northern most source is L1448 IRS3C, the middle source is L1448 IRS3A, 
and the southern-most source is L1448 IRS 3B. The blue-shifted emission is shown in
Figure 14a, while the red-shifted emission is shown in Figure 14b. The blue-shifted outflow
from IRS 3B becomes apparent at -4 \kms\ and the blue-shifted outflow
from IRS3C becomes visible at 1 \kms. The emission visible south of IRS3B, may be from 
L1448C. The red-shifted emission is apparent both IRS 3C and IRS3B at  7 \kms\ out to 17 \kms. There
is some red-shifted emission associated with IRS3A between 7 \kms\ and 9 \kms, but it is unclear
if this is outflow emission from this source. The contours are [-5, 5, 10, 15, 20, ...] $\times$ 
$\sigma$ and $\sigma$ = 0.1 Jy beam$^{-1}$; the beam is 2\farcs1 $\times$ 1\farcs7. 
The blue and red arrows denote the direction of the blueshifted
and redshifted outflows, respectively.}
\label{IRS3-outflows}
\end{figure}

\begin{figure}
\figurenum{14b}
\begin{center}
\includegraphics[scale=0.75]{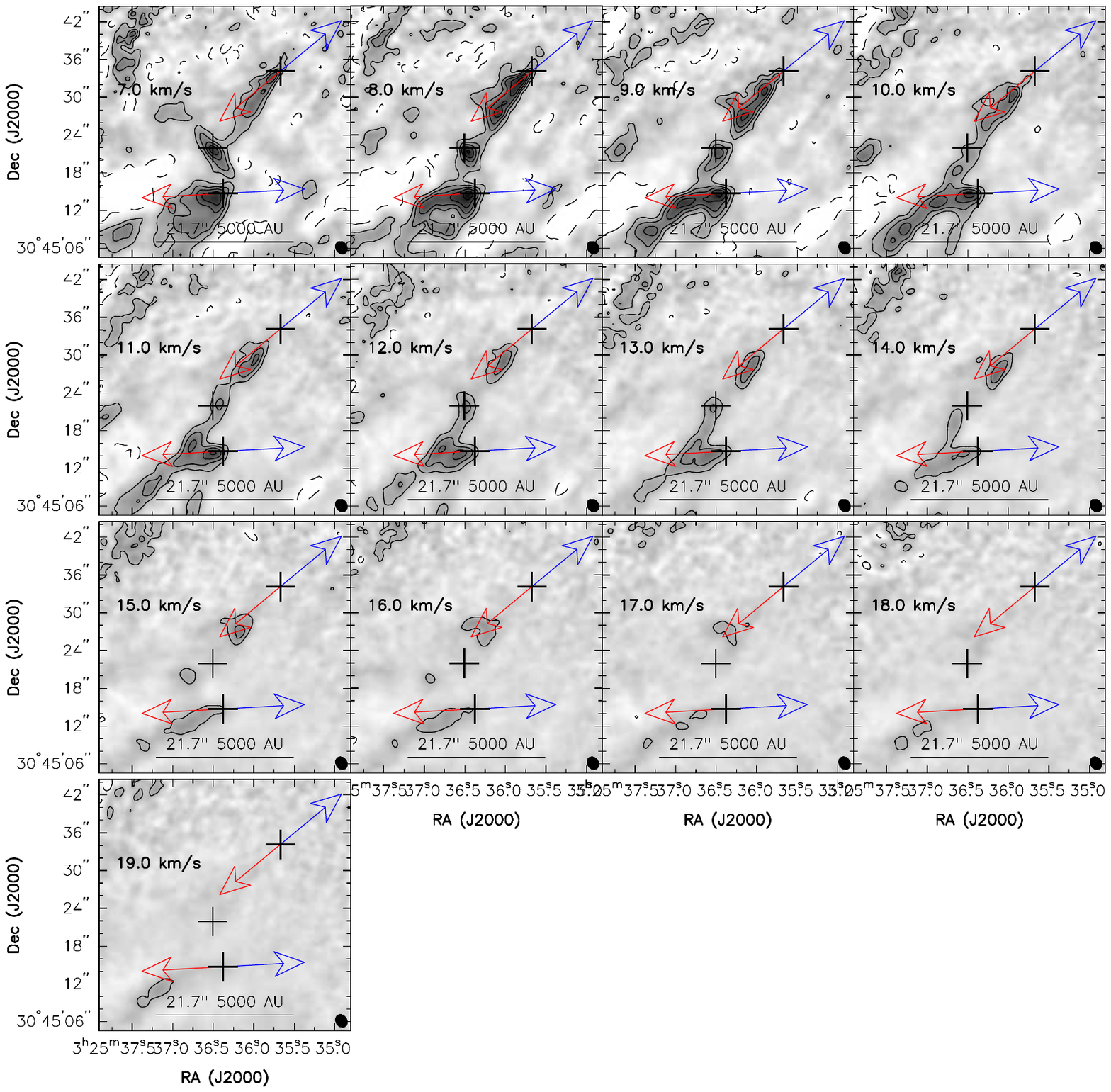}
\end{center}
\caption{}

\end{figure}

\begin{figure}
\begin{center}
\includegraphics[scale=0.75]{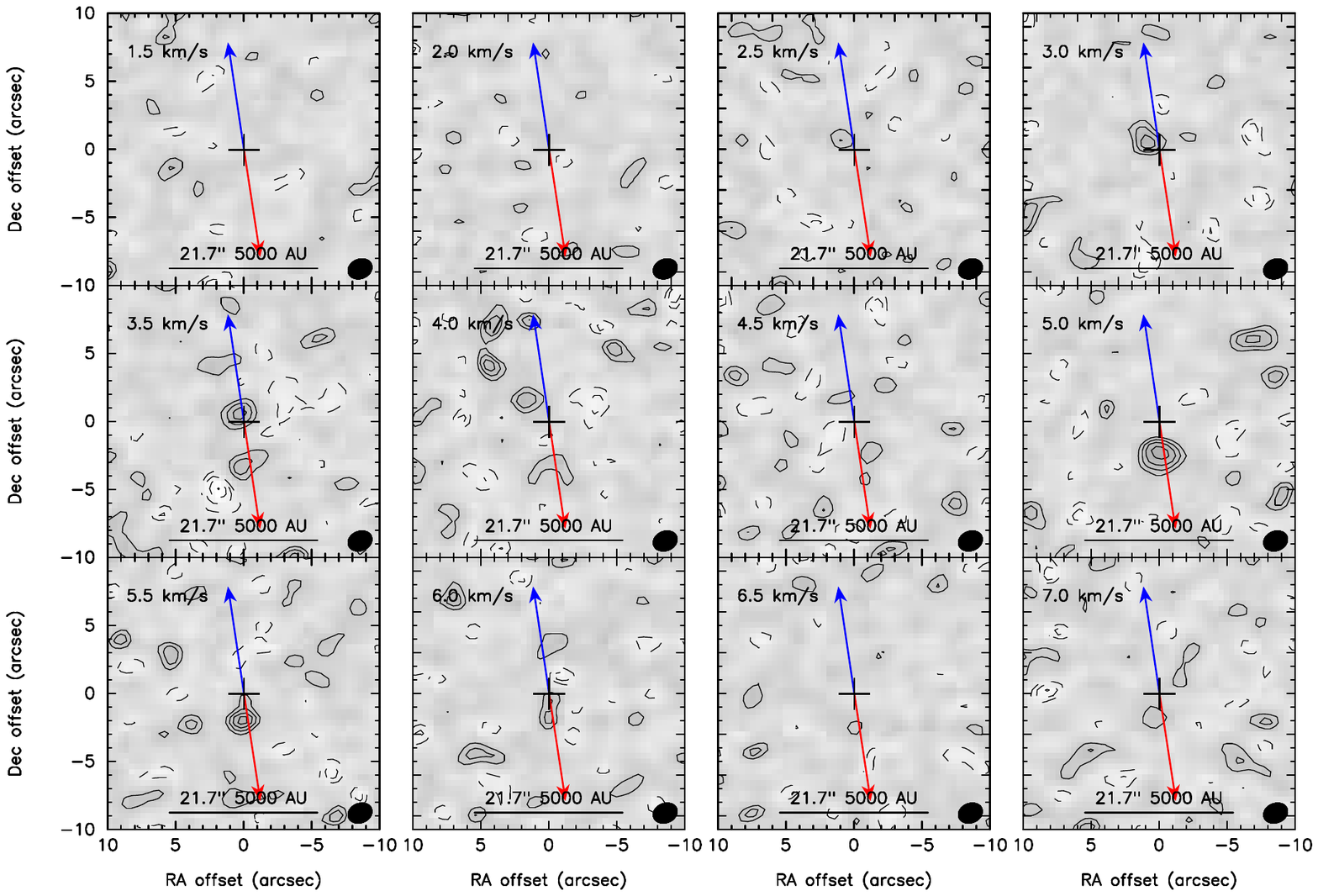}
\end{center}
\caption{Compact outflow from L1451-MMS. The outflow is detected to be in the same direction as
observed by \citet{pineda2011} and with a similar offset to the east of the 
1.3 mm continuum peak. The contours are 
$\pm$[2, 3, 4, 5] $\times$ $\sigma$ and $\sigma$=0.1 Jy beam$^{-1}$; 
the beam is 0\farcs92 $\times$ 0\farcs72. The blue and red arrows denote the direction of the blueshifted
and redshifted outflows, respectively.}
\label{L1451MMS-outflow}
\end{figure}

\begin{figure}
\begin{center}
\includegraphics[scale=0.75]{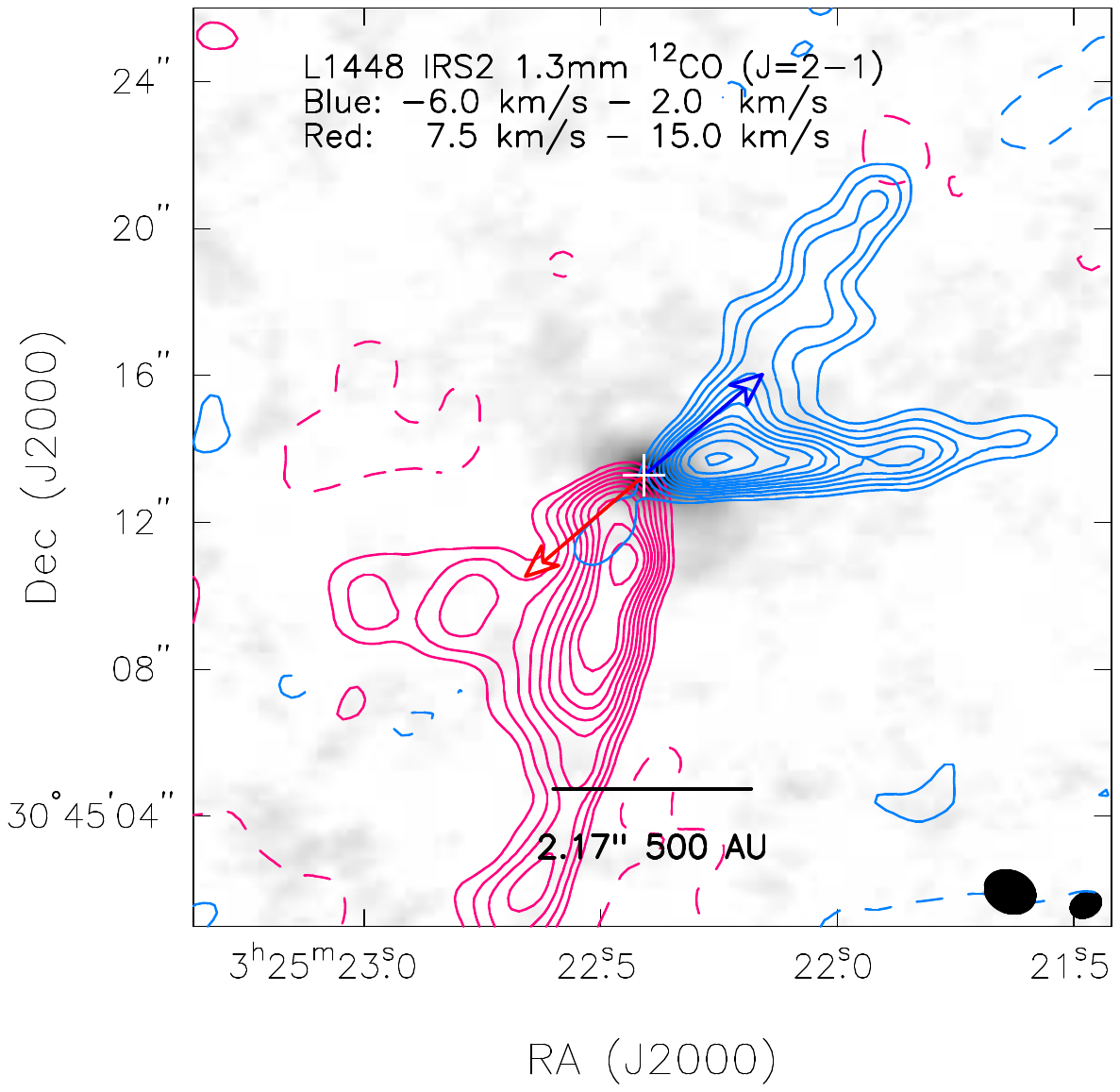}
\end{center}
\caption{Integrated intensity maps of $^{12}$CO emission toward L1448 IRS2 from the combined imaging of CARMA
and SMA datasets, overlaid on the CARMA 1.3 mm continuum map (grayscale). The emission maps reveal
a rather asymmetric appearance of the outflow cavities, opening wider to the southwest in
comparison to the northeast. Furthermore, the outflow vectors, which are centered on the
peak of the continuum emission, seem as if they would be better centered slightly to the
southwest, along the extended continuum emission. The contours are 
[-6, 6, 9, 12, 15, 18,...] $\times$ $\sigma$ and $\sigma$=2.47 K, 2.55 K for the red 
and blue-shifted integrated intensity maps, respectively. The beam for the CO emission is 
1\farcs42 $\times$ 1\farcs13 and the beam for the continuum is 0\farcs91 $\times$ 0\farcs68. 
The blue and red arrows denote the direction of the blueshifted
and redshifted outflows, respectively.}
\label{L1448IRS2-12CO}
\end{figure}

\begin{figure}
\begin{center}
\includegraphics[scale=0.75]{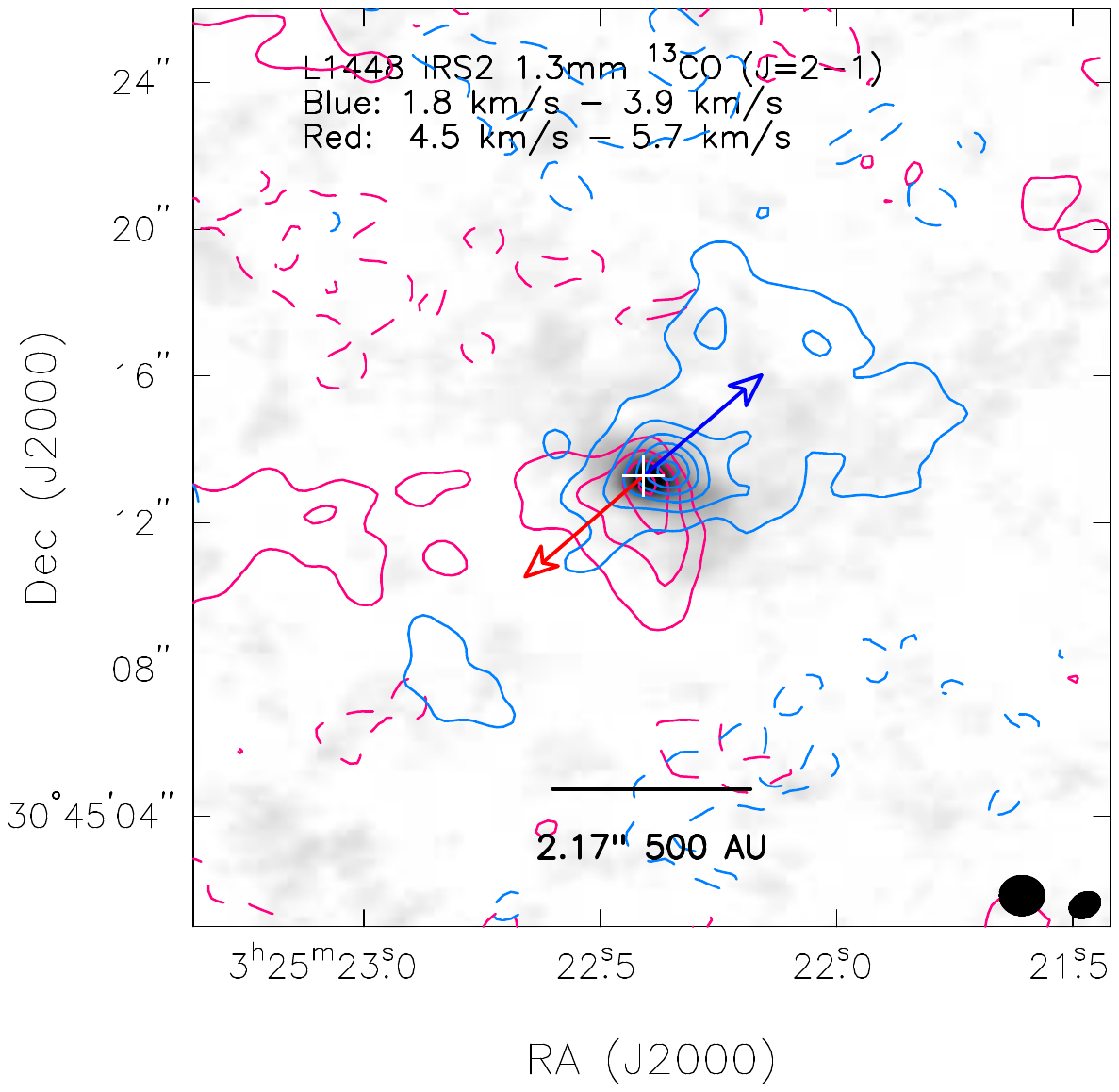}
\end{center}
\caption{Integrated intensity maps of $^{13}$CO emission toward L1448 IRS2 from the combined imaging of CARMA
and SMA datasets, overlaid on the CARMA 1.3 mm continuum map (grayscale). In the case of L1448 IRS2, the
$^{13}$CO emission is tracing the dense, lower velocity outflow. The positional offset of
the blue-shifted emission is consistent with that of the C$^{18}$O emission shown in Figure
\ref{IRS2-kinematics}. The contours are 
[-3, 3, 6, 9, 12, 15, 18,...] $\times$ $\sigma$ and $\sigma$=1.14 K, 1.44 K for the red 
and blue-shifted integrated intensity maps, respectively. The beam for the CO emission is 
1\farcs2 $\times$ 1\farcs06 and the beam for the continuum is 0\farcs91 $\times$ 0\farcs68. 
The blue and red arrows denote the direction of the blueshifted
and redshifted outflows, respectively.}
\label{L1448IRS2-13CO}

\end{figure}

\begin{figure}
\begin{center}
\includegraphics[scale=0.75]{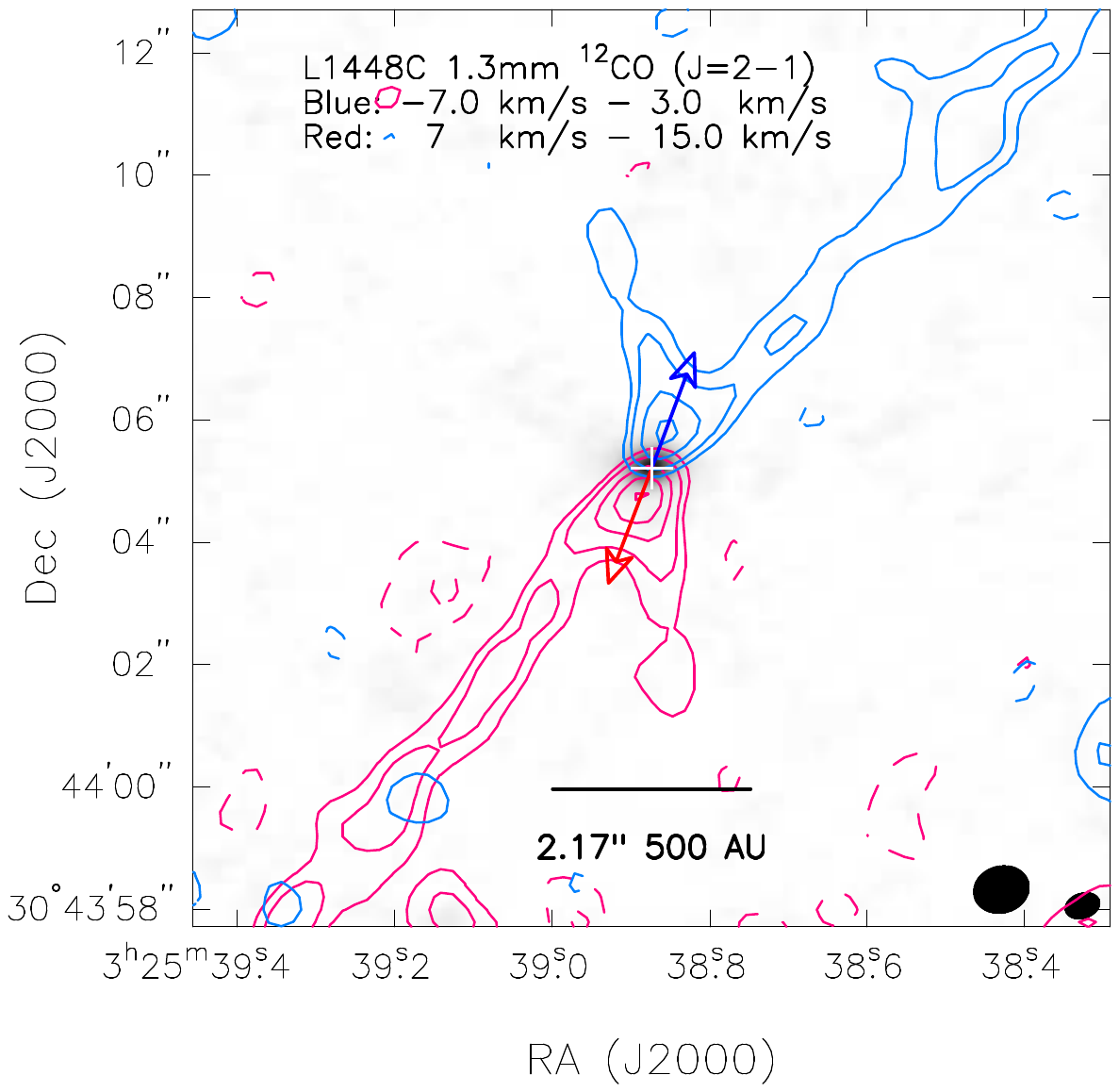}
\end{center}
\caption{Integrated intensity maps of $^{12}$CO emission toward L1448C from the CARMA
B and C-array data, overlaid on the CARMA 1.3 mm continuum map (grayscale). The contours are 
[-4, 4 6, 9, 12, 15, 18,...] $\times$ $\sigma$ and $\sigma$= 13.922 K, 15.39 K for the red 
and blue-shifted integrated intensity maps, respectively. The beam for the CO emission is 
0\farcs90 $\times$ 0\farcs75 and the beam for the continuum is 0\farcs58 $\times$ 0\farcs41. 
The blue and red arrows denote the direction of the blueshifted
and redshifted outflows, respectively.}
\label{L1448C-12CO}
\end{figure}

\begin{figure}
\begin{center}
\includegraphics[scale=0.6]{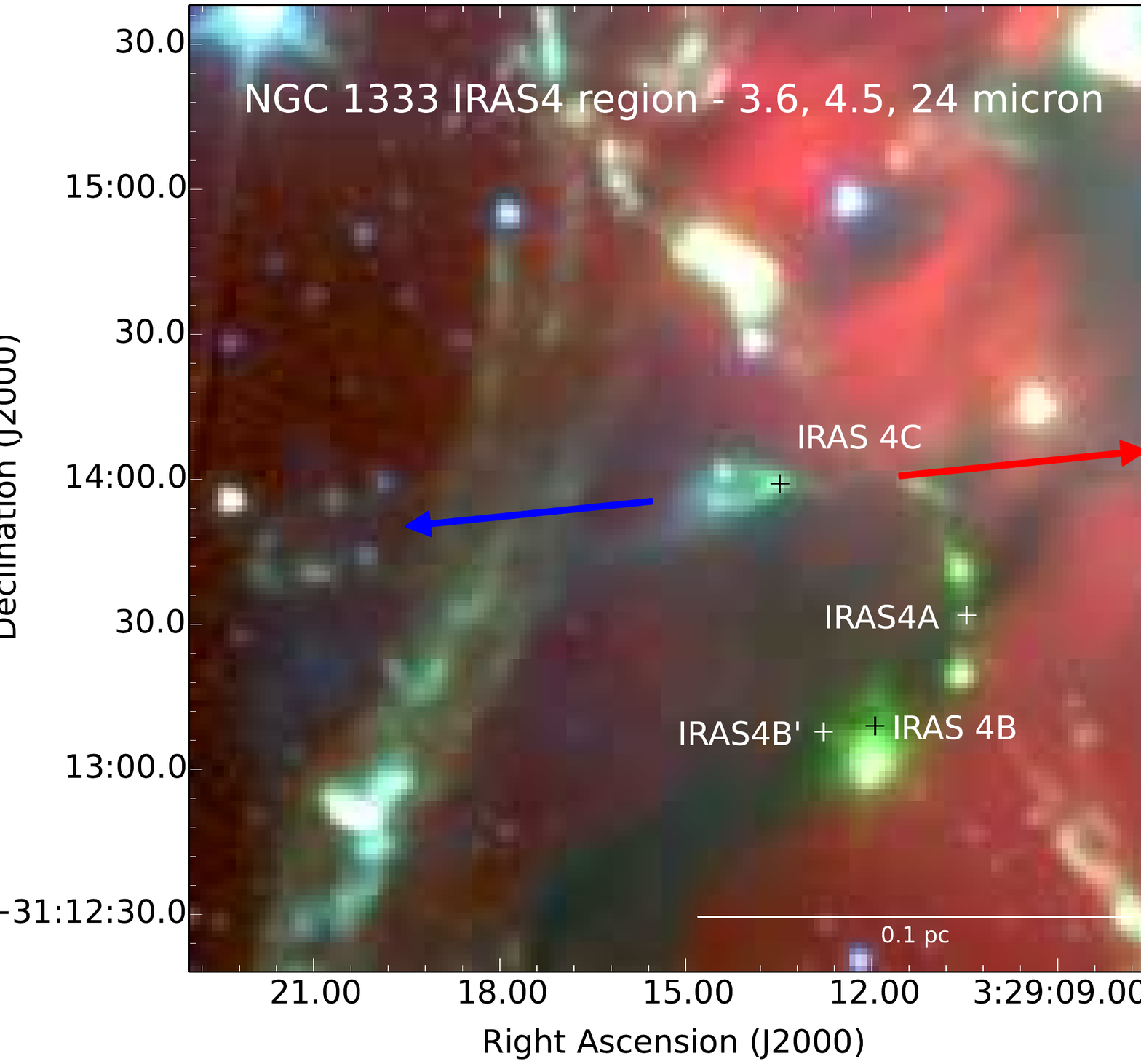}
\end{center}
\caption{\textit{Spitzer} IRAC 3.6 \micron, 4.5 \micron, and 8.0 \micron\ imaging of the NGC 1333 IRAS4 region. 
The blue and red vectors mark the outflow position angle adopted for Per-emb-14 (NGC 1333 IRAS4C).}
\label{IRAS4C-IRAC}
\end{figure}

\clearpage
\begin{deluxetable}{lllllllllll}
\tablewidth{0pt}
\rotate
\tabletypesize{\scriptsize}
\tablecaption{Source List}
\tablehead{
  \colhead{Source}  & \colhead{RA}      & \colhead{Dec}      & \colhead{Envelope Mass}   & \colhead{L$_{bol}$}    & \colhead{L$_{submm}$/L$_{bol}$}   & \colhead{T$_{bol}$}  & \colhead{Outflow PA} & \colhead{References} & \colhead{Other Names}\\
                    & \colhead{(J2000)} & \colhead{(J2000)}  & \colhead{($M_{\sun}$)}    & \colhead{($L_{\sun}$)} & \colhead{(\%)}            & \colhead{(K)}        & \colhead{(\degr)}    & \colhead{}  &\\

}
\startdata

L1451-MMS        & 3:25:10.25 & +30:23:55.0   & 0.36 (0.04) & $<$ 0.05    &     -       & $<$ 30 & 10 & 2, 8 & Per-bolo-2\\
L1448 IRS2       & 3:25:22.41 & +30:45:13.3   & 1.41 (0.14) & 3.6 (0.5) &  2.7 (0.4)  & 43 (2)  & 318 & 2, 3, 4, 5  & Per-emb-22\\
L1448 IRS3B      & 3:25:36.38 & +30:45:14.7   & 2.7 (0.27)  & 8.3 (0.8) &  3.4 (0.4)  & 57 (3)  & 285 & 2, 5, 6   & L1448 NB, Per-emb-33\\
L1448 IRS3C      & 3:25:35.67 & +30:45:34.1   & 1.1 (0.11)  & 1.4 (0.1) &  7.8 (0.9)  & 22 (1)  & 308 & 1, 2, 5   & L1448 NW\\
L1448 IRS3A      & 3:25:36.50 & +30:45:21.8   & 0.72 (0.07) & 4.3 (3.7) &     -       & 90 (18) & 155?   & 1, 2, 3, 6   & L1448 NA, Per-emb-33\\
L1448C           & 3:25:38.87 & +30:44:05.3   & 1.87 (0.19) & 9.2 (1.3) &  1.9 (0.3)  & 47 (2)  & 339 & 1, 2, 3, 5,11  & L1448-mm, Per-emb-26\\
L1448C-S         & 3:25:39.14 & +30:43:57.7  & 1.87 (0.19)\tablenotemark{a} & 0.68 (0.85) & - & 163 (51) & 40 & 2, 3, 10 & Per-emb-42\\
Per-emb-24\tablenotemark{\dagger}        & 3:28:45.31 & +31:05:41.6  & 0.19 (0.02) & 0.43 (0.01) & -         & 67 (10) & 101 & 1, 3 & IRAS 03256+3055\\
Per-emb-14       & 3:29:13.55 & +31:13:58.1  & 0.5 (0.05)  & 0.7 (0.08) & 6 (2)      & 31 (2)  & 95 & 1, 3, 5  & NGC 1333 IRAS4C\\
IRAS 03282+3035  & 3:31:20.94 & +30:45:30.2   & 0.96 (0.4)  & 1.3 (0.1) & 5.1 (0.7)  & 32 (2)  & 125 & 3, 5, 9 & Per-emb-5\\
IRAS 03292+3039  & 3:32:17.92 & +30:49:48.1   & 2.9 (0.1)   & 0.9 (0.1) & 9 (1)       & 27 (1)  & 127 & 2, 3, 7 & Per8, Per-emb-2\\
IC348-MMS        & 3:43:57.06 & +32:03:04.8  & 1.78 (0.04)  & 1.5 (0.1)  & 6.4 (0.9)  & 30 (2)  & 330 & 3, 5, 10 & Per-emb-11\\

\enddata
\tablecomments{References: (1) This work, (2) \citet{enoch2006}, (3) \citet{enoch2009}, (4) \citet{tobin2007}, 
(5) \citet{sadavoy2014}, (6) \citet{kwon2006}, (7) \citet{schnee2012}, (8) \citet{pineda2011},
(9) \citet{arce2006}, (10) \citet{pech2012}, (11) \citet{hirano2010}. 
The envelope mass for L1448 IRS3 is 4.47 $M_{\sun}$ as a whole
from the bolocam data and we assume that 60\% is in IRS3B, 16\% in IRS3A, and 24\% in IRS3C. These ratios come from the ratio of flux at 30 k$\lambda$,
for which the sources are all resolved from each other. }
\tablenotetext{\dagger}{This source is listed for the purposes of 
completeness only. It was a marginal detection in our first observation and 
we switched to Per-emb-14 for the remainder of our program.}
\tablenotetext{a}{Measurement combined with that of L1448C.}
\end{deluxetable}

\begin{deluxetable}{lllllllll}
\tablewidth{0pt}
\rotate
\tabletypesize{\scriptsize}
\tablecaption{CARMA 1.3 mm Observation Log}
\tablehead{
  \colhead{Sources} &\colhead{Config.}  & \colhead{Date} & \colhead{Track Length} & \colhead{Int. Time} & \colhead{Central Frequency} & \colhead{$\tau_{230GHz}$, RMS path, T$_{sys}$} & \colhead{Gain Calibrator/Flux Density} & \colhead{Flux Calibrator}\\
                    &                   & \colhead{(UT)} &  \colhead{(hrs)}       & \colhead{(mins)}    & \colhead{(GHz)}                  & \colhead{(nepers, $\mu$m, K)}          &  \colhead{(Name, Jy)}                  &
}
\startdata
%HH211 & C & 11 Dec 2009 & 7.9   & 5.0 &  90.175 & 0.3? &  0336+328, 1.6   &  Uranus      \\
%HH211 & B & 11 Dec 2009 & 7.9   & 5.0 &  90.175 & 0.3? &  0336+328, 1.6   &  Uranus      \\
%HH211 & A & 11 Dec 2009 & 7.9   & 5.0 &  90.175 & 0.3? &  0336+328, 1.6   &  Uranus      \\
%HH211 & D & 11 Dec 2009 & 7.9   & 5.0 &  90.175 & 0.3? &  0336+328, 1.6   &  Uranus      \\
L1448C & C & 11 Dec 2012 & 7.9   & 5.0 &  225.0491 & 0.21, 153, 350  &  0237+288, 1.6   &  Uranus      \\
L1448 IRS3A, IRS3B & ... & ... & ...  & ... &  ... & ... &  ...   &  ...     \\
L1448 IRS3C & ... & ... & ...  & ... & ... &  ...   &  ...  &  ...   \\
L1448 IRS2 & C & 04 Jan 2013 & 4.5   & 5.0 &  225.0491 & 0.13, 84, 300 &  0237+288, 1.6   &  Uranus      \\
L1451-MMS & ... & ... & ...  & ... &  ... & ... &  ...   &  ...     \\
Per-emb24 & ... & ... & ...   & ... & ... &  ...   &  ...   &  ...  \\
L1448 IRS2 & C & 05 Jan 2013 & 4.0   & 5.0 &  225.0491 & 0.23, 124, 400  &  0237+288, 1.6   &  Uranus      \\
L1451-MMS & ... & ... & ...  & ... &  ... & ... &  ...   &  ...     \\
Per-emb14 & ... & ... & ...   & ... & ... &  ...   &  ...  &  ...   \\
L1448C & B & 18 Jan 2013 & 7.5   & 3.0 &  225.0491 & 0.08, 95, 250 &  0237+288, 1.3   &  Uranus      \\
L1448 IRS3A, IRS3B & ... & ... & ...  &  ... & ... & ... &  ...       \\
L1448 IRS3C & ... & ... & ...   & ... & ... &  ...   &  ...     \\
0326+277 & ... & ... & ...  & 1.0 &  ... & ... &  ...   &  ...     \\
IRAS 03282+3035 & B & 20 Jan 2013 & 5.5   & 3.0 &  225.0491 & 0.075, 66, 200  &  0237+288, 1.3   &  Uranus      \\
IRAS 03292+3039 & ... & ... & ...  & ...  &  ... & ... &  ...   &  ...     \\
IC348-MMS & ... & ... & ...   & ... & ... &  ...   &  ...  &  ...   \\
0326+277 & ... & ... & ...  & 1.0 &  ... & ... &  ...   &  ...     \\
IRAS 03282+3035 & B & 21 Jan 2013 & 4.5   & 3.0 &  225.0491 & 0.08, 82, 250 &  0237+288, 1.3   &  Uranus      \\
IRAS 03292+3039 & ... & ... & ...  & ... &  ... & ... &  ...   &  ...     \\
IC348-MMS & ... & ... & ...   & ... & ... &  ...   &  ...  &  ...   \\
0326+277 & ... & ... & ...  & 1.0 &  ... & ... &  ...   &  ...     \\
L1448 IRS2 & B & 24 Jan 2013 & 7.0   & 3.0 &  225.0491 & 0.13, 58, 300  &  0237+288, 1.3   &  Uranus      \\
L1451-MMS & ... & ... & ...  & ... &  ... & ... &  ...   &  ...     \\
Per-emb-14 & ... & ... & ...  & ... & ... &  ...   &  ...  &  ...    \\
0326+277 & ... & ... & ...  & 1.0 &  ... & ... &  ...   &  ...     \\
Per-emb-14   & C & 16 Nov 2013 & 4.0   & 10.0 &  225.0491 & 0.35, 150, 600 &  0237+288, 1.3   &  Uranus      \\
Per-emb-14 & ... & 17 Nov 2013 & 4.0  & 10.0 & ... &  0.27, 142, 400   &  ...  &  ...    \\
IRAS 03282+3035 & C & 18 Nov 2013 & 4.5   & 5.0 &  225.0491 & 0.18, 138, 400 &  0237+288, 1.3   &  Uranus      \\
IRAS 03292+3039 & ... & ... & ...  & ... &  ... & ... &  ...   &  ...     \\
IC348-MMS & ... & ... & ...   & ... & ... &  ...   &  ...  &  ...   \\
\enddata
\end{deluxetable}

\begin{deluxetable}{lllllllll}
\tablewidth{0pt}
\rotate
\tabletypesize{\scriptsize}
\tablecaption{SMA Observation Log}
\tablehead{
  \colhead{Sources} &\colhead{Config.}  & \colhead{Date} & \colhead{Track Length} & \colhead{Antennas} & \colhead{Central Frequencies} & \colhead{T$_{sys}$} & \colhead{Gain Calibrator/Flux Density} & \colhead{Flux Calibrator}\\
                    &                   & \colhead{(UT)} &  \colhead{(hrs)}       & \colhead{}    & \colhead{(GHz)}                  & \colhead{(K)}          &  \colhead{(Name, Jy)}                  &
}
\startdata

L1448 IRS2 & Compact & 04 Nov 2007 & 10.0   & 7 &  225.434 & 300 &  3C84, 3.9   &  Uranus      \\
L1448 IRS2 & Compact & 06 Nov 2007 & 10.0   & 7 &  225.434 & 350 &  3C84, 3.9   &  Uranus      \\
L1448 IRS2 & Extended & 22 Sep 2013 & 9.0   & 5 &  225.434 (351.135)  & 190 (450)  &  3C84, 0237+288, 10.9, 1.4 (9.0, 1.3)   &  Uranus      \\
L1448 IRS2 & Extended & 28 Sep 2013 & 9.0   & 6 &  225.434 (351.135)   & 200 (525) &  3C84, 0237+288, 10.0, 1.4 (8.8, 1.25)    &  Uranus      \\

\enddata
\tablecomments{The SMA Extended configuration data were taken in dual 
receiver mode. The values in parentheses reflect the data 
for the 400 GHz receivers.}
\end{deluxetable}

\begin{deluxetable}{lllllllllll}
\tablewidth{0pt}
\rotate
\tabletypesize{\scriptsize}
\tablecaption{1.3 mm Continuum Properties}
\tablehead{
  \colhead{Source}  & \colhead{Resolution} & \colhead{Flux Density}      & \colhead{Peak Flux Density}   & \colhead{RMS Noise} & \colhead{Beam}       & \colhead{Beam PA} & \colhead{Deconvolved} & \colhead {Deconvolved} & \colhead {Outflow PA - }& \colhead{Mass} \\
  \colhead{}  & \colhead{} & \colhead{}      & \colhead{}   & \colhead{} & \colhead{}       & \colhead{} & \colhead{Size} & \colhead {PA} & \colhead{Continuum PA} & \colhead{} \\
                    &                      & \colhead{(mJy)}             & \colhead{(mJy beam$^{-1}$)}               & \colhead{(mJy beam$^{-1}$)}           & \colhead {(\arcsec)} & \colhead{(\degr)} & \colhead{(\arcsec)} & \colhead {(\degr)} & \colhead {(\degr)} & \colhead{($M_{\sun}$)}\\

}
\startdata

IC348MMS & low & 163.3 $\pm$ 3.6 & 75.5 & 1.00        & 0.44 $\times$ 0.32 & 107.87 & 0.40 $\times$ 0.35 & 78 $\pm$ 53  & 72 $\pm$ 53  & 0.104  $\pm$  0.002\\
IRAS 03282+3035 & low & 191.2 $\pm$ 3.4 & 99.3 & 1.10 & 0.44 $\times$ 0.32 & 106.22 & 0.39 $\times$ 0.30 & 24 $\pm$ 14  & 101 $\pm$ 14 & 0.121  $\pm$  0.002\\
L1448C & low & 160.4 $\pm$ 3.7 & 92.1 & 1.10          & 0.58 $\times$ 0.41 & 108.02 & 0.45 $\times$ 0.32 & 81 $\pm$ 19  & 78 $\pm$ 19  & 0.102  $\pm$  0.002\\
L1448 IRS2 & low & 77.9 $\pm$ 2.8 & 26.5 & 0.85       & 0.71 $\times$ 0.52 & 115.61 & 1.44 $\times$ 0.6  & 67 $\pm$ 5   & 71 $\pm$ 5  & 0.049  $\pm$  0.002\\
L1448 IRS3B & low & 751.8 $\pm$ 10.6 & 97.5 & 1.68    & 0.47 $\times$ 0.36 & 97.75  & 1.54  $\times$ 1.34& 70 $\pm$ 56  & 35 $\pm$ 56  & 0.477  $\pm$  0.007\\
L1448 IRS3C & low & 59.3 $\pm$ 2.9 & 35.2 & 1.26      & 0.58 $\times$ 0.41 & 107.76 & 0.35 $\times$ 0.12 & 23 $\pm$ 11  & 105 $\pm$ 11 & 0.038  $\pm$  0.002\\
L1451MMS & low & 27.0 $\pm$ 1.2 & 22.3 & 0.64         & 0.71 $\times$ 0.52 & 115.25 & 0.44 $\times$ 0.30 & 90 $\pm$ 24 & 90  & 0.017  $\pm$  0.001\\
IRAS 03292+3039 & low & 445.9 $\pm$ 5.6 & 48.7 & 1.07 & 0.44 $\times$ 0.32 & 106.88 & 1.2  $\times$ 0.88 & 176 $\pm$ 7  & 130 $\pm$ 7 & 0.283  $\pm$  0.004\\
Per-emb-14 & low & 56.8 $\pm$ 2.3 & 34.4 & 0.99       & 0.50 $\times$ 0.34 & 94.14  & 0.45 $\times$ 0.27 & 31 $\pm$ 13  & 64 $\pm$ 13  & 0.036  $\pm$  0.001\\
L1448 IRS3A & low & 74.1 $\pm$ 4.5 & 39.8 & 1.68      & 0.48 $\times$ 0.37 & 97.43  & 0.65 $\times$ 0.35 & 125 $\pm$ 7  & 30? $\pm$ 7 & 0.047  $\pm$  0.003\\
L1448C-S & low & 6.1 $\pm$ 1.3 & 7.0 & 1.10           & 0.58 $\times$ 0.41 & 108.02 & 0.53 $\times$ 0.39 & 79 $\pm$ 63  & 141 $\pm$ 63 & 0.004  $\pm$  0.001\\
\\
\hline
\\
IC348MMS & high & 105.8 $\pm$ 5.8 & 54.6 & 2.70       & 0.33 $\times$ 0.23 & 95.95  & 0.32 $\times$ 0.29   & 22 $\pm$ 82  & 128 $\pm$ 82 & 0.067  $\pm$  0.004\\
IRAS 03282+3035 & high & 149.6 $\pm$ 6.7 & 71.0 & 2.90 & 0.33 $\times$ 0.23 & 94.79  & 0.36 $\times$ 0.27  & 16 $\pm$ 29  & 109 $\pm$ 29 &  0.095  $\pm$  0.004\\
L1448C & high & 114.4 $\pm$ 3.1 & 77.7 & 1.04          & 0.39 $\times$ 0.28 & 100.59 & 0.23 $\times$ 0.16  & 76 $\pm$ 17  & 83 $\pm$ 17  & 0.073  $\pm$  0.002\\
L1448 IRS2 & high & 28.9 $\pm$ 2.6 & 13.7 & 1.28       & 0.42 $\times$ 0.29 & 96.42  & 0.86 $\times$ 0.26  & 70 $\pm$ 6   & 68 $\pm$ 6  & 0.018  $\pm$  0.002\\
L1448 IRS3B & high & 152.6 $\pm$ 6.2 & 54.3 & 1.4     & 0.33 $\times$ 0.23 & 95.46  & 0.34 $\times$ 0.29  & 54 $\pm$ 81  & 51 $\pm$ 81  & 0.097  $\pm$  0.004\\
L1448 IRS3C & high & 33.3 $\pm$ 3.1 & 23.5 & 1.80      & 0.33 $\times$ 0.22 & 95.40  & 0.33 $\times$ 0.06  & 29 $\pm$ 10  & 99 $\pm$ 10  & 0.021  $\pm$  0.002\\
L1451MMS & high & 17.5 $\pm$ 1.5 & 17.2 & 0.99        & 0.37 $\times$ 0.26 & 90.33  & point                & -   & -      &  0.011  $\pm$  0.001\\
IRAS 03292+3039 & high & 301.2 $\pm$ 7.6 & 27.2 & 1.70 & 0.34 $\times$ 0.23 & 95.58  & 1.1 $\times$  0.74  & 175 $\pm$ 7  & 131 $\pm$ 7  & 0.191  $\pm$  0.005\\
Per-emb-14 & high & 45.6 $\pm$ 2.6 & 26.3 & 1.25       & 0.35 $\times$ 0.24 & 269.46 & 0.40 $\times$ 0.1   & 25 $\pm$ 8   & 70 $\pm$ 8  & 0.029  $\pm$  0.002\\
L1448 IRS3A & high & 55.8 $\pm$ 3.5 & 29.0 & 1.4      & 0.34 $\times$ 0.23 & 94.82  & 0.51 $\times$ 0.22  & 137 $\pm$ 10 & 18? $\pm$ 12 & 0.035  $\pm$  0.002\\
L1448C-S & high & 7.5 $\pm$ 1.4 & 6.7 & 1.04           & 0.39 $\times$ 0.28 & 100.59 & point  & - & - & 0.005  $\pm$  0.001\\
\enddata
\tablecomments{Uncertainties in the flux densities and mass measurements are statistical only and do not take into account
the $\sim$20\% uncertainty in the absolute flux calibration. Deconvolved sizes and position angles are determined using the \textit{imfit} task in CASA. The
error in the relative angle between the deconvolved source position angle and outflow position angle assumes no error in the outflow position angle. }
\end{deluxetable}

\begin{deluxetable}{llllll}
\tablewidth{0pt}
\rotate
\tabletypesize{\scriptsize}
\tablecaption{1.3 mm Continuum at 100 k$\lambda$ and 50 k$\lambda$}
\tablehead{
  \colhead{Source}  & \colhead{100 k$\lambda$ Flux Density} &\colhead{100 k$\lambda$ Mass} & \colhead{50 k$\lambda$ Flux Density} &\colhead{50 k$\lambda$ Mass} &\colhead{Bolocam Flux\tablenotemark{a}}  \\
                    & \colhead{(mJy)} & \colhead{($M_{\sun}$)} & \colhead{(mJy)} & \colhead{($M_{\sun}$)} & \colhead{(mJy)}\\

}
\startdata
IRAS03282 & 162.0 $\pm$ 3.5 & 0.103 (0.098) $\pm$ 0.002 & 205.5 $\pm$ 5.1 & 0.130 (0.121) $\pm$ 0.003 & 540 $\pm$ 20\\
IC348MMS & 117.0 $\pm$ 3.7 & 0.074 (0.063) $\pm$ 0.002 & 166.5 $\pm$ 5.1 & 0.106 (0.083) $\pm$ 0.003 & 1010 $\pm$ 20\\
IRAS03292 & 156.5 $\pm$ 2.6 & 0.099 (0.080) $\pm$ 0.002 & 379.5 $\pm$ 5.3 & 0.241 (0.207) $\pm$ 0.003 & 1650 $\pm$ 40\\
L1448C & 111.0 $\pm$ 2.4 & 0.070 (0.053) $\pm$ 0.002 & 139.5 $\pm$ 2.9 & 0.088 (0.054) $\pm$ 0.002 & 1450 $\pm$ 40\\
L1448IRS3A & 38.4 $\pm$ 2.4 & 0.024 (0.018) $\pm$ 0.002 & 56.2 $\pm$ 3.0 & 0.036 (0.022) $\pm$ 0.002 & 560\tablenotemark{b} $\pm$70\\
L1448IRS3B & 112.3 $\pm$ 2.4 & 0.071 (0.047) $\pm$ 0.002 & 378.5 $\pm$ 3.0 & 0.240 (0.197) $\pm$ 0.002 & 2010\tablenotemark{b} $\pm$ 70\\
L1448IRS3C & 30.2 $\pm$ 2.4 & 0.019 (0.009) $\pm$ 0.001 & 46.9 $\pm$ 3.0 & 0.030 (0.010) $\pm$ 0.002 & 810\tablenotemark{b} $\pm$ 70\\
L1448IRS2 & 31.1 $\pm$ 1.8 & 0.020 (0.001) $\pm$ 0.001 & 59.2 $\pm$ 2.5 & 0.038 (0.0004) $\pm$ 0.002 & 1460 $\pm$ 50\\
L1451MMS & 18.4 $\pm$ 1.7 & 0.012 (0.011) $\pm$ 0.001 & 24.8 $\pm$ 1.4 & 0.016 (0.014) $\pm$ 0.001 & 110 $\pm$ 10\\
Per-emb-14 & 54.6 $\pm$ 2.2 & 0.035 (0.033) $\pm$ 0.001 & 67.9 $\pm$ 2.8 & 0.043 (0.041) $\pm$ 0.002 & 150\tablenotemark{b} $\pm$ 70\\
\enddata
\tablecomments{The mass values in parentheses are meant to reflect the estimated disk masses after removing the contribution
of envelope flux,estimated using the method of \citet{jorgensen2009}. Using this method, the envelope contribution
to the flux at 50 k$\lambda$ and 100 k$\lambda$ is $\sim$4\% and $\sim$ 2\%, respectively, and assuming a envelope density profile
with $\rho$ $\propto$ R$^{-1.5}$. These percentages are for the total envelope flux density as measured with the Bolocam
data from \citet{enoch2006}. The uncertainties in the flux densities and mass measurements are statistical only and do not take into account
the $\sim$20\% uncertainty in the absolute flux calibration. }
\tablenotetext{a}{Bolocam fluxes are scaled by (1.1 mm/1.3 mm)$^{\beta}$ = 0.71, assuming $\beta$ = 1.78 \citep{ossenkopf1994}, to better match the observed wavelength.}
\tablenotetext{b}{Estimated flux densities due to blended core.}
\end{deluxetable}

\end{document}